\documentclass[pdftex,twocolumn,epjc3_preprint,runningheads]{svjour3}

\usepackage[T1]{fontenc}
\usepackage{lmodern}
\usepackage{calc}
\usepackage{graphicx}
\usepackage{booktabs}
\usepackage{textcomp}
\usepackage{xspace}
\usepackage{relsize}
\usepackage{amssymb}
\usepackage{amsmath}
\usepackage{listings}
\usepackage{microtype}
\usepackage{multirow}
\usepackage{tabularx}
\usepackage{array}
\usepackage{placeins}
\usepackage{cuted}
\usepackage{soul} 
\usepackage{fixltx2e}
\usepackage[numbers,sort&compress]{natbib}
\usepackage[labelfont=bf,font=small]{caption}
\usepackage[skip=-2pt]{subcaption}
\usepackage[colorlinks,citecolor=blue,urlcolor=blue,linkcolor=blue,breaklinks=breakall]{hyperref}
\usepackage{breakurl}
\usepackage[dvipsnames]{xcolor}
\usepackage[clockwise,figuresright]{rotating}
\usepackage{siunitx}
\usepackage{tikz}

\usepackage{etoolbox}
\AfterEndEnvironment{strip}{\leavevmode}

\allowdisplaybreaks

\newcolumntype{L}{>{\raggedright\let\newline\\\arraybackslash\hspace{0pt}}X}
\newcolumntype{R}{>{\raggedleft\let\newline\\\arraybackslash\hspace{0pt}}X}
\newcolumntype{C}{>{\centering\let\newline\\\arraybackslash\hspace{0pt}}X}

\newcommand{\imperial}{Department of Physics, Imperial College London, Blackett Laboratory, Prince Consort Road, London SW7 2AZ, UK}
\newcommand{\nordita}{NORDITA, Roslagstullsbacken 23, SE-10691 Stockholm, Sweden}
\newcommand{\oslo}{Department of Physics, University of Oslo, N-0316 Oslo, Norway}
\newcommand{\adelaide}{Department of Physics, University of Adelaide, Adelaide, SA 5005, Australia}
\newcommand{\glasgow}{SUPA, School of Physics and Astronomy, University of Glasgow, Glasgow, G12 8QQ, UK}
\newcommand{\monash}{School of Physics and Astronomy, Monash University, Melbourne, VIC 3800, Australia}
\newcommand{\coepp}{Australian Research Council Centre of Excellence for Particle Physics at the Tera-scale}
\newcommand{\okc}{Oskar Klein Centre for Cosmoparticle Physics, AlbaNova University Centre, SE-10691 Stockholm, Sweden}
\newcommand{\su}{Department of Physics, Stockholm University, SE-10691 Stockholm, Sweden}
\newcommand{\mcgill}{Department of Physics, McGill University, 3600 rue University, Montr\'eal, Qu\'ebec H3A 2T8, Canada}
\newcommand{\ucla}{Physics and Astronomy Department, University of California, Los Angeles, CA 90095, USA}
\newcommand{\annecy}{LAPTh, Universit\'e de Savoie, CNRS, 9 chemin de Bellevue B.P.110, F-74941 Annecy-le-Vieux, France}
\newcommand{\harvard}{Department of Physics, Harvard University, Cambridge, MA 02138, USA}
\newcommand{\grappa}{GRAPPA, Institute of Physics, University of Amsterdam, Science Park 904, 1098 XH Amsterdam, Netherlands}
\newcommand{\sydney}{Centre for Translational Data Science, Faculty of Engineering and Information Technologies, School of Physics, The University of Sydney, NSW 2006, Australia}

\newcommand{\cernth}{Theoretical Physics Department, CERN, CH-1211 Geneva 23, Switzerland}
\newcommand{\lyon}{Univ Lyon, Univ Lyon 1, ENS de Lyon, CNRS, Centre de Recherche Astrophysique de Lyon UMR5574, F-69230 Saint-Genis-Laval, France}
\newcommand{\iuf}{Institut Universitaire de France, 103 boulevard Saint-Michel, 75005 Paris, France}
\newcommand{\zurich}{Physik-Institut, Universit\"at Z\"urich, Winterthurerstrasse 190, 8057 Z\"urich, Switzerland}
\newcommand{\krakow}{H.~Niewodnicza\'nski Institute of Nuclear Physics, Polish Academy of Sciences, 31-342  Krak\'ow, Poland}

\newcommand{\gambitacknospmare}{We warmly thank the Casa Matem\'aticas Oaxaca, affiliated with the Banff International Research Station, for hospitality whilst part of this work was completed, and the staff at Cyfronet, for their always helpful supercomputing support.  \GB has been supported by STFC (UK; ST/K00414X/1, ST/P000762/1), the Royal Society (UK; UF110191), Glasgow University (UK; Leadership Fellowship), the Research Council of Norway (FRIPRO 230546/F20), NOTUR (Norway; NN9284K), the Knut and Alice Wallenberg Foundation (Sweden; Wallenberg Academy Fellowship), the Swedish Research Council (621-2014-5772), the Australian Research Council (CE110001004, FT130100018, FT140100244, FT160100274), The University of Sydney (Australia; IRCA-G162448), PLGrid Infrastructure (Poland), Red Espa\~nola de Supercomputaci\'on (Spain; FI-2016-1-0021), Polish National Science Center (Sonata UMO-2015/17/D/ST2/03532), the Swiss National Science Foundation (PP00P2-144674), the European Commission Horizon 2020 Marie Sk\l{}odowska-Curie actions (H2020-MSCA-RISE-2015-691164), the ERA-CAN+ Twinning Program (EU \& Canada), the Netherlands Organisation for Scientific Research (NWO-Vidi 680-47-532), the National Science Foundation (USA; DGE-1339067), the FRQNT (Qu\'ebec) and NSERC/The Canadian Tri-Agencies Research Councils (BPDF-424460-2012).}

\makeatletter

\newcommand{\preprintnumber}[1]{\gdef\@preprintnumber{\begin{flushright}{#1}\end{flushright}}}

\g@addto@macro\bfseries{\boldmath}
\makeatother

\newcommand{\subparagraph}{} 
\usepackage{titlesec}
\titleformat*{\paragraph}{\bfseries}

\journalname{Eur. Phys. J. C}
\smartqed
\let\underscore\_
\renewcommand{\_}{\discretionary{\underscore}{}{\underscore}}

\makeatletter
\let\orgdescriptionlabel\descriptionlabel
\renewcommand*{\descriptionlabel}[1]{%
  \let\orglabel\label
  \let\label\@gobble
  \phantomsection
  \protected@edef\@currentlabel{#1}%
  \let\label\orglabel
  \orgdescriptionlabel{#1}%
}
\makeatother

\newcommand\postnewlinemarker{\hbox{\ensuremath{\hookrightarrow}}}
\newcommand\cpp[1]{{\lstinline!#1!}}  

\newcommand\yaml[1]{{\lstset{style=yaml}\lstinline!#1!\lstset{style=cpp}}}

\newcommand\term[1]{{\lstset{style=terminal}\lstinline!#1!\lstset{style=cpp}}}
\newcommand\fortran[1]{{\lstset{style=fortran}\lstinline!#1!\lstset{style=cpp}}}
\newcommand\py[1]{{\lstset{style=python}\lstinline!#1!\lstset{style=cpp}}}
\newcommand\customtilde{{\raisebox{0.2ex}{\scalebox{0.6}{\boldmath$\sim$}}}}
\newcommand\mathematica[1]{{\lstset{style=Mathematica}\lstinline!#1!\lstset{style=cpp}}}

\lstnewenvironment{lstlistingyaml}{\lstset{style=yaml}}{\lstset{style=cpp}}
\lstnewenvironment{lstlistingterm}{\lstset{style=terminal}}{\lstset{style=cpp}}
\lstnewenvironment{lstlistingfortran}{\lstset{style=fortran}}{\lstset{style=cpp}}
\lstnewenvironment{lstcpp}{\lstset{style=cpp}}{\lstset{style=cpp}}
\lstnewenvironment{lstcppalt}{\lstset{style=cppalt}}{\lstset{style=cpp}}
\lstnewenvironment{lstcppnum}{\lstset{style=cppnum}}{\lstset{style=cpp}}
\lstnewenvironment{lstyaml}{\lstset{style=yaml}}{\lstset{style=cpp}}
\lstnewenvironment{lstterm}{\lstset{style=terminal}}{\lstset{style=cpp}}
\lstnewenvironment{lsttermalt}{\lstset{style=terminalalt}}{\lstset{style=cpp}}
\lstnewenvironment{lsttext}{\lstset{style=text}}{\lstset{style=cpp}}
\lstnewenvironment{lstfortran}{\lstset{style=fortran}}{\lstset{style=cpp}}
\lstnewenvironment{lstpy}{\lstset{style=python}}{\lstset{style=cpp}}
\lstnewenvironment{lstmathematica}{\lstset{style=mathematica}}{\lstset{style=cpp}}

\newcommand{\tmpname}{}
\newcommand{\tmplistingname}{}
\makeatletter
\newif\ifATOlabelname
\lst@Key{labelname}{Listing}{\def\ATOlabelname{#1}\global\ATOlabelnametrue}
\makeatother
\lstnewenvironment{lstcpplabel}[1][]{
  \lstset{style=cpp,#1} 
  \ifATOlabelname
    \renewcommand{\tmpname}{\lstlistingname}
    \renewcommand{\tmplistingname}{\lstlistlistingname}
    \renewcommand{\lstlistingname}{\ATOlabelname}
    \renewcommand{\lstlistlistingname}{List of \lstlistingname s}
  \fi
}{
  \renewcommand{\lstlistingname}{\tmpname}
  \renewcommand{\lstlistlistingname}{\tmplistingname}
  \lstset{style=cpp}
}
\definecolor{solarized@base03}{HTML}{002B36}
\definecolor{solarized@base02}{HTML}{073642}
\definecolor{solarized@base01}{HTML}{586e75}
\definecolor{solarized@base00}{HTML}{657b83}
\definecolor{solarized@base0}{HTML}{839496}
\definecolor{solarized@base1}{HTML}{93a1a1}
\definecolor{solarized@base2}{HTML}{EEE8D5}
\definecolor{solarized@base3}{HTML}{FDF6E3}
\definecolor{solarized@yellow}{HTML}{B58900}
\definecolor{solarized@orange}{HTML}{CB4B16}
\definecolor{solarized@red}{HTML}{DC322F}
\definecolor{solarized@magenta}{HTML}{D33682}
\definecolor{solarized@violet}{HTML}{6C71C4}
\definecolor{solarized@blue}{HTML}{268BD2}
\definecolor{solarized@cyan}{HTML}{2AA198}
\definecolor{solarized@green}{HTML}{859900}
\definecolor{darkred}{HTML}{550003}
\definecolor{darkgreen}{HTML}{00AA00}

\newcommand\YAMLstringstyle{\footnotesize\color{solarized@green}\mdseries}
\newcommand\YAMLkeystyle{\footnotesize\color{solarized@blue}\ttfamily}
\newcommand\YAMLvaluestyle{\footnotesize\color{blue}\mdseries}
\newcommand\ProcessThreeDashes{\llap{\color{cyan}\mdseries-{-}-}}

\newcommand\CPPcommentstyle{\color{solarized@violet}\footnotesize\ttfamily}
\newcommand\CPPdirectivestyle{\color{solarized@magenta}\footnotesize\ttfamily}
\newcommand\termplainstyle{\footnotesize\ttfamily}

\newcommand\processLongMacroDelimiter
{%
\CPPdirectivestyle%
\#define%
}

\lstdefinestyle{cpp}
{
  language=C++,
  basicstyle=\footnotesize\ttfamily,
  basewidth={0.53em,0.44em}, 
  numbers=none,
  tabsize=2,
  breaklines=true,
  escapeinside={@}{@},
  showstringspaces=false,
  numberstyle=\tiny\color{solarized@base01},
  keywordstyle=\color{solarized@orange},
  stringstyle=\color{solarized@red}\ttfamily,
  identifierstyle=\color{solarized@blue},
  commentstyle=\CPPcommentstyle,
  directivestyle=\CPPdirectivestyle,
  emphstyle=\color{solarized@green},
  frame=single,
  rulecolor=\color{solarized@base2},
  rulesepcolor=\color{solarized@base2},
  literate={~} {\customtilde}1,
  moredelim=*[directive]\ \ \#,
  moredelim=*[directive]\ \ \ \ \#
}

\lstdefinestyle{cppalt}
{
  language=C++,
  basicstyle=\footnotesize\ttfamily,
  basewidth={0.53em,0.44em}, 
  numbers=none,
  tabsize=2,
  breaklines=true,
  escapeinside={*@}{@*},
  showstringspaces=false,
  numberstyle=\tiny\color{solarized@base01},
  keywordstyle=\color{solarized@orange},
  stringstyle=\color{solarized@red}\ttfamily,
  identifierstyle=\color{solarized@blue},
  commentstyle=\CPPcommentstyle,
  directivestyle=\CPPdirectivestyle,
  emphstyle=\color{solarized@green},
  frame=single,
  rulecolor=\color{solarized@base2},
  rulesepcolor=\color{solarized@base2},
  literate={~}{\customtilde}1,
  moredelim=**[is][\processLongMacroDelimiter]{BeginLongMacro}{EndLongMacro} 
}

\lstdefinestyle{cppnum}
{
  language=C++,
  basicstyle=\footnotesize\ttfamily,
  basewidth={0.53em,0.44em}, 
  numbers=none,
  tabsize=2,
  breaklines=true,
  escapeinside={@}{@},
  numberstyle=\tiny\color{solarized@base01},
  showstringspaces=false,
  numberstyle=\tiny\color{solarized@base01},
  keywordstyle=\color{solarized@orange},
  stringstyle=\color{solarized@red}\ttfamily,
  identifierstyle=\color{solarized@blue},
  commentstyle=\CPPcommentstyle,
  directivestyle=\CPPdirectivestyle,
  emphstyle=\color{solarized@green},
  frame=single,
  rulecolor=\color{solarized@base2},
  rulesepcolor=\color{solarized@base2},
  literate={~} {\customtilde}1,
  moredelim=*[directive]\ \ \#,
  moredelim=*[directive]\ \ \ \ \#
}

\lstdefinestyle{python}
{
  language=Python,
  basicstyle=\footnotesize\ttfamily,
  basewidth={0.53em,0.44em},
  numbers=none,
  tabsize=2,
  breaklines=true,
  escapeinside={@}{@},
  showstringspaces=false,
  numberstyle=\tiny\color{solarized@base01},
  keywordstyle=\color{blue},
  stringstyle=\color{orange}\ttfamily,
  identifierstyle=\color{darkred},
  commentstyle=\color{purple},
  emphstyle=\color{green},
  frame=single,
  rulecolor=\color{solarized@base2},
  rulesepcolor=\color{solarized@base2},
  literate = {~}{\customtilde}1
             {\ as\ }{{\color{blue}\ as\ \color{black}}}3
}

\lstdefinestyle{fortran}
{
  language=Fortran,
  basicstyle=\footnotesize\ttfamily,
  basewidth={0.53em,0.44em},
  numbers=none,
  tabsize=2,
  breaklines=true,
  escapeinside={@}{@},
  showstringspaces=false,
  numberstyle=\tiny\color{solarized@base01},
  keywordstyle=\color{blue},
  stringstyle=\color{orange}\ttfamily,
  identifierstyle=\color{Periwinkle},
  commentstyle=\color{purple},
  emphstyle=\color{green},
  morekeywords={and, or, true, false},
  frame=single,
  rulecolor=\color{solarized@base2},
  rulesepcolor=\color{solarized@base2},
  literate={~}{\customtilde}1
}

\lstdefinestyle{terminal}
{
  language=bash,
  basicstyle=\termplainstyle,
  numbers=none,
  tabsize=2,
  breaklines=true,
  escapeinside={@}{@},
  frame=single,
  showstringspaces=false,
  numberstyle=\tiny\color{solarized@base01},
  keywordstyle=\color{solarized@orange},
  stringstyle=\color{solarized@red}\ttfamily,
  identifierstyle=\color{black},
  commentstyle=\color{solarized@violet},
  emphstyle=\color{solarized@green},
  frame=single,
  rulecolor=\color{solarized@base2},
  rulesepcolor=\color{solarized@base2},
  morekeywords={gambit, cmake, make, mkdir},
  deletekeywords={test},
  literate = {\ gambit}{{\ }{\color{black}}gambit}7
             {/gambit}{{/}{\color{black}}gambit}6
             {gambit/}{{\color{black}}gambit{/}}6
             {/include}{{/}{\color{black}}include}8
             {cmake/}{{\color{black}}cmake/}6
             {.cmake}{{.}{\color{black}}cmake}6
             {~}{\customtilde}1
}

\lstdefinestyle{terminalalt}
{
  language=bash,
  basicstyle=\footnotesize\ttfamily,
  numbers=none,
  tabsize=2,
  breaklines=true,
  escapeinside={*@}{@*},
  frame=single,
  showstringspaces=false,
  numberstyle=\tiny\color{solarized@base01},
  keywordstyle=\color{solarized@orange},
  stringstyle=\color{solarized@red}\ttfamily,
  identifierstyle=\color{black},
  commentstyle=\color{solarized@violet},
  emphstyle=\color{solarized@green},
  frame=single,
  rulecolor=\color{solarized@base2},
  rulesepcolor=\color{solarized@base2},
  morekeywords={gambit, cmake, make, mkdir},
  deletekeywords={test},
  literate = {\ gambit}{{\ }{\color{black}}gambit}7
             {/gambit}{{/}{\color{black}}gambit}6
             {gambit/}{{\color{black}}gambit{/}}6
             {/include}{{/}{\color{black}}include}8
             {cmake/}{{\color{black}}cmake/}6
             {.cmake}{{.}{\color{black}}cmake}6
             {~}{\customtilde}1
}

\lstdefinestyle{text}
{
  language={},
  basicstyle=\footnotesize\ttfamily,
  identifierstyle=\color{black},
  numbers=none,
  tabsize=2,
  breaklines=true,
  escapeinside={*@}{@*},
  showstringspaces=false,
  frame=single,
  rulecolor=\color{solarized@base2},
  rulesepcolor=\color{solarized@base2},
  literate={~}{\customtilde}1
}

\lstdefinestyle{yaml}
{
  language=bash,
  escapeinside={@}{@},
  keywords={true,false,null},
  otherkeywords={},
  keywordstyle=\color{solarized@base0}\bfseries,
  basicstyle=\footnotesize\color{black}\ttfamily,
  identifierstyle=\YAMLkeystyle,
  sensitive=false,
  commentstyle=\color{solarized@orange}\ttfamily,
  morecomment=[l]{\#},
  morecomment=[s]{/*}{*/},
  stringstyle=\YAMLstringstyle\ttfamily,
  moredelim=**[s][\YAMLkeystyle]{,}{:},   
  moredelim=**[l][\YAMLvaluestyle]{:},    
  morestring=[b]',
  morestring=[b]",
  literate =    {---}{{\ProcessThreeDashes}}3
                {>}{{\textcolor{solarized@red}\textgreater}}1
                {|}{{\textcolor{solarized@red}\textbar}}1
                {\ -\ }{{\mdseries\color{black}\ -\ \negmedspace}}3
                {\}}{{{\color{black} \}}}}1
                {\{}{{{\color{black} \{}}}1
                {[}{{{\color{black} [}}}1
                {]}{{{\color{black} ]}}}1
                {~}{\customtilde}1,
  breakindent=0pt,
  breakatwhitespace,
  columns=fullflexible
}

\lstdefinestyle{mathematica}
{
  language={Mathematica},
  basicstyle=\footnotesize\ttfamily,
  basewidth={0.53em,0.44em},
  numbers=none,
  tabsize=2,
  breaklines=true,
  escapeinside={@}{@},
  numberstyle=\tiny\color{black},
  showstringspaces=false,
  numberstyle=\tiny\color{solarized@base01},
  keywordstyle=\color{solarized@orange},
  stringstyle=\color{solarized@red}\ttfamily,
  identifierstyle=\color{solarized@orange}\ttfamily,
  commentstyle=\color{solarized@gray}\ttfamily,
  directivestyle=\color{solarized@orange}\ttfamily,
  emphstyle=\color{solarized@green},
  frame=single,
  rulecolor=\color{solarized@base2},
  rulesepcolor=\color{solarized@base2},
  literate={~} {\customtilde}1,
  moredelim=*[directive]\ \ \#,
  moredelim=*[directive]\ \ \ \ \#,
  mathescape=true
}

\newcommand{\doublecross}[2]{\hyperref[#2]{\textbf{#1}}}
\newcommand{\doublecrosssf}[2]{\hyperref[#2]{\textbf{\textsf{#1}}}}

\newcommand{\startglossary}{\section{Glossary}\label{glossary}Here we explain some terms that have specific technical definitions in \GB.\begin{description}}
\newcommand{\finishglossary}{\end{description}}

\newcommand{\bcode}{\begin{lstlisting}}
\newcommand{\ecode}{\end{lstlisting}}

\newcommand{\eV}{\ensuremath{\text{e}\mspace{-0.8mu}\text{V}}\xspace}

\newcommand{\GeV}{\text{G\eV}\xspace}
\newcommand{\TeV}{\text{T\eV}\xspace}

\newcommand{\MSbar}{$\MSBar$\xspace}
\newcommand{\MSBar}{\overline{MS}}

\newcommand{\gambit}{\textsf{GAMBIT}\xspace}

\newcommand{\colliderbit}{\textsf{ColliderBit}\xspace}
\newcommand{\flavbit}{\textsf{FlavBit}\xspace}

\newcommand{\scannerbit}{\textsf{ScannerBit}\xspace}

\newcommand{\GB}{\gambit}

\newcommand{\pythiaeight}{\textsf{Pythia\,8}\xspace}

\newcommand{\higgsbounds}{\textsf{HiggsBounds}\xspace}

\newcommand{\higgssignals}{\textsf{HiggsSignals}\xspace}
\newcommand{\ds}{\textsf{DarkSUSY}\xspace}
\newcommand{\darksusy}{\ds}

\newcommand{\micromegas}{\textsf{micrOMEGAs}\xspace}

\newcommand\flexiblesusy{\FlexibleSUSY}
\newcommand\FlexibleSUSY{\textsf{FlexibleSUSY}\xspace}

\newcommand\SOFTSUSY{\textsf{SOFTSUSY}\xspace}

\newcommand\SUSYHIT{\textsf{SUSY-HIT}\xspace}

\newcommand\susyhit{\SUSYHIT}
\newcommand\gmtwocalc{\textsf{GM2Calc}\xspace}
\newcommand\SARAH{\textsf{SARAH}\xspace}

\newcommand\superiso{\textsf{SuperIso}\xspace}

\newcommand\nulike{\textsf{nulike}\xspace}
\newcommand\gamLike{\textsf{gamLike}\xspace}
\newcommand\gamlike{\gamLike}

\newcommand\MultiNest{\textsf{MultiNest}\xspace}
\newcommand\multinest{\MultiNest}

\newcommand\diver{\textsf{Diver}\xspace}
\newcommand\ddcalc{\textsf{DDCalc}\xspace}

\newcommand\beq{\begin{equation}}
\newcommand\eeq{\end{equation}}

\renewcommand{\url}[1]{\href{#1}{#1}}

\newcommand\msf{m_{\tilde{f}}}
\newcommand\msfsq{m_{\tilde{f}}^2}
\newcommand\MHusq{M_{H_u}^2}
\newcommand\MHdsq{M_{H_d}^2}

\begin{document}

\preprintnumber{CERN-TH-2017-169, CoEPP-MN-17-11, NORDITA 2017-081}

\title{A global fit of the MSSM with GAMBIT}

\author
{
The GAMBIT Collaboration:
Peter Athron\thanksref{inst:a,inst:b,e1} \and
Csaba Bal\'azs\thanksref{inst:a,inst:b} \and
Torsten Bringmann\thanksref{inst:c} \and
Andy Buckley\thanksref{inst:d} \and
Marcin Chrz\k{a}szcz\thanksref{inst:e,inst:f} \and
Jan Conrad\thanksref{inst:g,inst:h} \and
Jonathan M.~Cornell\thanksref{inst:i} \and
Lars A.~Dal\thanksref{inst:c} \and
Joakim Edsj\"o\thanksref{inst:g,inst:h} \and
Ben Farmer\thanksref{inst:g,inst:h} \and
Paul Jackson\thanksref{inst:k,inst:b} \and
Abram Krislock\thanksref{inst:c} \and
Anders Kvellestad\thanksref{inst:m,e2} \and
Farvah Mahmoudi\thanksref{inst:n,inst:o,e5} \and
Gregory D.\ Martinez\thanksref{inst:p} \and
Antje Putze\thanksref{inst:r} \and
Are Raklev\thanksref{inst:c} \and
Christopher Rogan\thanksref{inst:s} \and
Aldo Saavedra\thanksref{inst:t,inst:b} \and
Christopher Savage\thanksref{inst:m} \and
Pat Scott\thanksref{inst:q,e3} \and
Nicola Serra\thanksref{inst:e} \and
Christoph Weniger\thanksref{inst:u} \and
Martin White\thanksref{inst:k,inst:b,e4}
}

\institute{%
  \monash\label{inst:a} \and
  \coepp\label{inst:b} \and
  \oslo\label{inst:c} \and
  \glasgow\label{inst:d} \and
  \zurich\label{inst:e} \and
  \krakow\label{inst:f} \and
  \okc\label{inst:g} \and
  \su\label{inst:h} \and
  \mcgill\label{inst:i} \and
  \adelaide\label{inst:k} \and
  \nordita\label{inst:m} \and
  \lyon\label{inst:n} \and
  \cernth\label{inst:o} \and
  \ucla\label{inst:p} \and
  \annecy\label{inst:r} \and
  \harvard\label{inst:s} \and
  \sydney\label{inst:t} \and
  \imperial\label{inst:q} \and
  \grappa\label{inst:u}
}

\thankstext{e1}{peter.athron@coepp.org.au}
\thankstext{e2}{anders.kvellestad@nordita.org}
\thankstext{e3}{p.scott@imperial.ac.uk}
\thankstext{e4}{martin.white@adelaide.edu.au}
\thankstext[*]{e5}{Also \iuf.}

\titlerunning{MSSM7 global fit with GAMBIT}
\authorrunning{The GAMBIT Collaboration}

\date{Received: date / Accepted: date}

\maketitle

\begin{abstract}
We study the seven-dimensional Minimal Supersymmetric Standard Model (MSSM7) with the new \gambit software framework, with all parameters defined at the weak scale. Our analysis significantly extends previous weak-scale, phenomenological MSSM fits, by adding more and newer experimental analyses, improving the accuracy and detail of theoretical predictions, including dominant uncertainties from the Standard Model, the Galactic dark matter halo and the quark content of the nucleon, and employing novel and highly-efficient statistical sampling methods to scan the parameter space. We find regions of the MSSM7 that exhibit co-annihilation of neutralinos with charginos, stops and sbottoms, as well as models that undergo resonant annihilation via both light and heavy Higgs funnels.  We find high-likelihood models with light charginos, stops and sbottoms that have the potential to be within the future reach of the LHC.  Large parts of our preferred parameter regions will also be accessible to the next generation of direct and indirect dark matter searches, making prospects for discovery in the near future rather good.

\end{abstract}

\tableofcontents

\section{Introduction}
\label{intro}

The most straightforward supersymmetric extension of the Standard
 Model (SM) of elementary particles is the Minimal Supersymmetric
 Standard Model (MSSM)~\cite{arXiv:1506.08277}.  The MSSM can help
 deal with many of the theoretical and experimental shortcomings of
 the SM.  Most notably it stabilises the electroweak
 scale \cite{Dimopoulos:1981yj,Witten:1981nf,Dimopoulos:1981zb,Sakai:1981gr,Kaul:1981hi}
 with respect to large corrections from new
 physics at the Planck scale \cite{Weinberg:1975gm, Weinberg:1979bn,
 Gildener:1976ai,Susskind:1978ms,tHooft:1980xss}, allows the
 unification of gauge
 couplings \cite{Ellis:1990wk,Langacker:1991an,Amaldi:1991cn,Anselmo:1991uu},
 provides a dark matter (DM) candidate that can fit the observed relic
 abundance \cite{Ellis:1983ew,Jungman:1995df} and predicts a light
 Higgs boson, in accordance with the 2012
 discovery \cite{Aad:2012tfa,Chatrchyan2012}. This has prompted a
 vast number of investigations of this model, with recent literature
 studying precision corrections to the Higgs
 mass \cite{Bagnaschi:2014rsa, arXiv:1504.05200, Lee:2015uza, arXiv:1601.01890, Bahl:2016brp, Athron:2016fuq, Bagnaschi:2017xid, Staub:2017jnp, Passehr:2017ufr, Bahl:2017aev} and other aspects of Higgs physics \cite{arXiv:1608.02573, arXiv:1608.00638,
 arXiv:1512.00437, arXiv:1511.08461, arXiv:1511.07853,
 arXiv:1511.06002, arXiv:1506.08462, arXiv:1504.06932,
 arXiv:1504.06625, arXiv:1504.04308,
 arXiv:1502.05653}, DM \cite{Profumo:2016zxo,
 Roy:2016zst, arXiv:1602.08103, arXiv:1602.01030, arXiv:1602.00590,
 arXiv:1601.04718, IC79_SUSY, arXiv:1511.05964, arXiv:1511.05386,
 arXiv:1510.06295, arXiv:1510.05378, arXiv:1510.04291,
 arXiv:1510.03498, arXiv:1510.03460, arXiv:1510.02470,
 arXiv:1510.02473, arXiv:1509.09159, arXiv:1509.05076,
 arXiv:1508.04383, arXiv:1508.04373, arXiv:1507.06164,
 arXiv:1412.4789, arXiv:1507.05584, arXiv:1507.04644,
 arXiv:1505.04595, Crivellin:2015oha, arXiv:1504.05554,
 arXiv:1504.05091, arXiv:1504.00915, arXiv:1504.00504,
 arXiv:1503.07142, arXiv:1503.03478, Catalan:2015cna,
 arXiv:1502.06000, arXiv:1502.05703, arXiv:1502.05672,
 arXiv:1502.05406, arXiv:1412.8698}, the matter-antimatter asymmetry
 of the Universe \cite{arXiv:1512.09172, arXiv:1508.04144,
 arXiv:1508.00011}, vacuum stability \cite{arXiv:1606.08356}, cosmic
 inflation \cite{arXiv:1503.08867, arXiv:1407.4110, arXiv:1405.4125,
 arXiv:1312.3623, arXiv:1305.1066, arXiv:1304.5202, arXiv:1303.5351,
 arXiv:1205.2815}, and various measurements of
 precision \cite{Kobakhidze:2016mdx, gm2calc, arXiv:1507.05836,
 arXiv:1505.01987, arXiv:1504.05500, arXiv:1503.08703,
 arXiv:1503.08219, arXiv:1503.06850} and
 flavour \cite{arXiv:1509.05414, arXiv:1504.00930} observables. MSSM
 predictions for observations related to the above also yield the most
 important constraints on the theory.

Indeed the MSSM is one of the best-studied extensions to the SM to date.  The
latest literature covers topics ranging from global fits to the MSSM
with different choices for the number of dimensions in the electroweak
scale parameterisation~\cite{arXiv:1608.02489, arXiv:1511.06205,
arXiv:1510.08840, arXiv:1507.07008, Mastercode15,
arXiv:1506.02499, arXiv:1504.03260}, to studies of various aspects of
its more constrained versions defined at a high
scale \cite{Han:2016gvr, Ellis:2016tjc, Kitano:2016dvv, arXiv:1603.08834,
arXiv:1509.08838, arXiv:1509.07031, arXiv:1509.02530,  Fittinocoverage,
Kowalska:2015kaa, arXiv:1505.04702, arXiv:1503.06186}. Theoretical
considerations, such as naturalness
\cite{Dutta:2016sva, arXiv:1604.02102, arXiv:1603.09347, arXiv:1602.01699, arXiv:1512.05781, arXiv:1509.02929, arXiv:1506.05962, arXiv:1504.05403, arXiv:1503.04137, arXiv:1503.01473, arXiv:1502.06893, arXiv:1501.05698}, and
gauge unification within the MSSM
\cite{Ellis:2017erg, arXiv:1512.09148, arXiv:1508.04176, arXiv:1506.05850, arXiv:1505.04950}, have also been a concern.
Due to the steady stream of results from the LHC, the implications of collider searches for the MSSM have been a particularly active field of study
\cite{Barr:2016sho, Beenakker:2016lwe, arXiv:1605.09502, SUSYAI, arXiv:1604.07438, arXiv:1511.09284, arXiv:1510.06652, arXiv:1509.00473, arXiv:1507.06726, arXiv:1507.06706, arXiv:1507.04006, Bechtle:2015nta, arXiv:1506.02148, arXiv:1506.00665, arXiv:1505.03729, arXiv:1505.02996, arXiv:1504.03689, arXiv:1504.01726, arXiv:1504.00927, arXiv:1503.00414, arXiv:1502.05712, arXiv:1502.05044, arXiv:1502.03734, arXiv:1501.05307}.

The most general version of the MSSM has a very large number of parameters. Assuming only $CP$ conservation, there are in total 63 free parameters, coming mostly from soft supersymmetry-breaking terms. There are two distinct ways to approach the exploration of the MSSM. The first is to take inspiration from the apparent unification of the gauge couplings at a high scale, defining a small set of unified parameters at that scale --- a Grand Unified Theory (GUT) --- and then run them down to the electroweak scale in order to compare with experiment. This is done for example in the four-parameter Constrained MSSM (CMSSM) with common mass parameters for the scalar and fermion soft masses~\cite{Nilles:1983ge}, and various generalisations of the CMSSM, such as the Non-Universal Higgs Mass models 1 and 2 (NUHM1 and NUHM2), which split the Higgs soft masses from the other scalar masses~\cite{Matalliotakis:1994ft,Olechowski:1994gm,Berezinsky:1995cj,Drees:1996pk,Nath:1997qm}.  We have carried out extensive global fits to these GUT-motivated models in a companion paper to this one~\cite{CMSSM}.

The other approach is to remain agnostic about the high-scale properties of the theory, and to define all the parameters at an energy near the electroweak scale.  This is the so-called `phenomenological MSSM' (pMSSM \cite{Djouadi:1998di}). Given the impracticality of studying the complete parameter space, it is necessary to make some physically-motivated assumptions and simplifications in order to focus upon a relevant and tractable lower-dimensional subspace of the full model.

In this paper, we perform the first global fit of the weak-scale phenomenological MSSM to make use of the \gambit global-fitting framework~\cite{gambit}. Our work improves upon previous pMSSM studies on the following fronts:
\begin{enumerate}
\item The larger number of observables in our composite likelihood function, including: sparticle searches at LEP, Run I and Run II of the LHC, observables and constraints on the Higgs sector from LEP, the Tevatron and the LHC, direct and indirect dark matter searches from a multitude of experiments, and a large number of flavour and precision observables.
\item Experimental likelihoods reconstructed from event rates, where applicable, including: Monte Carlo simulation of LHC observables, kinematical acceptance-based event rates for LEP sparticle searches, indirect DM search likelihoods computed at the level of single events, and direct DM search limits based on sophisticated simulation of the relevant experiments.
\item The use of the \gambit hierarchical model database and statistical framework, for an advanced treatment of uncertainties from dominant SM nuisance parameters across different observables, and a consistent theoretical and statistical treatment of all likelihoods.
\item The use of \diver~\cite{ScannerBit}, a new scanner based on differential evolution, which improves sampling performance compared to techniques used previously, and allows us to more accurately locate the high-likelihood regions.
\item The results in this paper are based on a public, open-source software, which allows for the reproduction and extension of our results by the interested reader.\footnote{\url{gambit.hepforge.org}}
\end{enumerate}

We begin in Sec.~\ref{sec:phys} by introducing the models, parameters and nuisances over which we scan, followed by our adopted priors and sampling methodology.  In Sec.~\ref{sec:lnL}, we then give a brief summary of the observables and likelihoods that we employ.  We present our main results in Sec.\ \ref{sec:results} and their implications for future searches for the MSSM in Sec.~\ref{sec:future}, and conclude in Sec.~\ref{sec:conc}.

All \GB input files, generated likelihood samples and best-fit benchmarks for this paper are publicly available online through \textsf{Zenodo} \cite{the_gambit_collaboration_2017_801640}.

\section{Models and scanning framework}
\label{sec:phys}

\subsection{Model definitions and parameters}

\GB makes no fundamental distinction between parameters of BSM theories and nuisance parameters, scanning over each on an equal footing.  Here we sample simultaneously from four different models: a 7-parameter phenomenological MSSM, and three models describing constraints on different areas of known physics relevant for calculating observables in the MSSM.  These nuisance models respectively describe the SM, the Galactic DM halo, and nuclear matrix elements for different light quark flavours (relevant for direct detection of DM).

\subsubsection{MSSM7}

\begin{table}
\begin{center}
\begin{tabular}{l c c c}
\hline
Parameter & Minimum & Maximum & Priors    \\
\hline
$A_{u_3}(Q)$               & $-$10\,TeV& 10\,TeV & flat, hybrid  \\
$A_{d_3}(Q)$               & $-$10\,TeV& 10\,TeV & flat, hybrid  \\
$\MHusq(Q)$                & $-(10\,\mathrm{TeV})^2$ & $(10\,\mathrm{TeV})^2$  & flat, hybrid \\
$\MHdsq(Q)$                & $-(10\,\mathrm{TeV})^2$ & $(10\,\mathrm{TeV})^2$  & flat, hybrid \\
$\msfsq(Q)$                & 0 & $(10\,\mathrm{TeV})^2$  & flat, hybrid \\
$M_2(Q)$                   & $-$10\,TeV& 10\,TeV & split; flat, hybrid  \\
$\tan\beta(m_Z)$           & 3         & 70      & flat          \\
\hline
$\mathrm{sgn}(\mu)$     & \multicolumn{2}{c}{$+$}    & fixed         \\
$Q$                     & \multicolumn{2}{c}{1\,TeV} & fixed         \\
\hline
\end{tabular}
\caption{\label{tab:param} MSSM7 parameters, ranges and priors adopted in the scans of this paper.  For a parameter $x$ of mass dimension $n$, the ``hybrid'' prior is flat where $|x| < (100\,\mathrm{GeV})^n$, and logarithmic elsewhere.  The ``split hybrid'' prior for $M_2$ refers to the fact that we carried out every scan twice: once with a hybrid prior over $0\le M_2\le 10$\,TeV, and again with a hybrid prior over $-10\,\mathrm{TeV}\le M_2\le 0$.  In addition to the priors listed here, we also carry out additional scans of fine-tuned regions associated with specific relic density mechanisms, where we restrict models to mass spectra that satisfy various conditions.  See text for details. \label{tab:SUSY_parameters}}
\end{center}
\end{table}

The most general formulation of the $CP$-conserving MSSM is given by the \gambit model \textsf{MSSM63atQ}.  Full details of the Lagrangian can be found in Sec.\ 5.4.3 of Ref.\ \cite{gambit}. This model has 63 free, continuous MSSM parameters: 3 gaugino masses $M_1$, $M_2$ and $M_3$, 9 parameters each from the trilinear coupling matrices $\mathbf{A}_u, \mathbf{A}_d$ and $\mathbf{A}_e$, 6 real numbers associated with each of the matrices of squared soft masses $\mathbf{m}^\mathbf{2}_Q$, $\mathbf{m}^\mathbf{2}_u$, $\mathbf{m}^\mathbf{2}_d$, $\mathbf{m}^\mathbf{2}_L$ and $\mathbf{m}^\mathbf{2}_e$, and three additional parameters describing the Higgs sector. We choose to work with the explicit mass terms $m^2_{H_u}$ and $m^2_{H_d}$ for the two Higgs doublets.   By swapping the Higgs bilinear couplings $b$ and $\mu$ for the ratio of vacuum expectation values for the up-type and down-type Higgs fields $\tan\beta\equiv v_\mathrm{u}/v_\mathrm{d}$, and demanding that the model successfully effect Electroweak Symmetry Breaking, we can reduce the remaining continuous freedom to a single parameter ($\tan\beta$).  This leaves only a free sign for $\mu$, which constitutes an additional (64th) discrete  parameter.  In this definition, $\tan\beta$ is specified at the scale $m_Z$, and all other parameters are defined at some other generic scale $Q$, usually taken to be near to the weak scale.

This parameter set is currently too large to explore in a global fit, and in any case much of the phenomenology can be captured in smaller models that incorporate simplifying assumptions.  In this first paper, we explore the \textsf{MSSM7atQ}, a 7-parameter subspace of the \textsf{MSSM63atQ}.  Inspired by GUT theories, we set
\begin{align}
\frac{3}{5}\cos^2\theta_\mathrm{W}M_1 = \sin^2\theta_\mathrm{W}M_2 = \frac{\alpha}{\alpha_\mathrm{s}}M_3,
\label{eq:GUT_relation}
\end{align}
at the scale $Q$.  We assume that all entries in $\mathbf{A}_u, \mathbf{A}_d$ and $\mathbf{A}_e$ are zero except for $(\mathbf{A}_u)_{33}=A_{u_3}$ and $(\mathbf{A}_d)_{33}=A_{d_3}$. We take all of the off-diagonal entries in $\mathbf{m}^\mathbf{2}_Q$, $\mathbf{m}^\mathbf{2}_u$, $\mathbf{m}^\mathbf{2}_d$, $\mathbf{m}^\mathbf{2}_L$ and $\mathbf{m}^\mathbf{2}_e$ to be zero, so as to suppress flavour-changing neutral currents.  By setting all remaining mass matrix entries to a universal squared sfermion mass $m^2_{\tilde f}$, we reduce the final list of free parameters to $M_2$, $A_{u_3}$, $A_{d_3}$, $m^2_{\tilde f}$, $m^2_{H_u}$, $m^2_{H_d}$ and $\tan\beta$ (plus the input scale $Q$ and the sign of $\mu$).  The MSSM7 has been studied in significant work in the previous literature, e.g.\ \cite{BergstromGondolo96,Mandic:2000jz,darksusy,BMSSM,IC22Methods,Adrian-Martinez:2013ayv}.

We assume that $R$-parity is conserved, making the lightest supersymmetric particle (LSP) absolutely stable, and discard all parameter combinations where the LSP is not a neutralino.  This choice is discussed in more detail in Sec.\ 2.1.1 of the companion paper \cite{CMSSM}.

In Table~\ref{tab:param}, we give the parameter ranges over which we scan the MSSM7 in this paper.  We choose to define all parameters other than $\tan\beta$ at $Q=1$~TeV, and investigate positive $\mu$ (for a definition of $\mu$ please see the superpotential given in Sec.\ 5.4.3 of Ref.\ \cite{gambit}.).  We intend to return to the $\mu<0$ branch of this model in future work, where we compare with less constrained subspaces of the \textsf{MSSM63atQ}.

\subsubsection{Nuisance parameters}
\label{sec:nuisance}

We make use of three different nuisance models in our scans: the SM as defined in SLHA2 \cite{Allanach:2008qq,gambit}, a model of the Galactic DM halo that follows a truncated Maxwell-Boltzmann velocity distribution \cite{gambit,DarkBit}, and a model containing nuclear matrix elements required for calculating DM-nucleon couplings \cite{gambit,DarkBit}.  We vary a total of five parameters across these models: the strong coupling at scale $m_Z$ in the \MSbar scheme, the top pole mass, the local density of DM, and the strange and up/down quark contents of the nucleon.  The treatment of these models and parameters here is identical to the treatment in the companion paper \cite{CMSSM}, where we scan the nuisance parameters using flat priors, and impose constraints on them within the likelihood function.  We refer the reader to Secs.\ 2.1.2--2.1.4 and Sec.\ 3.1 of Ref.\ \cite{CMSSM} for details.  For ease of reference however, here we reproduce in Table\ \ref{tab:nuisance} the parameter ranges and values used in our scans.

\begin{table}[tp]
\begin{center}
\begin{tabular}{l@{\hspace{-5mm}}c@{\,}r}
\hline
Parameter & & Value($\pm$Range) \\
\hline
\textbf{Varied} & & \\
Strong coupling & $\alpha_s^{\MSBar}(m_Z)$      & $0.1185(18)$   \\
Top quark pole mass  & \phantom{$^{\MSBar}$}$m_t$\phantom{$^{\MSBar}$}  &  $173.34(2.28)$\,GeV\\
Local DM density & \phantom{$^{\MSBar}$}$\rho_0$\phantom{$^{\MSBar}$} &  0.2--0.8\,GeV\,cm$^{-3}$\\
Nuclear matrix el. (strange)  & \phantom{$^{\MSBar}$}$\sigma_s$\phantom{$^{\MSBar}$} & $43(24)$\,MeV \\
Nuclear matrix el. (up + down) & \phantom{$^{\MSBar}$}$\sigma_l$\phantom{$^{\MSBar}$} & $58(27)$\,MeV \\
\hline
\textbf{Fixed} & & \\
Electromagnetic coupling & $1/\alpha^{\MSBar}(m_Z)$        & $127.940$       \\
Fermi coupling $\times$ $10^{5}$ & \phantom{$^{\MSBar}$}$G_{F,5}$\phantom{$^{\MSBar}$} & $1.1663787$ \\
Z pole mass  & \phantom{$^{\MSBar}$}$m_Z$\phantom{$^{\MSBar}$}  &  $91.1876$\,GeV\\
Bottom quark mass & $m_b^{\MSBar}(m_b)$    & $4.18$\,GeV \\
Charm quark mass & $m_c^{\MSBar}(m_c)$ & $1.275$\,GeV \\
Strange quark mass & $m_s^{\MSBar}(2\,\text{GeV})$  & $95$\,MeV\\
Down quark mass & $m_d^{\MSBar}(2\,\text{GeV})$  &   $4.80$\,MeV  \\
Up quark mass & $m_u^{\MSBar}(2\,\text{GeV})$     & $2.30$\,MeV \\
$\tau$ pole mass  & \phantom{$^{\MSBar}$}$m_\tau$\phantom{$^{\MSBar}$}  &  $1.77682$\,GeV\\
CKM Wolfenstein parameters: & \phantom{$^{\MSBar}$}$\lambda$\phantom{$^{\MSBar}$} & 0.22537 \\
 & \phantom{$^{\MSBar}$}$A$\phantom{$^{\MSBar}$} & 0.814 \\
 & \phantom{$^{\MSBar}$}$\bar\rho$\phantom{$^{\MSBar}$} & 0.117 \\
 & \phantom{$^{\MSBar}$}$\bar\eta$\phantom{$^{\MSBar}$} & 0.353 \\
Most probable halo speed & \phantom{$^{\MSBar}$}$v_0$\phantom{$^{\MSBar}$} & 235\,km\,s$^{-1}$ \\
Local disk circular velocity & \phantom{$^{\MSBar}$}$v_{\rm rot}$\phantom{$^{\MSBar}$} & 235\,km\,s$^{-1}$ \\
Local escape velocity & \phantom{$^{\MSBar}$}$v_{\rm esc}$\phantom{$^{\MSBar}$} & 550\,km\,s$^{-1}$ \\
Up contribution to proton spin & \phantom{$^{\MSBar}$}$\Delta^{(p)}_u$\phantom{$^{\MSBar}$} & 0.842 \\
Down contrib. to proton spin & \phantom{$^{\MSBar}$}$\Delta^{(p)}_d$\phantom{$^{\MSBar}$} & $-$0.427 \\
Strange contrib. to proton spin & \phantom{$^{\MSBar}$}$\Delta^{(p)}_s$\phantom{$^{\MSBar}$} & $-$0.085 \\
\hline\end{tabular}
\caption{Standard Model, dark matter halo and nuclear nuisance parameters and ranges.  We vary each of the parameters in the first section of the table simultaneously with MSSM7 parameters in all of our scans, employing flat priors on each.  The parameters listed in the second section of the table are constant in all scans. Further details and references for the chosen values can be found in Secs.\ 2.1.2--2.1.4 of the companion paper \cite{CMSSM}.} \label{tab:nuisance}
\end{center}
\end{table}

\subsection{Scanning methodology}
\label{sec:scan_alg}

Our scanning methodology is similar to that detailed in Sec.\ 2.2 of the companion paper \cite{CMSSM}; here we give only a short summary, focussing in particular on any points of difference.

We use a number of different priors, two different samplers\footnote{Here we use `sampler' and `scanner' synonymously.}, and a range of different sampler configurations to scan the parameter space of the MSSM7.  We then combine the results of all scans into a single set of likelihood samples, and use it to generate profile likelihoods for the MSSM7 parameters and observables.  Using multiple scanners, priors and sampling settings allows for accurate determination of both the highest-likelihood point and profile likelihood contours.  We do not consider Bayesian posteriors in the current paper, as our preliminary investigations indicate that Bayesian results in weak-scale MSSM models are dominated by the choice of priors.  This suggests that a careful choice of prior (based on e.g. fine-tuning considerations) is needed for later interpretation, which is beyond the scope of the current paper.

We primarily rely on the differential evolution scanner \diver \textsf{1.0.0}~\cite{ScannerBit}, but also perform some supplementary scans with the nested sampler \multinest \textsf{3.10} \cite{MultiNest}. For a performance comparison of \diver, \multinest and other modern scanning algorithms, please see Sec.\ 11 of Ref.\ \cite{ScannerBit}.  Unlike in the companion paper, we repeat all \diver scans with both self-adaptive jDE \cite{Brest06} and its $\lambda$jDE variant \cite{ScannerBit}.  The $\lambda$jDE algorithm is more aggressive in finding the maximum likelihood, whereas jDE ensures a more thorough exploration of large regions of degenerate likelihood.  Carrying out jDE-only scans (as opposed to retaining $\lambda$jDE for all scans, as in Ref.\ \cite{CMSSM}) is more beneficial in the MSSM7 than in the NUHM2 or its subspaces, as the additional freedom of the MSSM7 means that individual regions satisfying the bound set by the DM relic density, whilst more numerous, occupy a more fragmented and isolated volume of the parameter space than in `smaller' models like the CMSSM.

We carry out scans with effectively logarithmic priors using both \diver and \multinest, where all parameters except $\tan\beta$ follow so-called ``hybrid'' priors.  The hybrid prior we use for parameter $x$ of mass dimension $n$ is flat in the region $|x| < (100\,\mathrm{GeV})^n$, and logarithmic for larger $|x|$.  We supplement these scans with additional \diver runs using flat priors on the parameters most sparsely sampled at large values in log-prior scans ($A_{u_3}$, $A_{d_3}$ and $\MHusq$), and additional \multinest scans using flat priors on all dimensionful parameters.  To further improve the completeness of the sampling across the parameter space, we also subdivide every run into separate scans for $M_2 > 0$ and $M_2 < 0$.

As we show in Sec.\ \ref{sec:results}, chargino co-annihilation is the dominant mechanism for producing an acceptable DM abundance in large parts of the allowed parameter space. To ensure that we also explore the narrow parameter regions where sfermion co-annihilation or resonant annihilation is responsible for depleting the relic density to or below the observed value, we perform four `directed' scans with different additional conditions on the particle spectrum:\begin{enumerate}
\item \textit{squark co-annihilation} -- a cut on the mass of the lightest squark $m_{\tilde q_1} < 1.5\,m_{\tilde \chi_1^0}$,
\item \textit{slepton co-annihilation} -- a cut on the mass of the lightest slepton $m_{\tilde l_1} < 1.75\,m_{\tilde \chi_1^0}$,
\item \textit{heavy Higgs funnel} -- a cut on the mass of the CP-odd Higgs boson $1.8\,m_{\tilde \chi_1^0} < m_A < 2.2\,m_{\tilde \chi_1^0}$, and
\item \textit{light Higgs funnel} -- a cut on the lightest neutralino mass $25 < m_{\tilde \chi_1^0} / \mathrm{GeV} < 85$.
\end{enumerate}
For all four scans, we also require that $m_{\tilde \chi_1^\pm} > 1.05\,m_{\tilde \chi_1^0}$, to avoid parameter combinations where chargino co-annihilation dominates.  We carry out these scans exclusively with \diver, using pure jDE. In these directed scans, we allow $M_2$ to vary freely over its full positive and negative range (Table\ \ref{tab:SUSY_parameters}).

The mass conditions effectively act as priors, allowing us to obtain a starting population of points roughly in the right area, before refining the fit to the best-fit likelihood region corresponding to each mechanism for depleting the relic density.  Note that the bounds that we use for this process are quite generous compared to the actual \textit{definitions} of these regions that we use later in this paper, as they are only designed to direct the scanner to the correct vicinity of parameter space in which to look for good fits, not to act as cut for physical interpretation.  To generate an initial population of parameter points satisfying the cut in question, we simply draw randomly from the overall parameter space, and assign zero likelihood to all parameter combinations that do not satisfy the cut.  We then take the set of resulting points satisfying the cut, and use them with \diver to discover new points with higher likelihood values, continuing until the algorithm indicates convergence (suggesting that the best fit has been found, to within some tolerance).  The looser mass cut on sleptons compared to squarks ensures that we are able to generate an initial population within the required cut in a reasonable time.

Taking into account all sampling configurations, priors and parameter space partitionings, this leads to a total of 2 $\mathrm{sgn}(M_2)$ $\times$ 2 priors $\times$ (1 \multinest + 2 \diver configurations) + 4 directed scans = 16 scans. Each of the 16 scans took on the order of a few days to run on 2400 modern (Intel Core i7) cores. In total, these scans resulted in $1.8 \times 10^8$ valid samples, of which $1.76 \times 10^7$ ($1.37 \times 10^7$) were within the 2D $95\%$ ($68\%$) CL region of the global best-fit point. After the scans we ran all samples through a postprocessing step, using the \textsf{postprocessor} scanner of \scannerbit~\cite{ScannerBit}, in order to correct for a bug in the flavour likelihood and apply Run II LHC searches which had just recently become available in \colliderbit~\cite{ColliderBit}. Following this postprocessing step $1.67 \times 10^7$ ($2 \times 10^5$) of the original samples ended within the $95\%$ ($68\%$) CL region of the global best-fit point. 

For the main \diver scans, we set the population size \fortran{NP} to 19\,200, and the convergence threshold \fortran{convthresh} to $10^{-5}$; for the directed scans, we instead set \fortran{NP} = 6000 and \fortran{convthresh} = $10^{-4}$.  For \multinest, we use \fortran{nlive} = 5000 live points and a stopping tolerance of \fortran{tol} = 0.1. These are relatively loose choices, but this results from the fact that we are only using \multinest to bulk out the dominant likelihood mode of each scan rather than to locate the best fit point with high accuracy (the later is performed by \diver).

\section{Observables and likelihoods}
\label{sec:lnL}

We compute a wide range of collider, flavour, DM and precision observables, and combine them with results from the latest experimental searches to produce an extensive set of likelihood components, which we then use to construct a composite likelihood function for driving the samplers in this paper.  The composite likelihood includes \begin{itemize}
\item the DM relic density measurement $\Omega_{\rm c}h^2 = 0.1188\pm0.0010$ from \textit{Planck} \cite{Planck15cosmo} (implemented as an upper limit, so as to permit an additional non-neutralino DM component)
\item \textit{Fermi}-LAT {\tt Pass 8} dwarf limits on DM annihilation \cite{LATdwarfP8},
\item IceCube 79-string limits on DM annihilation in the Sun \cite{IC79,IC79_SUSY},
\item direct DM searches by LUX~\cite{LUX2013,LUX2016,LUXrun2}, Panda-X \cite{PandaX2016}, PICO~\cite{PICO60,PICO2L}, XENON100~\cite{XENON2013}, SuperCDMS~\cite{SuperCDMS} and SIMPLE~\cite{SIMPLE2014},
\item the anomalous magnetic moment of the muon \cite{PDG10,gm2exp},
\item MSSM contributions to the mass of the $W$ boson,
\item 59 different flavour observables measured by LHCb, Babar and Belle,
\item 13 different ATLAS and CMS analyses from Run I and Run II (as in the companion paper \cite{CMSSM}, we apply the Run II searches as a postprocessing step),
\item Higgs limits, mass measurements and signal strengths from LEP and the LHC, and
\item limits from LEP on sparticle production and decay in 18 different channels.
\item nuisance likelihoods associated the local density of DM \cite{DarkBit}, the quark contents of the nucleon \cite{DarkBit}, the top mass \cite{SDPBit} and the strong coupling \cite{SDPBit}.
\end{itemize}
The theoretical treatments, experimental data and final likelihood functions for all these observables match those described in Sec.\ 3 of the companion paper \cite{CMSSM}, so we refer the reader to that paper, and the \GB module papers \cite{DarkBit,ColliderBit,SDPBit,FlavBit}, for details.  The only exception is the DM relic density calculation, which we perform for the MSSM7 with \micromegas \textsf{3.6.9.2} \cite{micromegas} (with settings \cpp{fast} = 1, \cpp{Beps} = $10^{-5}$), rather than \darksusy \textsf{5.1.3} \cite{darksusy}, because the former is faster for some highly degenerate sfermion co-annihilation models.

The observables that we include draw on many other external software packages: \ddcalc \textsf{1.0.0} \cite{DarkBit}, \flexiblesusy \textsf{1.5.1}\footnote{\flexiblesusy gets model-dependent information from \SARAH \cite{Staub:2008uz,Staub:2010jh} and uses some numerical routines from \SOFTSUSY \cite{Allanach:2001kg,Allanach:2013kza}.} \cite{Athron:2014yba}, \gamlike \textsf{1.0.0} \cite{DarkBit}, \gmtwocalc \textsf{1.3.0} \cite{gm2calc}, \higgsbounds \textsf{4.3.1} \cite{Bechtle:2008jh,Bechtle:2011sb,Bechtle:2013wla}, \higgssignals \textsf{1.4} \cite{HiggsSignals}, \nulike \textsf{1.0.0} \cite{IC22Methods,IC79_SUSY}, \pythiaeight \textsf{8.212} \cite{Sjostrand:2014zea}, \superiso \textsf{3.6} \cite{Mahmoudi:2007vz,Mahmoudi:2008tp,Mahmoudi:2009zz} and \susyhit \textsf{1.5} \cite{Djouadi:1997yw,Muhlleitner:2003vg,Djouadi:2006bz,Butterworth:2010ym}.

\section{Results}
\label{sec:results}

\begin{table*}
\small
\center
\sisetup{round-mode = places, round-precision = 3,round-integer-to-decimal,exponent-product = {\hspace{-0.5mm}\cdot\hspace{-0.5mm}} }
\begin{tabular}{l S[table-format=+3.3] S[table-format=+3.3e1] S[table-format=+3.3e1] S[table-format=+3.3e1] S[table-format=+3.3e1] S[table-format=+3.3e1] S[table-format=+3]}
Likelihood term & {\ \ Ideal} & \multicolumn{1}{l}{{$A/H$-funnel}} & \multicolumn{1}{l}{{$\tilde{b}$ co-ann.}} & \multicolumn{1}{l}{{$\tilde{t}$ co-ann.}} & \multicolumn{1}{l}{{$\tilde{\chi}_1^{\pm}$ co-ann.}} & \multicolumn{1}{l}{{$Z/h$-funnel}} & \multicolumn{1}{l}{{$\Delta\ln\mathcal{L}_\mathrm{BF}$}} \\
\hline
LHC sparticle searches & 0 & 0.000 & 0.000 & 0.000 & 0.000 & 0.000 & 0\\
LHC Higgs & -37.7339 &  -38.657  &  -38.647  &  -39.050  &  -38.347  &  -38.593  &  .6131 \\
LEP Higgs & 0 &  0.000  &  0.000  &  0.000  &  0.000  &  0.000 & 0 \\
ALEPH selectron & 0 &  0.000  & 0.000  & 0.000  & 0.000  & 0.000 & 0 \\
ALEPH smuon & 0 &  0.000  &  0.000  &  0.000  &  0.000  &  0.000 & 0  \\
ALEPH stau & 0 &  0.000  &  0.000  &  0.000  &  0.000  &  0.000 & 0 \\
L3 selectron & 0 &  0.000  &  0.000  &  0.000  &  0.000  &  0.000 & 0  \\
L3 smuon & 0 & 0.000  &  0.000  &  0.000  &  0.000  &  0.000 & 0 \\
L3 stau & 0 & 0.000  &  0.000  &  0.000  &  0.000  &  0.000 & 0 \\
L3 neutralino leptonic & 0 & 0.000  & 0.000  & 0.000  & 0.000  & 0.000 & 0  \\
L3 chargino leptonic & 0 &  0.000  &  0.000  &  0.000  &  0.000  &  0.000 & 0 \\
OPAL chargino hadronic & 0 &  0.000  &  0.000  &  0.000  &  0.000  &  0.000 & 0 \\
OPAL chargino semi-leptonic & 0 &  0.000  &  0.000  &  0.000  &  0.000  &  0.000 & 0 \\
OPAL chargino leptonic & 0 &  0.000  &  0.000  &  0.000  &  0.000  &  0.000 & 0 \\
OPAL neutralino hadronic & 0 &  0.000  &  0.000  &  0.000  &  0.000  &  0.000 & 0  \\
$B_{(s)}\to \mu^+\mu^-$ & 0 &  -2.033  &  -2.024  &  -2.021  &  -1.998  &  -1.997 & 1.998  \\
Tree-level B and D decays & 0 &  -15.318  &  -15.284  &  -15.287  &  -15.315 &  -15.333 & 15.315 \\
$B^0\to K^{*0}\mu^+\mu^-$ & -184.260 &  -194.316  &  -195.283  &  -193.103  &  -194.734  &  -195.551 & 10.474  \\
$B\to X_s\gamma$ & 9.799 &  8.030  &  8.710  &  6.978  &  8.334  &  8.795 & 1.465  \\
$a_\mu$ & 20.266 &  14.027  &  14.114  &  14.299  &  14.269  &  14.090 & 5.997  \\
$W$ mass & 3.281 &  3.081  & 2.813  & 2.778  & 3.096  & 2.643 & .185 \\
Relic density & 5.989 &  5.989  &  5.989  &  5.989  &  5.989  &  5.989 & 0 \\
PICO-2L & -1 &  -1.000  &  -1.000  &  -1.000  &  -1.000  &  -1.000 & 0 \\
PICO-60 F & 0 &  0.000  &  0.000  &  0.000  &  0.000  &  -0.001 & 0  \\
SIMPLE 2014 & -2.972 &  -2.972  &  -2.972  &  -2.972  &  -2.972  &  -2.972 & 0  \\
LUX 2015  & -0.64 &  -0.657  &  -0.693  &  -0.670  &  -0.660  &  -0.650 & .020  \\
LUX 2016 & -1.467 &   -1.501  &  -1.574  &  -1.527  &  -1.506  &  -1.487 & .039 \\
PandaX 2016  & -1.886 &  -1.909  &  -1.960  &  -1.927  &  -1.912 &  -1.899 & .026  \\
SuperCDMS 2014 & -2.248 &  -2.248  &  -2.248  &  -2.248  &  -2.248  &  -2.248 & 0 \\
XENON100 2012 & -1.693 &   -1.684  &  -1.667  &  -1.678  &  -1.683  &  -1.688 & .010  \\
IceCube 79-string & 0.0 &  -0.032  &  0.000  &  0.000  &  -0.069  &  0.000 & .069  \\
$\gamma$ rays (Fermi-LAT dwarfs) & -33.244 &  -33.374  &  -33.367  &  -33.363  &  -33.371  &  -33.255 & .127  \\
$\rho_0$ & 1.142 &  1.139  &  1.115  &  1.138 &  1.142 &  1.141 & 0 \\
$\sigma_s$ and $\sigma_l$  & -6.115 &  -6.115  &  -6.117  &  -6.115  &  -6.128  &  -6.116 & .013 \\
$\alpha_s(m_Z)({\MSBar})$ & 6.500 &  6.493  &  6.427  &  6.409  &  6.496  &  6.457 & .004 \\
Top quark mass & -0.645 &  -0.647  &  -0.687  &  -0.645  &  -0.654  &  -0.751 & .009 \\
\hline
Total & -226.927 &  -263.704  & -264.354 & -264.016 & -263.272 & -264.426  &  36.345 \\
\hline
&  &  &  &  & \\
Quantity & & \multicolumn{1}{l}{{$A/H$-funnel}} & \multicolumn{1}{l}{{$\tilde{b}$ co-ann.}} & \multicolumn{1}{l}{{$\tilde{t}$ co-ann.}} & \multicolumn{1}{l}{{$\tilde{\chi}_1^{\pm}$ co-ann.}} & \multicolumn{1}{l}{{$Z/h$-funnel}} \\
\cmidrule{1-7}
$A_{d_3}$ (1\,TeV) &  &  9582.567  &  9669.750  &  9706.338  &  9376.461  &  1639.611 \\
$A_{u_3}$ (1\,TeV) &  &  -9389.783  &  2957.229  &  2197.287  &  2923.877  &  3660.585 \\
$M_2$ (1\,TeV) & &  3768.368  &  2404.020  &  1498.770  &  2469.296  &  2032.136 \\
$\tan\beta$ ($m_Z$) &  &  7.133  &  11.862  &  12.743  &  46.632 &  19.058 \\
$m_{H_u}^2$ (1\,TeV) & &  -1.2713e+07  &  -2.4900e+06  &  -9.7574e+05  &  -7.8301e+05  &  -6.0770e+05 \\
$m_{H_d}^2$ (1\,TeV) & &  3.7481e+05  &  1.0453e+07  &  7.8235e+06  &  2.7289e+07 &  3.1889e+06 \\
$m_{\tilde{f}}^2$ (1\,TeV) & &  9.6798e+07  &  9.2289e+06  &  3.0064e+06  &  1.3518e+07 &  9.5741e+06 \\
$m_t$ & &  173.289  &  173.120  &  173.325  &  173.445 &  172.990 \\
$\alpha_s(m_Z)({\MSBar})$ & &  0.119  &  0.119  &  0.119  &  0.119 &  0.119 \\
$\rho_0$ & &  0.409  &  0.372  &  0.390  &  0.399 &  0.406 \\
$\sigma_s$ &  &  42.966  &  43.242  &  42.916  &  44.101  &  42.591 \\
$\sigma_l$ &  &  57.987  &  57.442  &  58.265  &  58.773  &  58.095 \\
\cmidrule{1-7}
$M_1 (M_\mathrm{SUSY})$ & &  2002.225  &  1242.861  &  767.869  &  1283.505 &  1053.133 \\
$\mu (M_\mathrm{SUSY})$ & &  367.156  &  1477.923  &  987.697  &  253.479 &  69.449 \\
$m_{\tilde{t}_1}$ & &  9012.999  &  1237.689  &  759.551  &  2440.084  &  2132.455 \\
$m_{\tilde{\tau}_1}$ & &  9845.047  &  3034.359  &  1730.209  &  3698.869  &  3097.127 \\
$m_{A}$ & &  793.380  &  3567.851  &  2956.071  &  5348.470 &  1804.886 \\
$m_{h}$ & &  125.099  &  125.088  &  123.988  &  124.731  &  126.427 \\
$m_{\tilde{\chi}_1^0}$ & &  379.116  &  1233.050  &  759.524  &  258.939  &  69.247 \\
$($\%bino, \%Higgsino$)$ & & \multicolumn{1}{l}{{\,$(0,100)$}} & \multicolumn{1}{l}{{\ \ $(98,2)$}} & \multicolumn{1}{l}{{\ \ $(98,2)$}} & \multicolumn{1}{l}{{\,$(0,100)$}} & \multicolumn{1}{l}{{\,$(0,100)$}} \\
$m_{\tilde{\chi}_2^0}$ & &  -381.804  &  -1491.708  &  994.456  &  -262.754  &  -73.665 \\
$($\%bino, \%Higgsino$)$ & & \multicolumn{1}{l}{{\,$(0,100)$}} & \multicolumn{1}{l}{{\,$(0,100)$}} & \multicolumn{1}{l}{{\ \ $(2,97)$}} & \multicolumn{1}{l}{{\,$(0,100)$}} & \multicolumn{1}{l}{{\,$(0,100)$}} \\
$m_{\tilde{\chi}_1^\pm}$ & &  380.734  &  1488.287  &  990.571  &  261.179  &  71.618 \\
$($\%wino, \%Higgsino$)$ & & \multicolumn{1}{l}{{\,$(0,100)$}} & \multicolumn{1}{l}{{\ \ $(1,99)$}} & \multicolumn{1}{l}{{\ \ $(2,98)$}} & \multicolumn{1}{l}{{\,$(0,100)$}} & \multicolumn{1}{l}{{\,$(0,100)$}} \\
$m_{\tilde{g}}$ & &  12370.525  &  7920.520  &  5006.746  &  8104.365 &  6711.215 \\
$\Omega h^2$ & &  1.537e-2  &  3.89e-2  &  1.046e-2  & 8.027e-3  &  8.382e-4 \\
\cmidrule{1-7}
\end{tabular}
\caption{\label{tab:mssm7-bf-1} Best-fit points in the $A/H$-funnel, $\tilde{b}$ co-annihilation, $\tilde{t}$ co-annihilation, $\tilde{\chi}_1^{\pm}$ co-annihilation and $Z/h$ funnel regions. For each point, we show the individual likelihood contributions, parameter values (including nuisance parameters) and derived quantities crucial for interpreting the mass spectrum. Other SM and astrophysical parameters are set to the fixed values given in Table~\ref{tab:nuisance}.  SLHA1 and SLHA2 files for the best-fit points can be found in the online data associated with this paper \cite{the_gambit_collaboration_2017_801640}.}
\end{table*}

\begin{figure}[tbp]
\centering
\includegraphics[width=\columnwidth]{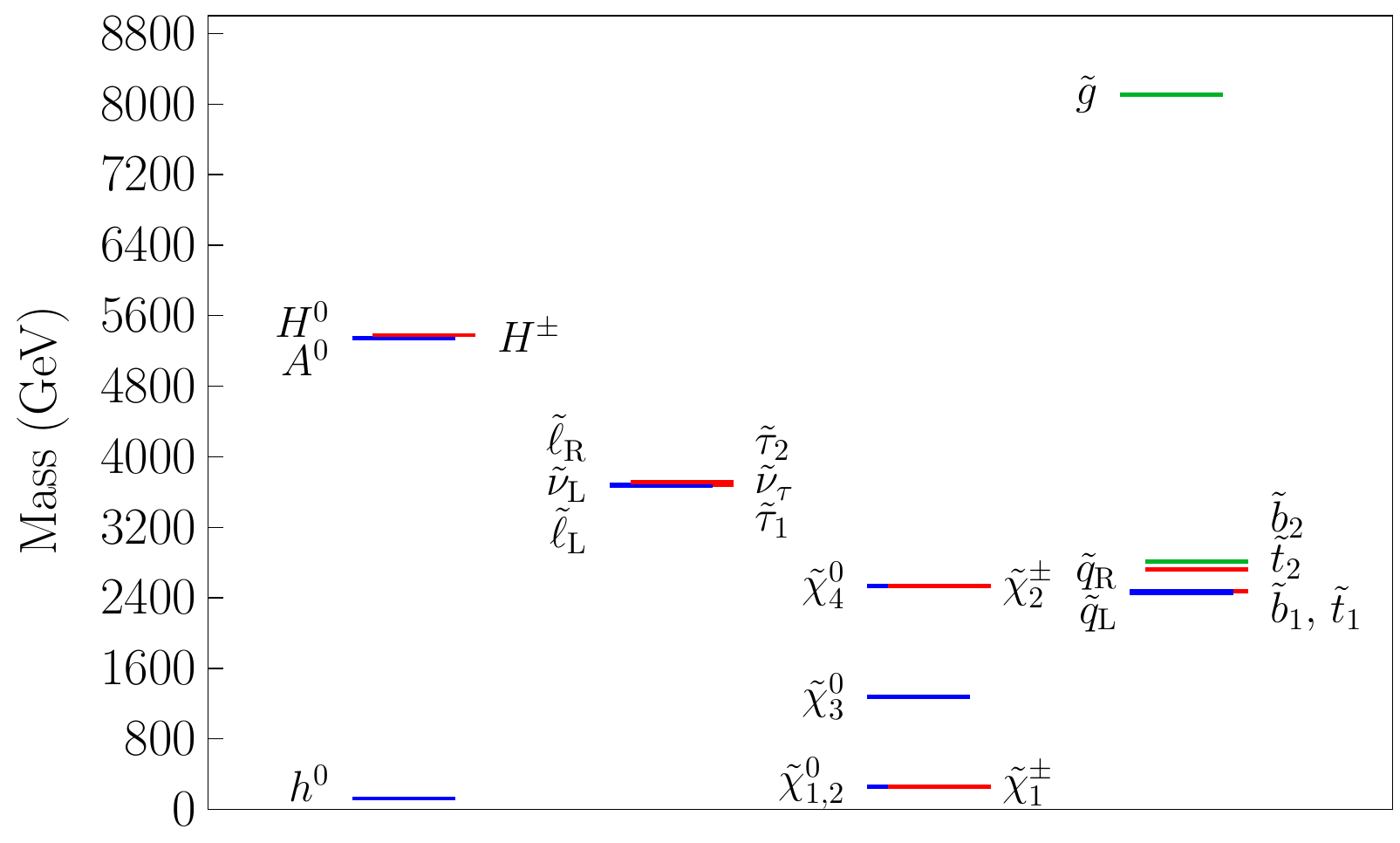}
\caption{Sparticle mass spectrum of the best-fit point.}
\label{fig:bf-spectrum}
\end{figure}

\subsection{Best fits}
In much of the parameter space of the MSSM7 (and indeed the MSSM more generally), the annihilation cross-section of heavy neutralino DM is so small that the thermal relic density greatly exceeds the value measured by \textit{Planck}.  Such models are robustly ruled out.  The only way for a model to respect this upper limit is to exhibit one or more specific mechanisms for depleting the thermal abundance, typically associated with co-annihilation with another supersymmetric species, or resonant annihilation via a neutral boson `funnel'.

Five such mechanisms play a role within the final 95\% confidence level (CL) regions of our scans.  In the following discussion, we classify samples and colour regions according which mechanism(s) they display:
\begin{itemize}
\item chargino co-annihilation: $\tilde\chi_1^0$ $\ge50\%$ Higgsino,
\item stop co-annihilation: $m_{\tilde{t}_1} \leq 1.2\,m_{\tilde\chi^0_1}$,
\item sbottom co-annihilation: $m_{\tilde{b}_1} \leq 1.2\,m_{\tilde\chi^0_1}$,
\item $A/H$ funnel: $1.6\,m_{\tilde\chi^0_1} \leq m_\textrm{heavy} \leq 2.4\,m_{\tilde\chi^0_1}$,
\item $h/Z$ funnel: $1.6\,m_{\tilde\chi^0_1} \leq m_\textrm{light} \leq 2.4\,m_{\tilde\chi^0_1}$,
\end{itemize}
where `heavy' may be $H^0$ or $A^0$, and `light' may be $h^0$ or $Z^0$, and a parameter combination qualifies as a member of a region if either condition is satisfied.  Indeed, this is the strategy we adopt in general: if a model fulfils one of these conditions, we include it in the region, even if it ends up becoming a member of multiple regions, and even if some dominate over others.  For clarity, we do not make any attempt to identify hybrid regions, or determine which of the mechanisms dominates (as to do so would require assumptions about relative temperature dependences and interferences between different partial annihilation rates). The union of these regions contain the full set of models allowed at 95\% CL.

In Table \ref{tab:mssm7-bf-1}, we show the details of the best-fit point in each of these five regions, breaking down the final log-likelihood into contributions from the different observables included in the fit.  The overall best fit occurs in the chargino co-annihilation region, where the lightest two neutralinos and the lightest chargino are all dominantly Higgsino, and thus highly degenerate in mass.  All pairwise annihilations and co-annihilations between any of these three species can thus contribute significantly to the final relic density in this region.  In Fig.\ \ref{fig:bf-spectrum} we give a visual representation of the mass spectrum of this point, where one can see clearly that we have some very light neutralinos and charginos in this model. The masses are around 260\,GeV, making them potentially interesting targets for future LHC searches (Sec.\ \ref{sec:LHC}).

We also define a so-called `ideal' reference likelihood in Table \ref{tab:mssm7-bf-1}.  This is the best likelihood that a model could realistically achieve were it to predict all observed quantities precisely, and predict no additional contribution beyond the expected background in searches that have produced only limits.  We compute this for most likelihood components by assuming that the model prediction is either equal to the measured value or the background-only prediction.  For some highly composite observables, where many different channels enter and the SM or background-only prediction can in principle be improved upon by introducing a BSM contribution, we take the ideal case to be the best fit achievable in a more general, effective phenomenological framework.  The two likelihoods that we apply this treatment to are those associated with LHC measurements of Higgs properties, and the angular observables of the $B^0\to K^{*0}\mu^+\mu^-$ decay observed by LHCb.  In the former case, we take the ideal likelihood to be the best fit obtainable by independently varying the mass, width and branching fractions of a single Higgs in order to fit the LHC data contained in \higgssignals.  For the latter, we take the ideal likelihood from the best-fit point that we found in a flavour EFT global fit, discussed in Sec.\ 6.2 of the \flavbit paper \cite{FlavBit}.

The log-likelihood difference between the global best fit and the ideal case is $\Delta\ln\mathcal{L}_\mathrm{BF} = 36.345$.  Compared to the CMSSM ($\Delta\ln\mathcal{L}_\mathrm{BF} = 36.820$; 4 BSM parameters + 1 sign), NUHM1 ($\Delta\ln\mathcal{L}_\mathrm{BF} = 36.702$; 5 BSM parameters + 1 sign) and NUHM2 ($\Delta\ln\mathcal{L}_\mathrm{BF} = 36.362$; 6 BSM parameters + 1 sign) \cite{CMSSM}, we see a fairly mild improvement from moving to the MSSM7 (7 BSM parameters).  It is possible to use $\Delta\ln\mathcal{L}_\mathrm{BF}$ to estimate the overall goodness of fit, but this requires knowledge of the effective number of degrees of freedom (see Sec.\ 4.1 of Ref.\ \cite{CMSSM} for details).  Guessing this to be between 30 and 50 (on the basis of the number of active observables in the fit) would lead to a $p$-value of between $2\times 10^{-5}$ and $0.02$.  Comparing the specific case of e.g.\ 37 degrees of freedom to the equivalent calculation for the NUHM2, NUHM1 and CMSSM with 38, 39 and 40 degrees of freedom, respectively, the $p$-value for the MSSM7 would be $5.9\times 10^{-4}$, compared to $5.9\times 10^{-4}$ (NUHM2), $7.1\times 10^{-4}$ (NUHM1) and $9.4\times 10^{-4}$ (CMSSM).  This comparison is not entirely fair, given that we have allowed $\mathrm{sgn}(\mu)$ to vary in the other three theories but not in the MSSM7.  Nonetheless, it does suggest that the likelihood improvement in the MSSM7 is not sufficient to overcome the $p$-value penalty arising from the larger number of free parameters compared to the three GUT-scale models.

\begin{figure*}[tbh]
  \centering
  \includegraphics[width=0.32\textwidth]{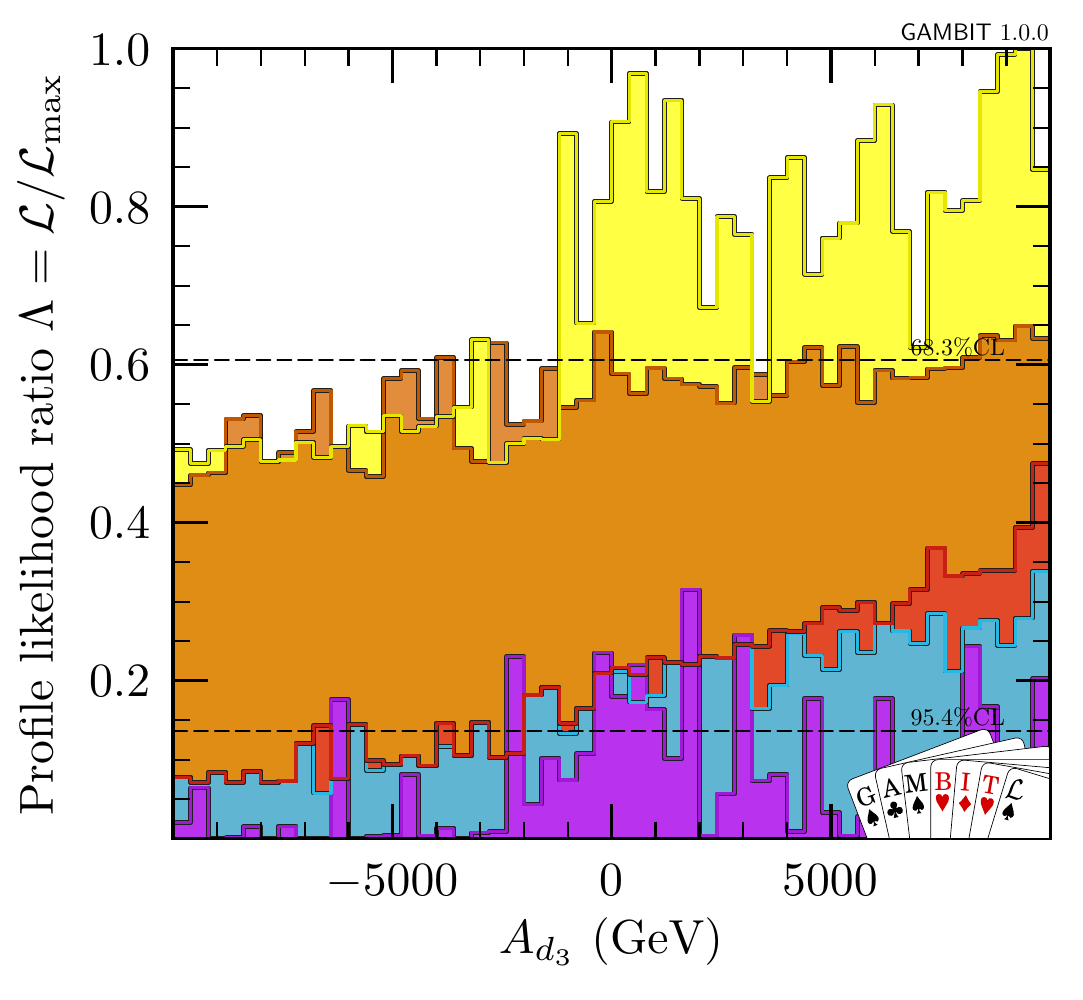}
  \includegraphics[width=0.32\textwidth]{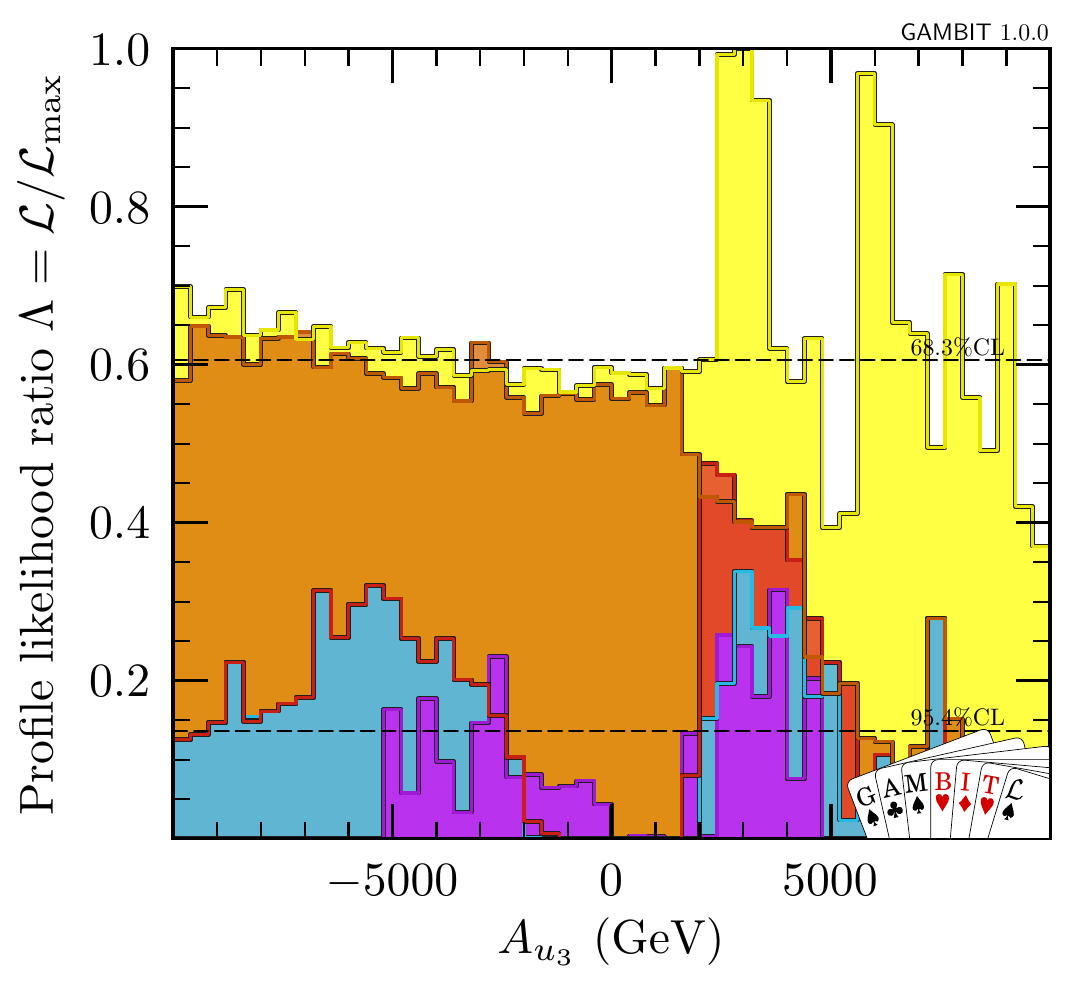}
  \includegraphics[width=0.32\textwidth]{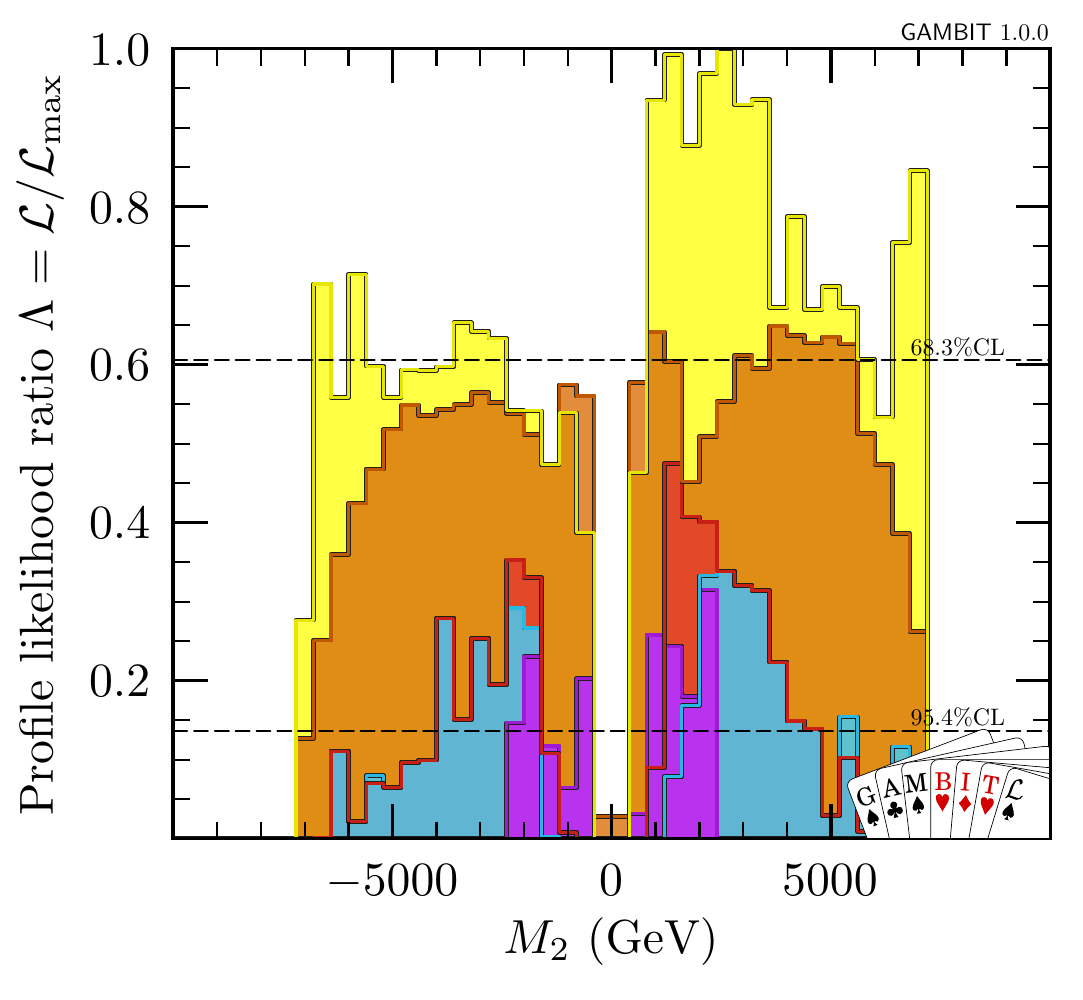}\\
  \includegraphics[width=0.32\textwidth]{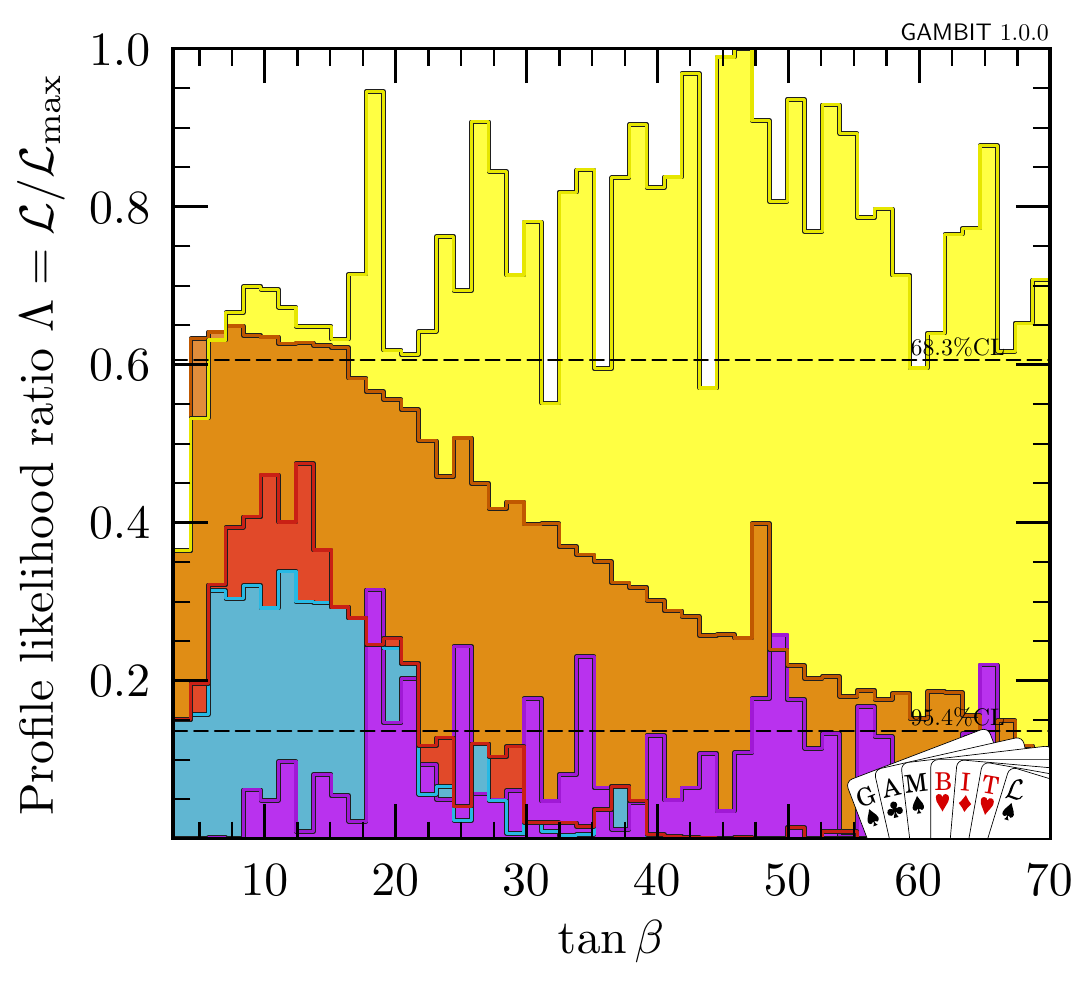}
  \includegraphics[width=0.32\textwidth]{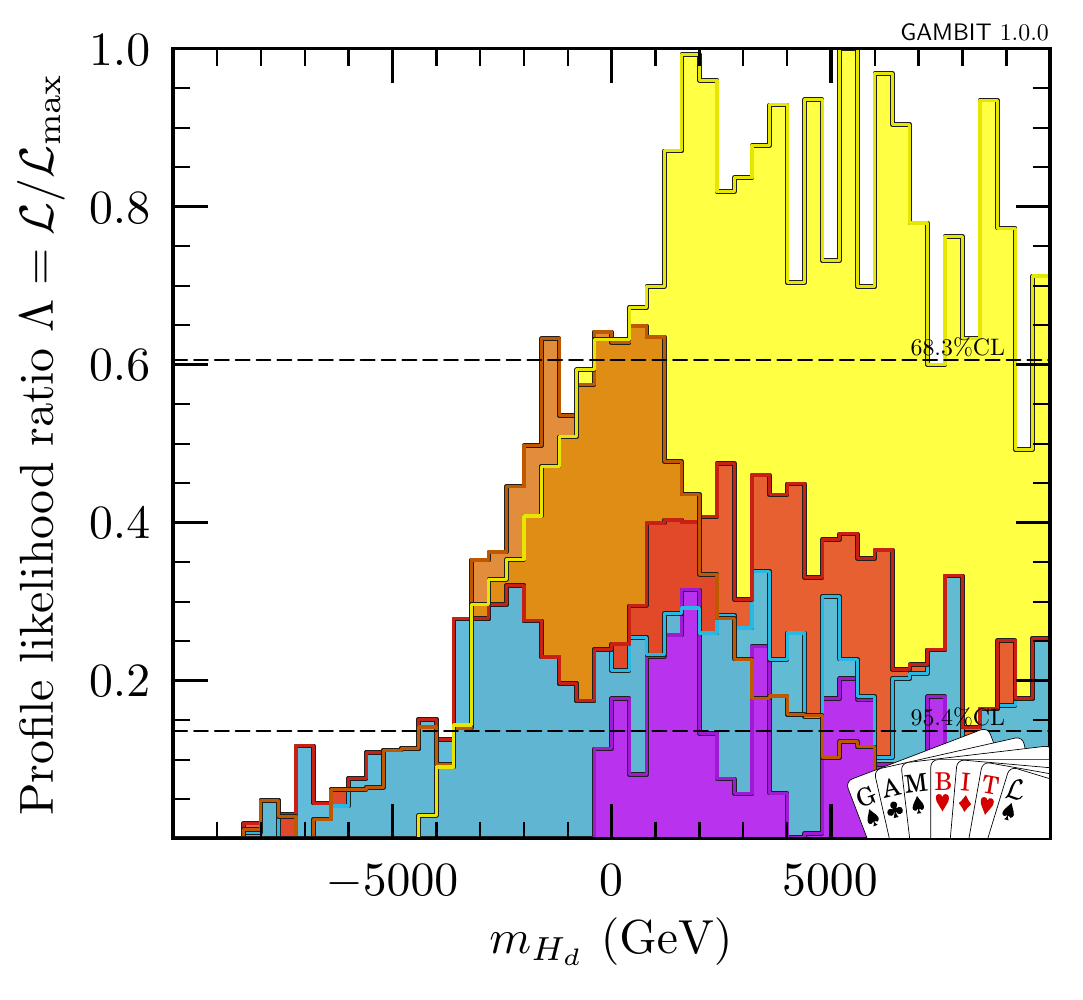}
  \includegraphics[width=0.32\textwidth]{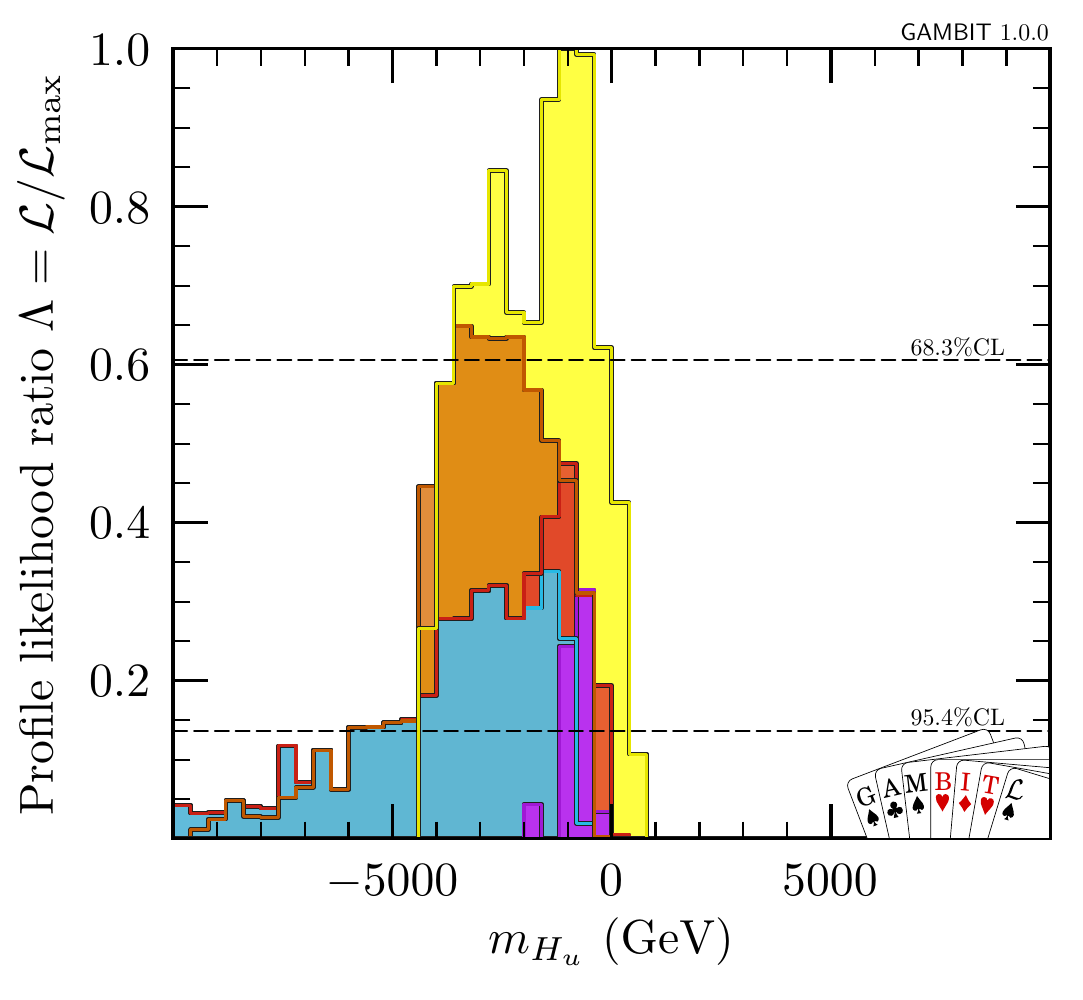}\\
  \includegraphics[width=0.32\textwidth]{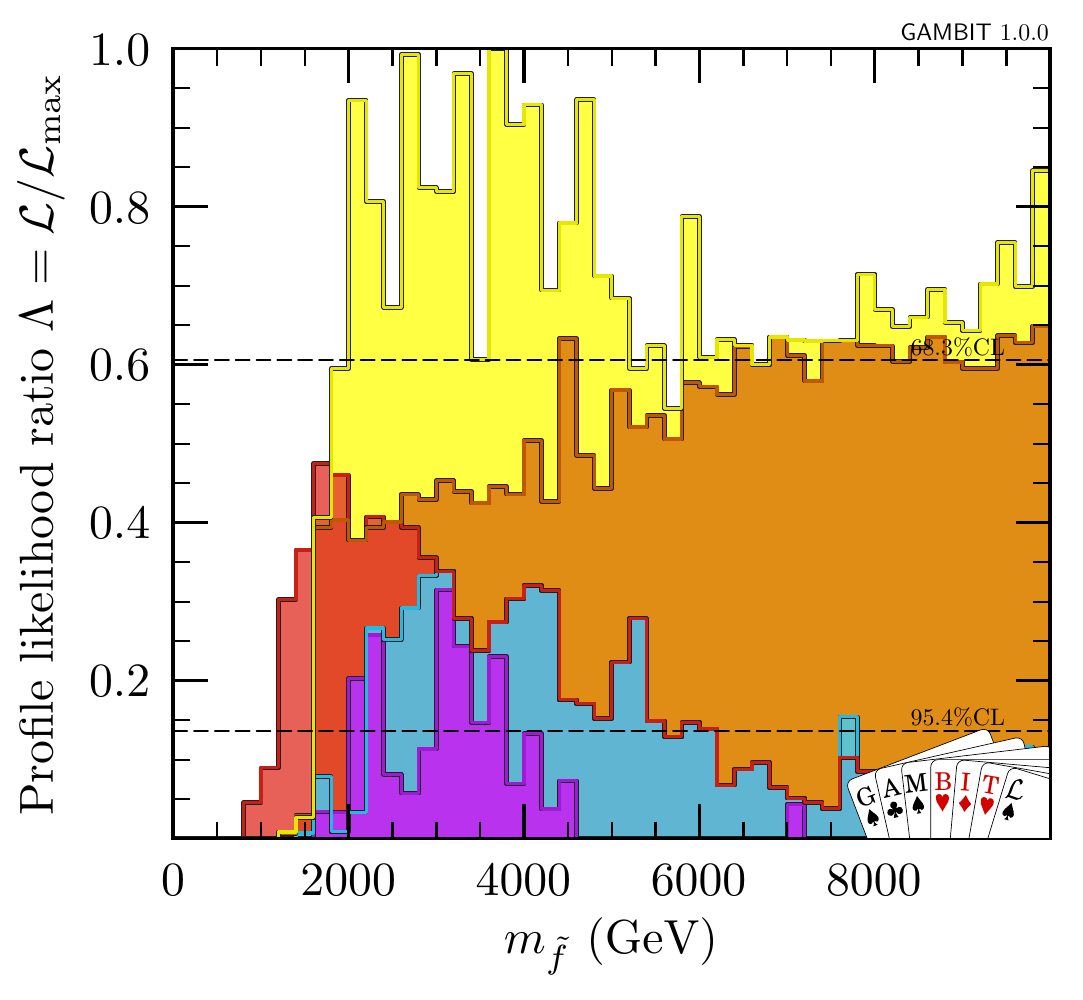}
  \includegraphics[width=0.32\textwidth]{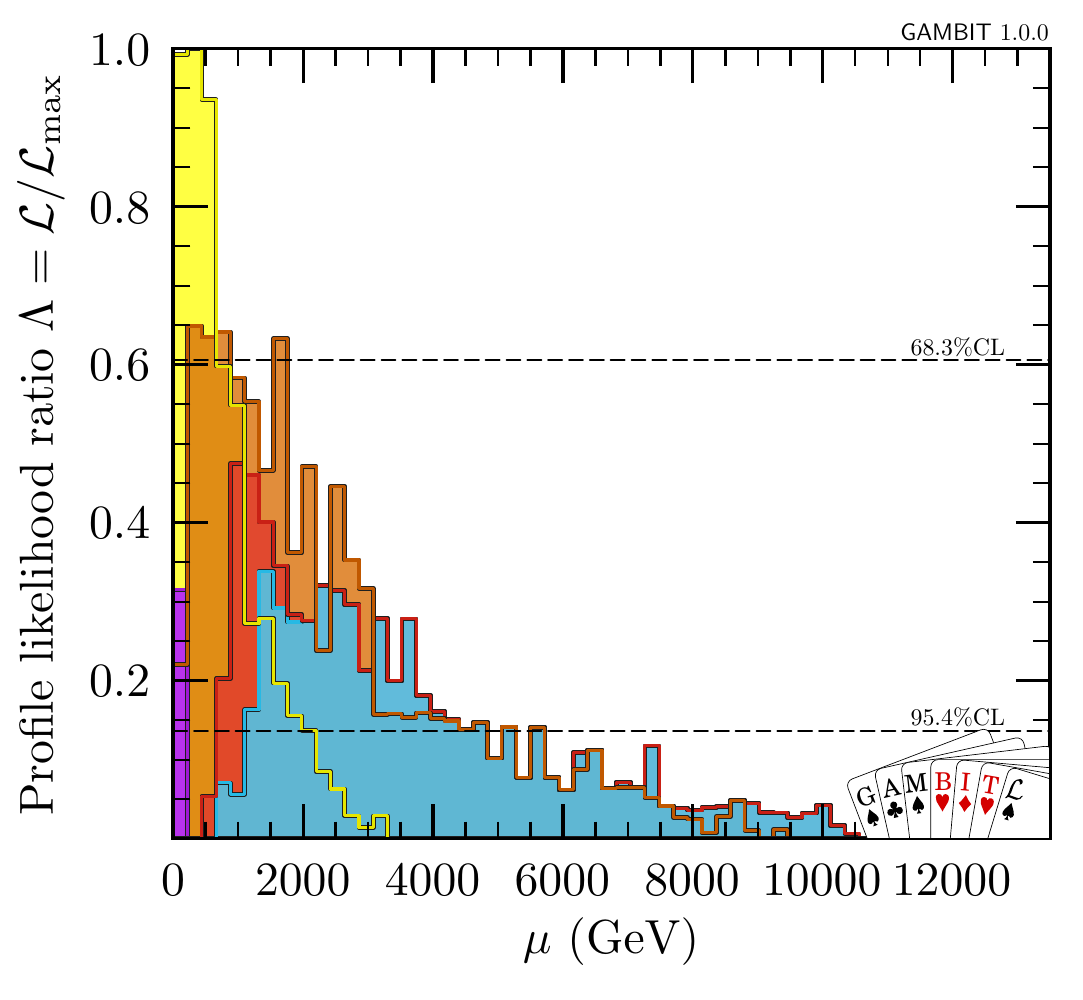}
  \includegraphics[width=0.32\textwidth]{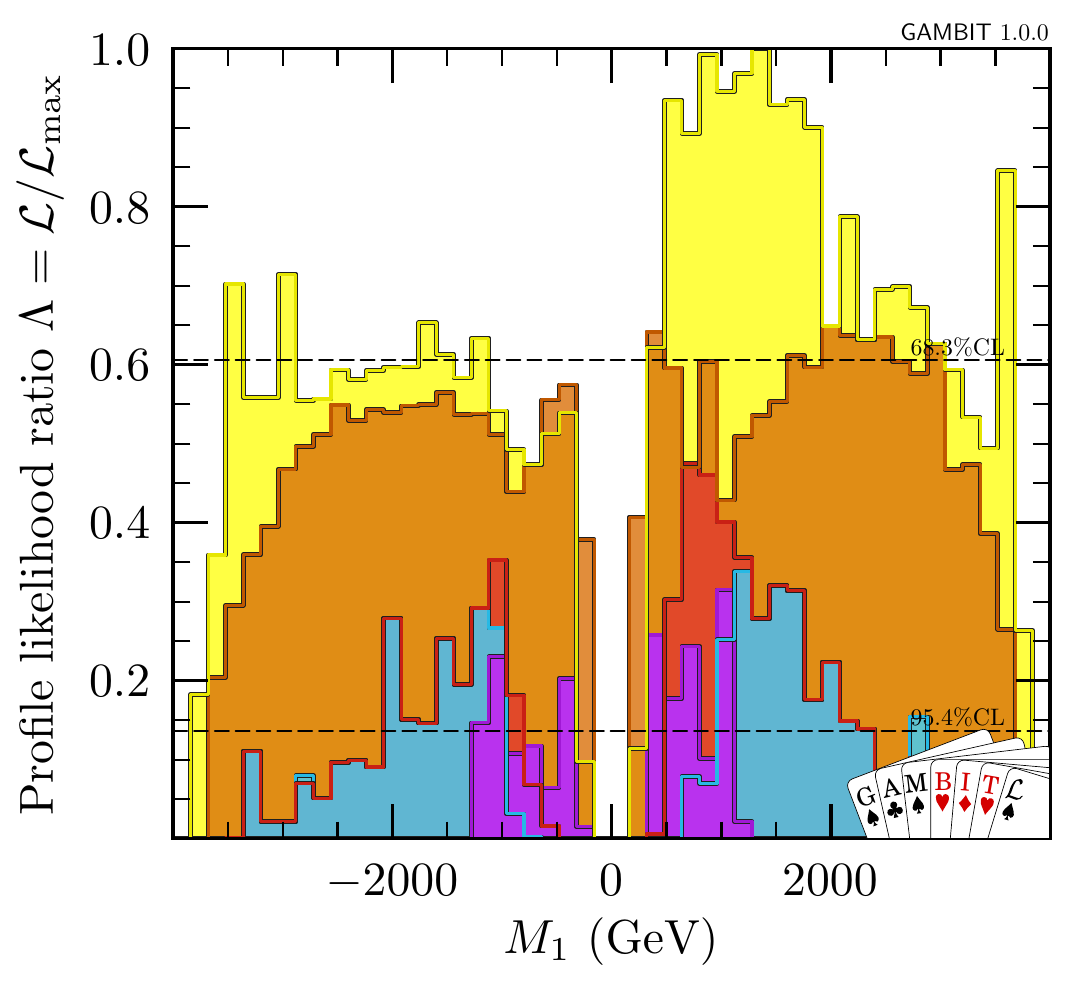}\\
  \includegraphics[height=4mm]{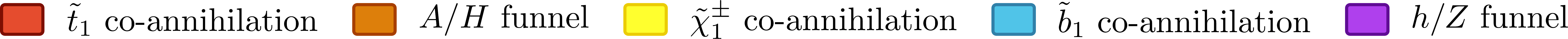}
  \caption{1D profile likelihood ratio for the input parameters $A_{d_3}$, $A_{u_3}$, $M_2$, $\tan\beta$, $m_{H_d}$, $m_{H_u}$ and $m_{\tilde{f}}$, as well as the derived parameters $M_1$ and $\mu$.}
  \label{fig:1d_like_parameters}
\end{figure*}

\begin{figure*}[tbh]
  \centering
  \includegraphics[width=0.49\textwidth]{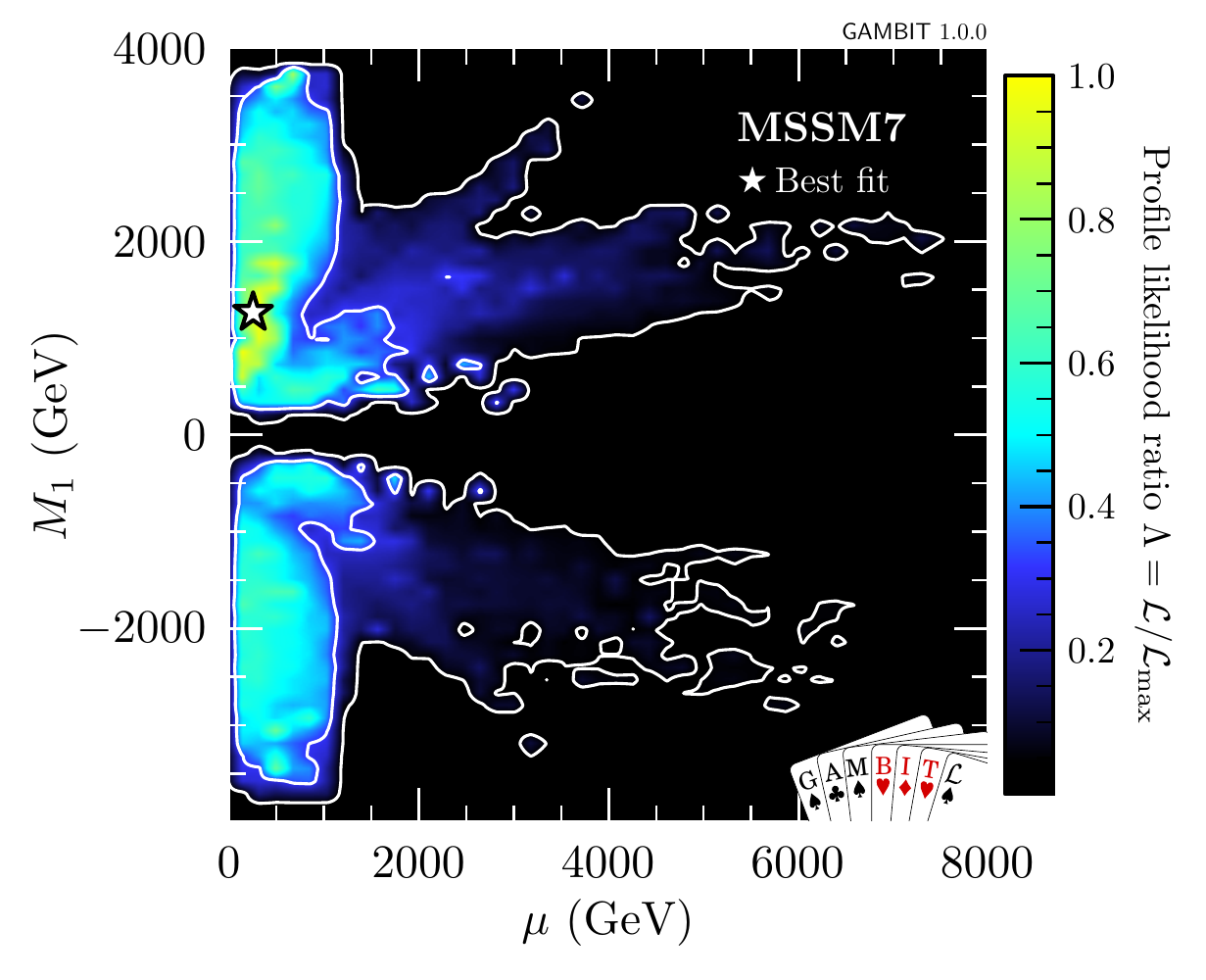}
  \includegraphics[width=0.49\textwidth]{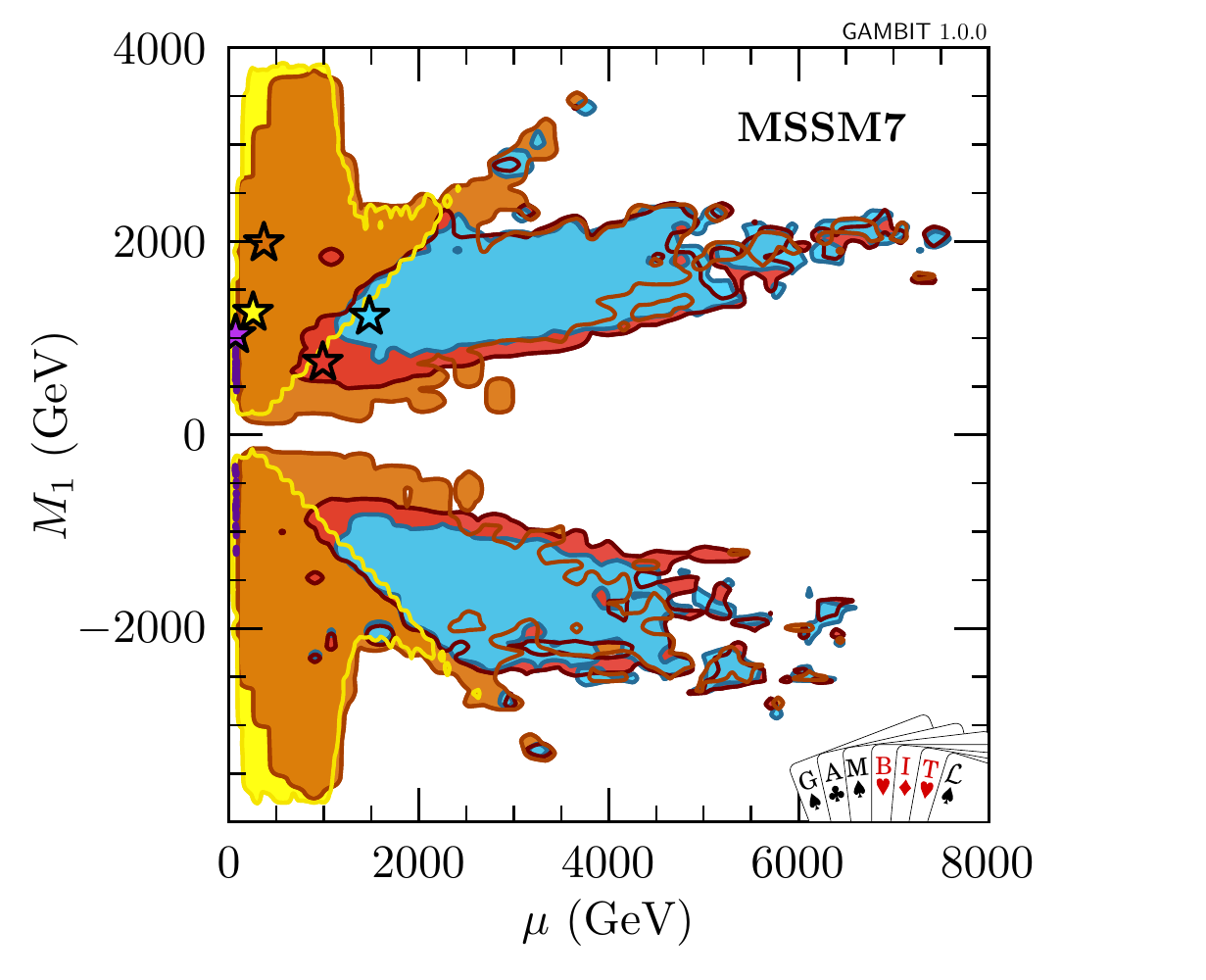}\\
  \includegraphics[width=0.49\textwidth]{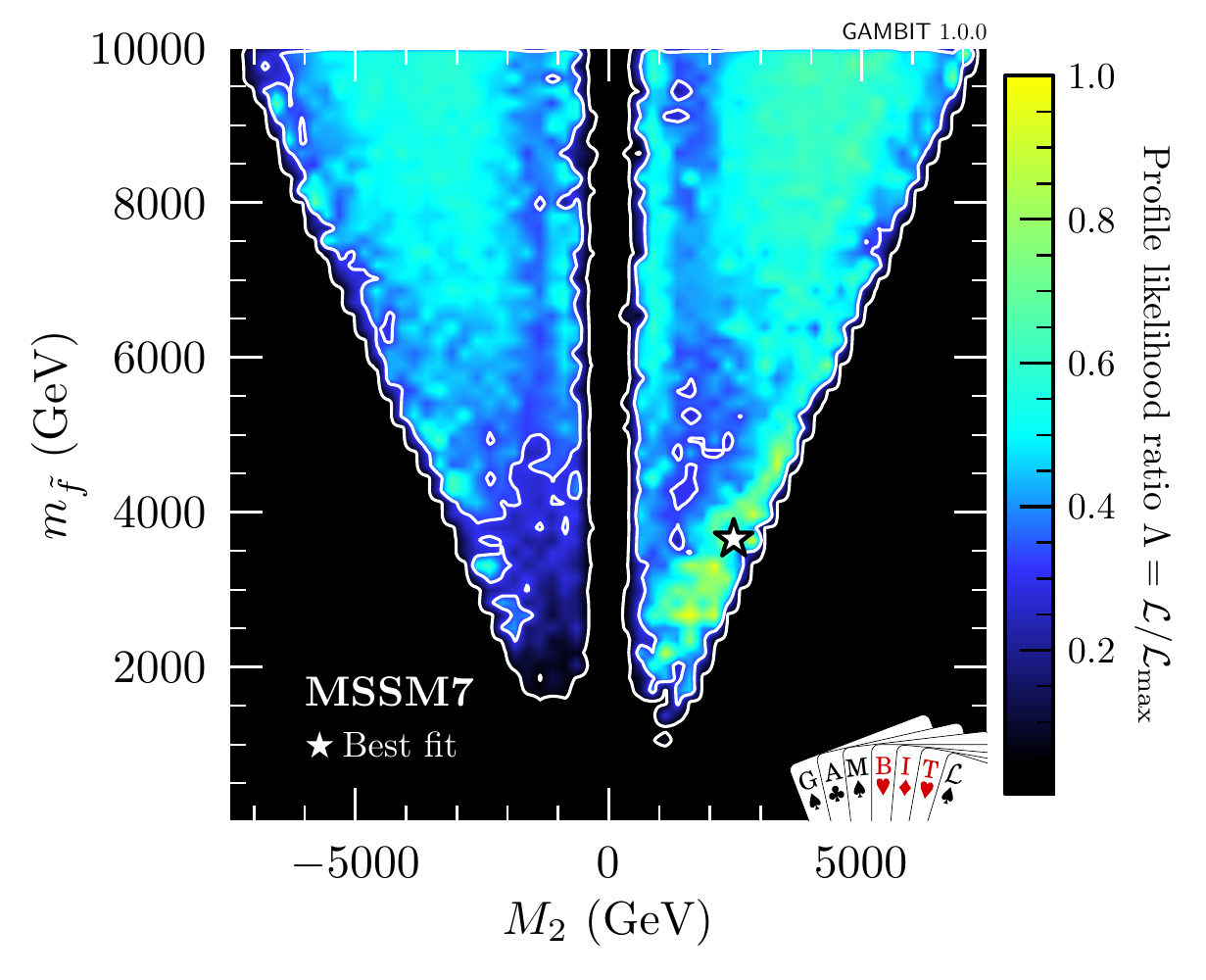}
  \includegraphics[width=0.49\textwidth]{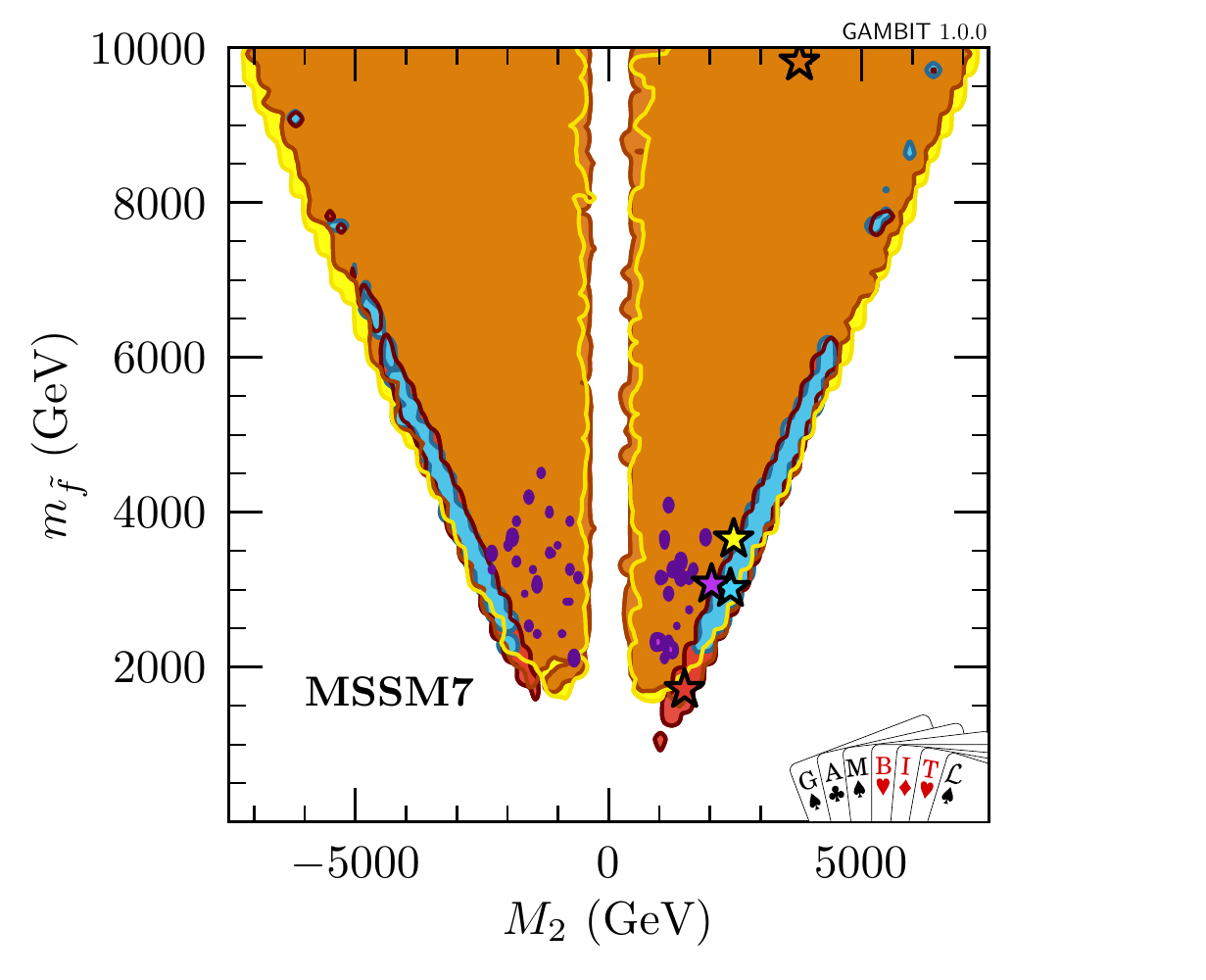}\\
  \includegraphics[height=4mm]{figures/rdcolours5.pdf}
  \caption{
  \textit{Left:} Joint profile likelihoods in the $\mu$--$M_1$ (top) and $M_2$--$m_{\tilde{f}}$ planes (bottom). Stars indicate the point of highest likelihood in each plain, and white contours correspond to the $1\sigma$ and $2\sigma$ CL regions with respect to the best-fit point.
  \textit{Right:} Coloured regions indicating in which parts of the $2\sigma$ best-fit region different co-annihilation and funnel mechanisms contribute to keeping the relic density low. The best-fit point in each region is indicated by a star with the corresponding colour.
  }
  \label{fig:2d_param_planes}
\end{figure*}

As with the CMSSM, NUHM1 and NUHM2, the individual contributions to $\Delta\ln\mathcal{L}_\mathrm{BF} = 36.345$ come almost entirely from known anomalies unexplainable in either the SM or MSSM7.  In particular, these include the $D$ and $D^*$ meson decay ratios $R_D$ and $R_{D^*}$ (contained in the tree-level $B$ and $D$ decay likelihood; see Sec.\ 5.1 of Ref.\ \cite{FlavBit}), the magnetic moment of the muon ($a_\mu$; see Sec.\ 4.2.2 of Ref.\ \cite{SDPBit}) and the angular observables of the electroweak penguin decay $B^0\to K^{*0}\mu^+\mu^-$ (see Sec.\ 5.2 of Ref.\ \cite{FlavBit}).  We refer the reader to Secs.\ 4.1--4.3 of the companion paper \cite{CMSSM} for further discussion.

The best fits possible by relic density mechanisms other than chargino co-annihilation are not drastically worse than the global best fit.  The best models in the four other regions all lie within $\Delta\ln\mathcal{L} < 1.2$ of the global optimum.  Compared to the best fit with chargino co-annihilation, the best stop co-annihilation model has the light stop ($m_{\tilde{t}_1} = 760$\,GeV) needed to co-annihilate with the neutralino, and therefore a light sfermion spectrum more generally, due to the universality of $m_{\tilde{f}}$ at the weak scale.  This point thus ends up being penalised by both the LHC Higgs likelihood and $B\to X_s\gamma$, but advantaged by $B^0\to K^{*0}\mu^+\mu^-$.  In contrast, the best-fit sbottom co-annihilation point has a heavier spectrum, with all sfermions masses above 1\,TeV, and hence only suffers on $B^0\to K^{*0}\mu^+\mu^-$ relative to the global best fit.  Both the light and heavy funnel best fits are hybrids with chargino co-annihilation, showing light charginos and neutralinos.  The best-fit $A/H$ funnel point is only marginally worse than the global best, improving slightly on it in terms of $B^0\to K^{*0}\mu^+\mu^-$ but losing out due to slightly worse fits to $a_\mu$, $B\to X_s\gamma$ and the LHC Higgs likelihood.  The spectrum of the best fit in the $Z/h$ funnel region is split, with heavy sfermions and gluinos, but light charginos and neutralinos.  The latter lead to significant SUSY loop corrections to the $W$ self energy.  This model is also slightly worse than the best fit in terms of $B^0\to K^{*0}\mu^+\mu^-$ and $a_\mu$, but recovers somewhat by making a smaller contribution to $B\to X_s\gamma$.

\subsection{Preferred regions}

We begin by giving the 1D profile likelihoods for each of our input parameters in Fig.\ \ref{fig:1d_like_parameters}.  For simplicity, we refer to $\msf \equiv ({\msfsq})^{1/2}$ rather than the input parameter $\msfsq$.  We also give 1D profile likelihoods for the derived parameters $\mu$ and $M_1$.  The GUT-inspired relation (Eq.\ \ref{eq:GUT_relation}) means that $M_1 \approx 0.48 M_2 \approx 0.18 M_3$, while $|\mu|$ is determined from the EWSB conditions. With $M_1 < M_2$ it follows that $M_1$ and $\mu$ are the main mass parameters controlling the composition of the lightest neutralino. In our results we show $M_1$ and $\mu$ at the scale where the spectrum is calculated, $Q = \sqrt{m_{\tilde{t}_1}m_{\tilde{t}_2}} \equiv M_\mathrm{SUSY}$. Due to the central role played by the $\mu$ parameter, it is more instructive to discuss the results connected to the Higgs sector in terms of $\mu$ and $m_{A^0}$ than $m_{H_u}^2$ and $m_{H_d}^2$.

In Fig.\ \ref{fig:1d_like_parameters} and throughout this paper, we show profile likelihood regions coloured according to the different co-annihilation and funnel mechanisms contributing to keeping the neutralino relic density low enough to evade the \textit{Planck} bound.  These are: chargino co-annihilation (yellow), stop co-annihilation (red), sbottom co-annihilation (blue), the $A/H$ funnel (orange) and the $h/Z$ funnel (purple); the definitions of these classifications can be found in the previous subsection.

Figure \ref{fig:2d_param_planes} shows the 2D joint profile likelihood for $M_1$ and $\mu$ (top) and $M_2$ and $m_{\tilde{f}}$ (bottom).  The edges of the coloured regions here correspond to 95\% CL relative to the best fit of the entire sample, not relative to the best fits of each coloured region.  Here we see that the parameter space allowed at 95\% CL encompasses three distinct regions, each expressing a different composition for the lightest neutralino and chargino:
\begin{description}
\item[\textbf{Region 1.}] $\mu < |M_1|$. $\tilde{\chi}_1^0$ and $\tilde{\chi}_1^\pm$ are mainly Higgsino.
\item[\textbf{Region 2.}] $\mu \approx |M_1|$.  $\tilde{\chi}_1^0$ is a Higgsino/bino mixture and $\tilde{\chi}_1^\pm$ is dominantly Higgsino.
\item[\textbf{Region 3.}] $\mu > |M_1|$.  $\tilde{\chi}_1^0$ is bino. As $\mu$ increases, $\tilde{\chi}_1^\pm$ remains Higgsino-dominated up to $\mu \approx 2|M_1| \approx M_2$, after which the wino component dominates.
\end{description}
Due to Eq.\ \ref{eq:GUT_relation}, a purely wino-dominated $\tilde{\chi}_1^0$ is not possible in the MSSM7.

For Regions 1 and 2, the masses of the lightest chargino and the two lightest neutralinos are nearly degenerate, and all very close to $\mu$.  The neutralino relic density is therefore depleted by all pairwise annihilations and co-annihilations between the three species, which we collectively refer to simply as `chargino co-annihilation'. In Region 1, where the lightest neutralino is essentially a pure Higgsino, the relic density constraint implies $\mu \lesssim 1.2$~\TeV. The $A/H$-funnel also contributes across most of Regions 1 and 2, except in the case of very low $\mu$ or $\mu \ll |M_1|$, where the dependence of $m_{A^0}$ on $|\mu|$ makes it difficult to satisfy the funnel relation $m_{A^0} \sim 2m_{\tilde\chi^0_1}$.

In Region 3, a small mass difference between the lightest neutralino and chargino is no longer automatic. The dominant relic density mechanisms in this parameter region are stop and sbottom co-annihilation, supported by annihilation through the $A/H$ funnel.  The tuning required in the former to get the lightest neutralino and lightest squark nearly degnerate in mass shows up as strongly-correlated bands in the $M_2$--$\msf$ plane (lower panels of Fig.\ \ref{fig:2d_param_planes}). Because the MSSM7 employs a common sfermion soft-mass parameter $m_{\tilde{f}}^2$ at the input scale ($Q = 1$~\TeV in our case), mass splittings among different sfermions are mostly generated by varying amounts of mixing. In comparison, the contribution from RGE running from $Q = 1$~\TeV to $Q = M_\mathrm{SUSY}$, which splits $m_{\tilde{f}}^2$ into individual soft masses, is generally subdominant.

In the tree-level stop mass matrix the off-diagonal element is $v y_t(A_{u_3} \sin\beta - \mu \cos\beta)$, while it is $v y_{b,\tau}(A_{d_3} \cos\beta - \mu \sin\beta)$ in the sbottom and stau mass matrices, where $y_{t,b,\tau}$ are the fermion Yukawa couplings and $v \approx 246$~\GeV. Because increased left-right mixing reduces the mass of the lighter of the two mass eigenstates, the large top Yukawa ensures that $\tilde{t}_1$ is the lightest sfermion across most of the allowed parameter space (including for models that exhibit sbottom co-annihilation). With $3 \leq \tan\beta \leq 70$ the terms $A_{u_3} \sin\beta$ (stop) and $\mu \sin\beta$ (sbottom and stau) dominate the sfermion mixing in large regions of parameter space. The dependence on large $\mu$ to obtain a sbottom mass significantly lower than the mass set by the common $\msf$ parameter explains why the sbottom co-annihilation region does not extend as far to small $\mu$ as the stop co-annihilation region in Fig.\ \ref{fig:2d_param_planes}. Also, since $y_{b} \approx 2.5 y_{\tau}$, the lightest stau remains heavier than the lightest sbottom in the regions of parameter space with large mixing for the down-type sfermions, which explains the absence of any region dominated by stau co-annihilation in our results.

The requirement that all sfermions are heavier than the lightest neutralino excludes large regions of parameter space at $\msf \lesssim 1.3|M_2| \approx 2.6 |M_1|$ in the bottom panels of Fig.\ \ref{fig:2d_param_planes}. The steep slope of the exclusion boundary can broadly be understood as a consequence of the $\mu$-dependent mixing in the sfermion sector. The region close to the boundary at $\msf \approx 1.3|M_2|$ is part of Regions 2 and 3 ($\mu \gtrsim |M_1|$) in the $\mu$--$M_1$ plane, so that increasing $M_2$ in this region pushes up both $M_1$ and $\mu$. To keep the lightest sfermion heavier than the neutralino, $\msf$ must therefore increase enough to compensate \textit{both} the increase in neutralino mass from $M_1$ and the potential decrease in the light sfermion mass due to the $\mu$-dependent left-right mixing. We come back to this interplay between the parameters of the neutralino and sfermion sector when discussing the $\mu$--$\tan\beta$ plane in Fig.\ \ref{fig:2d_param_planes_2}.

The region of small $|M_1|$ in the upper row of Fig.\ \ref{fig:2d_param_planes} (and therefore also small $|M_2|$ in the lower row) is strongly constrained by LHC limits.  Direct LHC searches are also strongly constraining at low $m_{\tilde{f}}$ (lower panels).  Gluino searches are particularly effective, as Eq.\ (\ref{eq:GUT_relation}) implies that the gluino mass parameter is $M_3 \approx 7M_1$ at a scale of 1\,TeV.  Given that simplified gluino mass limits reach up to $1.9$\,TeV \cite{ATLAS-CONF-2016-078}, this disfavours bino masses in the MSSM7 of up to $\sim$300\,GeV.  Indeed, this is the main reason that we do not find the same preference for very light binos observed in Ref.\ \cite{MasterCodeMSSM10}, where each of the gaugino masses was allowed to vary independently.  The common sfermion mass parameter means that, for light 3rd generation squarks, the 1st and 2nd generation squarks are not necessarily decoupled. Thus, LHC searches for 1st and 2nd generation squarks also constrain how far down towards low neutralino masses (small $\mu$ or $|M_1|$; upper panels of Fig.\ \ref{fig:2d_param_planes}) and low sfermion masses (lower panels of Fig.\ \ref{fig:2d_param_planes}) the stop and sbottom co-annihilation regions extend.  Measurements of the $125$~\GeV Higgs, limits from DM direct detection experiments, flavour physics and precision measurements of the $W$ mass also contribute to disfavouring low gaugino masses in our fits.

At low sfermion masses, we also see a weak preference for positive $M_2$, stemming from the $(g-2)_\mu$ likelihood. Because we assume $\mu > 0$ for our model, having $M_2$ (and thus $M_1$) positive ensures a positive SUSY contribution to $(g-2)_\mu$.

\begin{figure*}[tbh]
  \centering
  \includegraphics[width=0.49\textwidth]{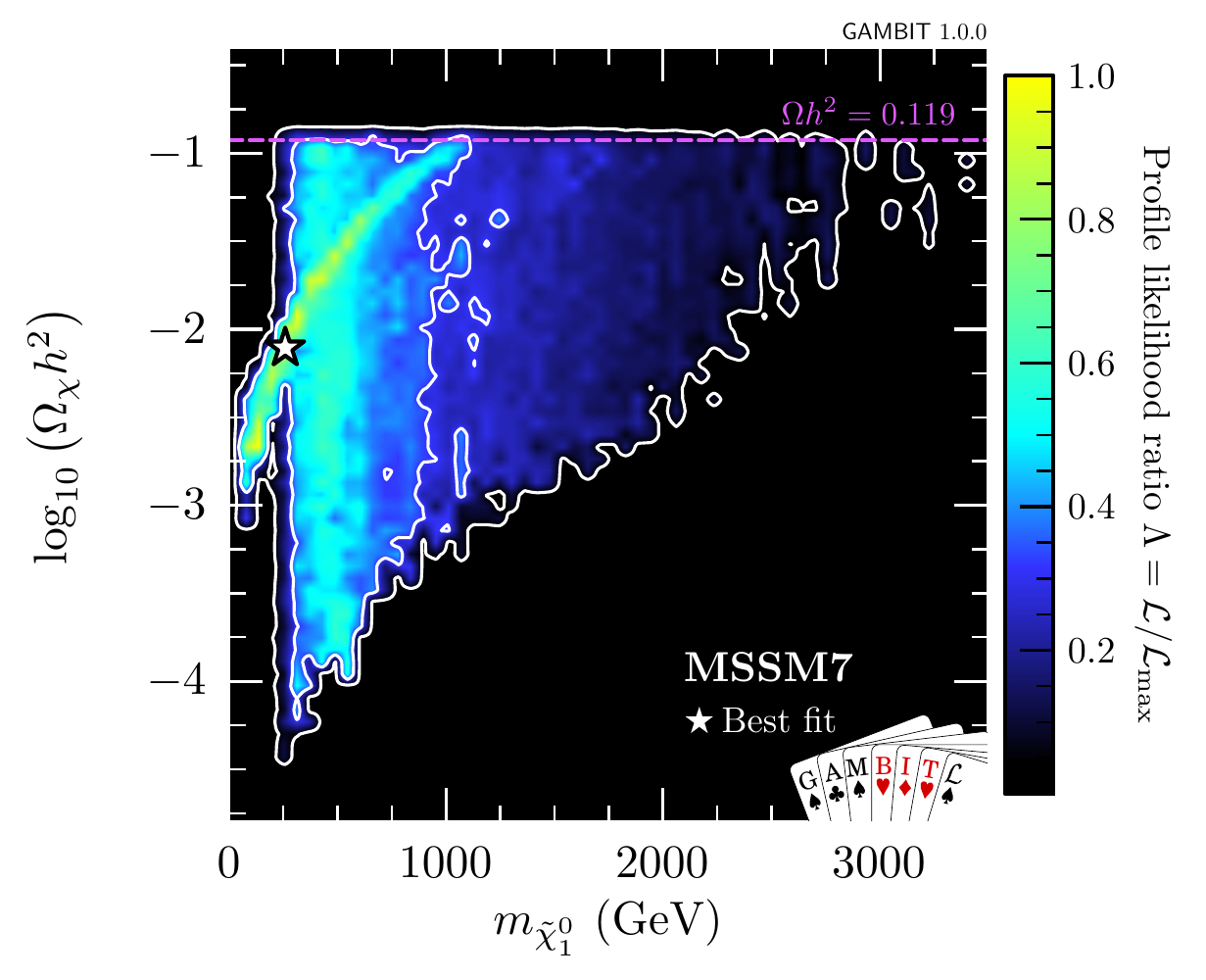}
  \includegraphics[width=0.49\textwidth]{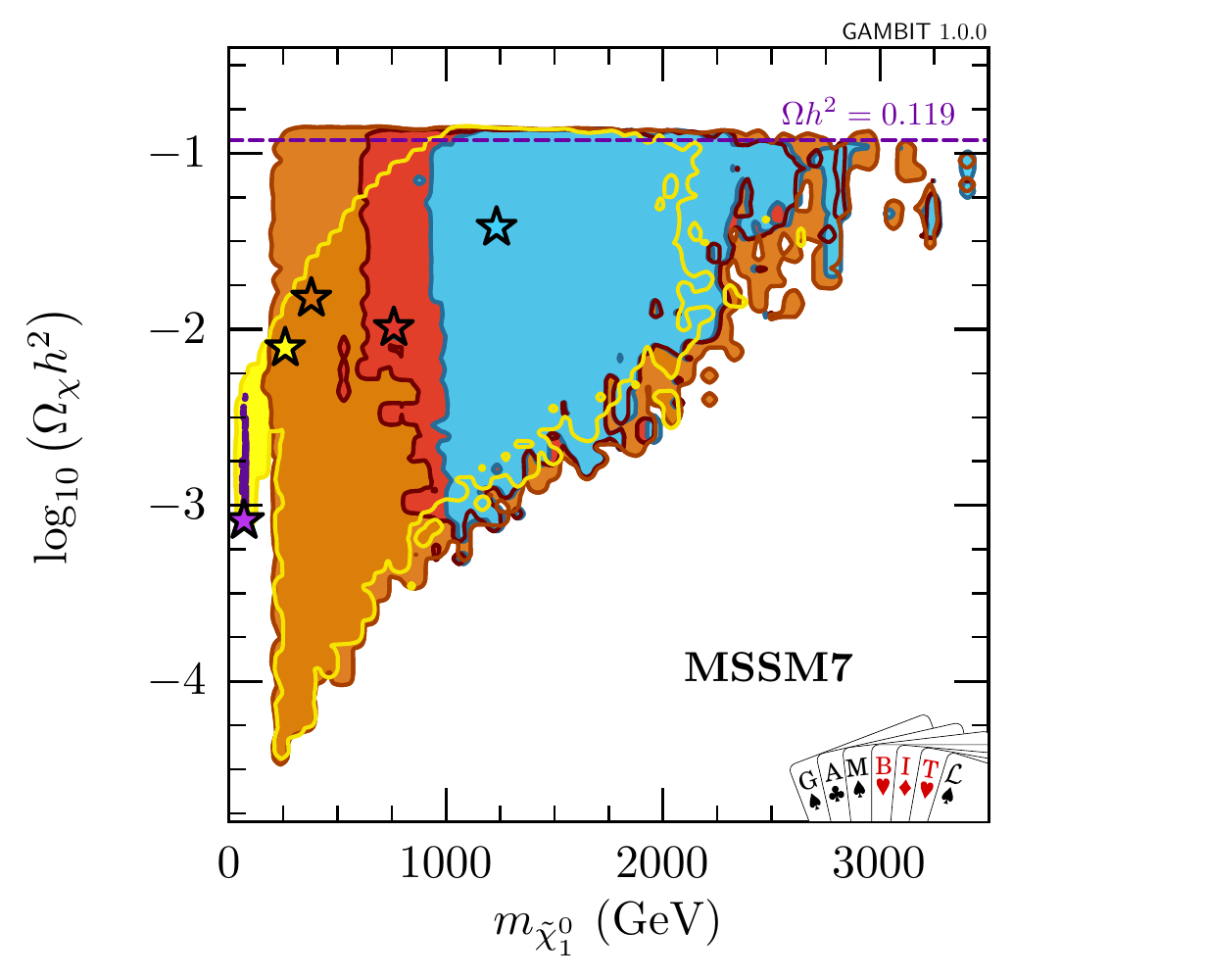}\\
  \includegraphics[width=0.49\textwidth]{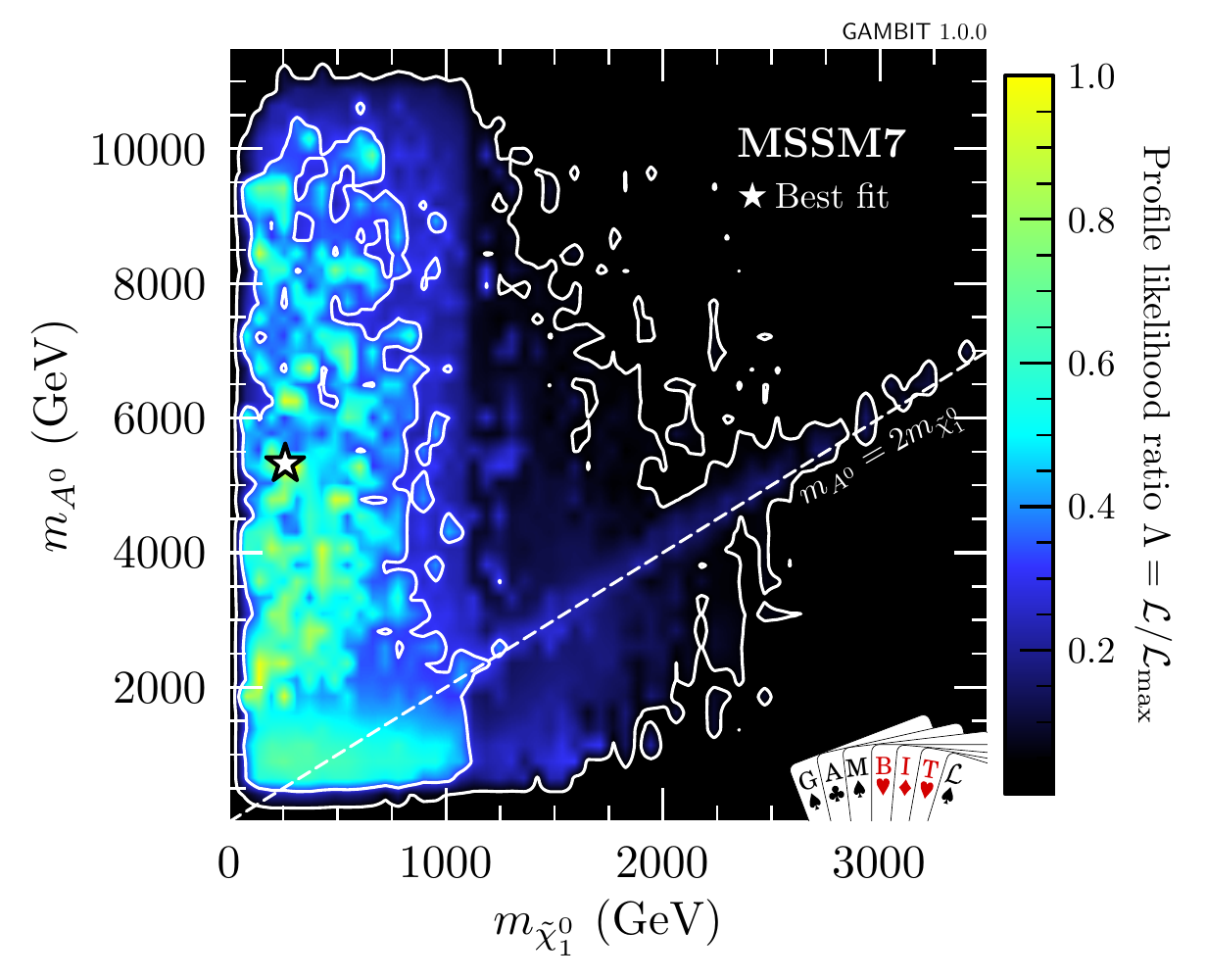}
  \includegraphics[width=0.49\textwidth]{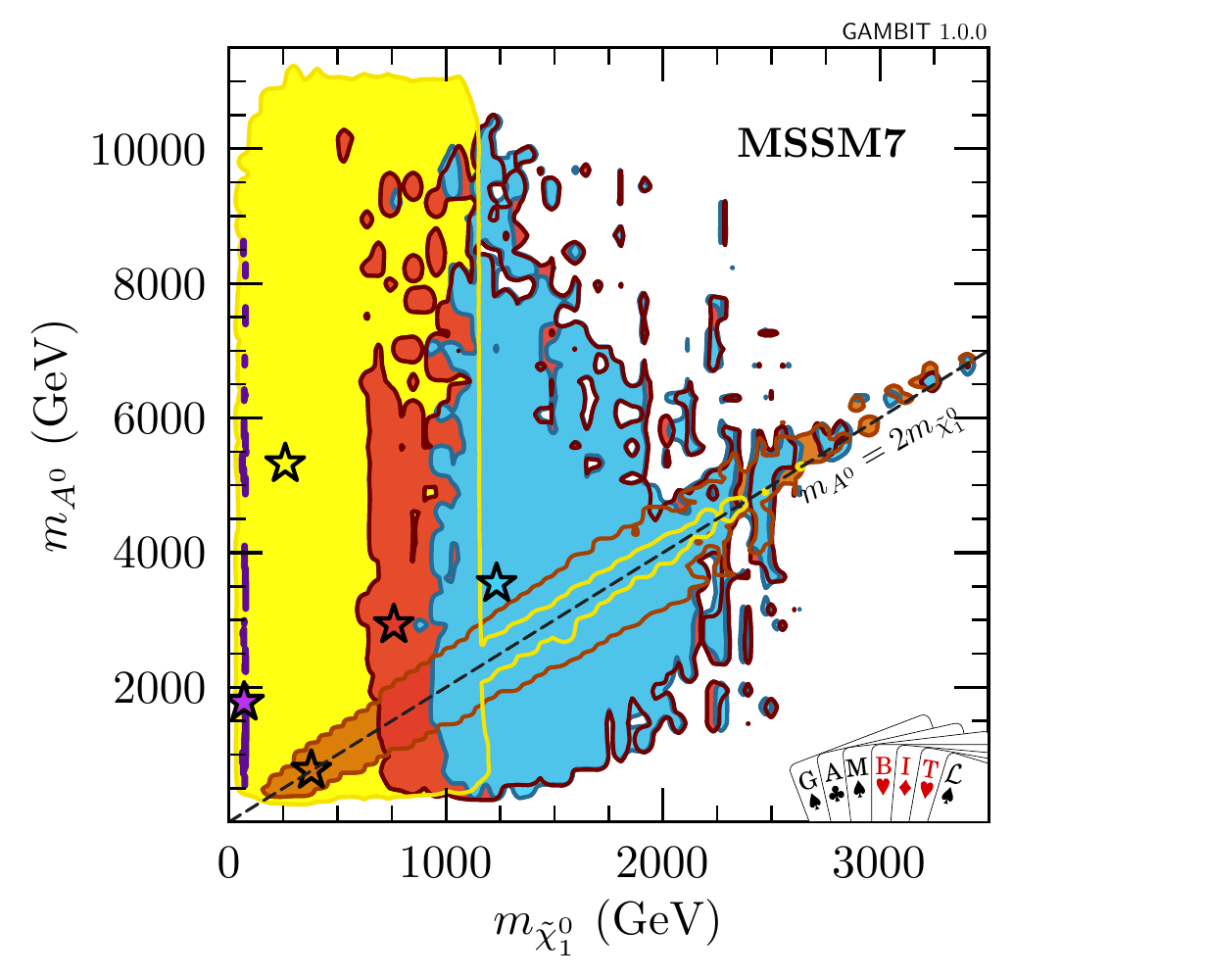}\\
  \includegraphics[height=4mm]{figures/rdcolours5.pdf}
  \caption{\textit{Left:} Joint profile likelihoods in the mass of the lightest neutralino and its relic density $\Omega h^2$ (top), and in the masses of the lightest neutralino and the CP-odd Higgs $A^0$ (bottom). \textit{Right:} Coloured regions indicating in which parts of the $2\sigma$ best-fit region different co-annihilation and funnel mechanisms contribute to the relic density. The best-fit point in each region is indicated by a star with the corresponding colour.}
  \label{mchi_vs_Oh2}
\end{figure*}

In Fig.\ \ref{mchi_vs_Oh2} we explore the impacts of the relic density constraint on the MSSM7 in more detail, investigating the profile likelihood of $\Omega h^2$ and $m_A$ as a function of the mass of the neutralino.  The behaviour of Higgsino DM follows a relatively well-known pattern, seen also in the CMSSM and NUHM \cite{CMSSM}: Higgsino DM co-annihilates steadily less efficiently as the neutralino mass increases, passing through the observed value of the relic density at $m_{\tilde\chi_1^0}\sim1.2$\,TeV.  At higher masses, exceeding the observed relic density can only be avoided by resorting (whether in full or in part) to the heavy Higgs funnel or another co-annihilation mechanism --- in this case, stop and/or sbottom co-annihilation.  This can be seen in the lower-right panel of Fig.\ \ref{mchi_vs_Oh2}, where above $m_{\tilde\chi_1^0}\sim1.2$\,TeV, the chargino co-annihilation region only exists along the funnel line $m_{A^0}\sim 2m_{\tilde\chi^0_1}$.

At $m_{\tilde\chi_1^0}\lesssim1.2$\,TeV, the efficiency of Higgsino co-annihilation makes for sub-dominant Higgsino DM, as seen in the diagonal chargino co-annihilation region in the upper-right panel of Fig.\ \ref{mchi_vs_Oh2}.  This can be tempered by fine-tuning the Higgsino-bino mixture, bringing up the relic density to the observed value, but such combinations are now very strongly constrained by direct detection, where mixed gaugino-Higgsino DM maximises both the spin-dependent and spin-independent neutralino-nucleon scattering cross-sections.

\begin{figure*}[tbh]
  \centering
  \includegraphics[width=0.49\textwidth]{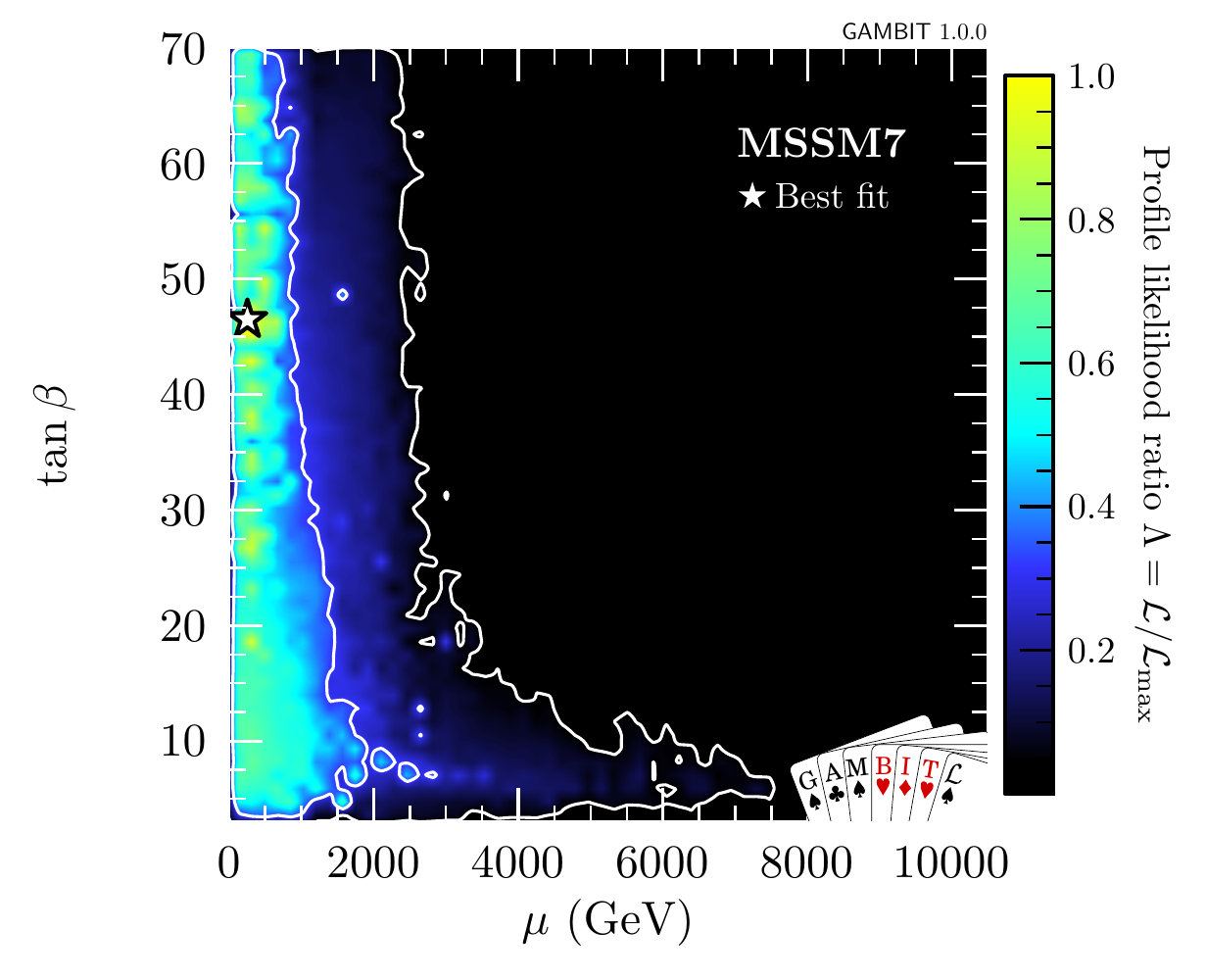}
  \includegraphics[width=0.49\textwidth]{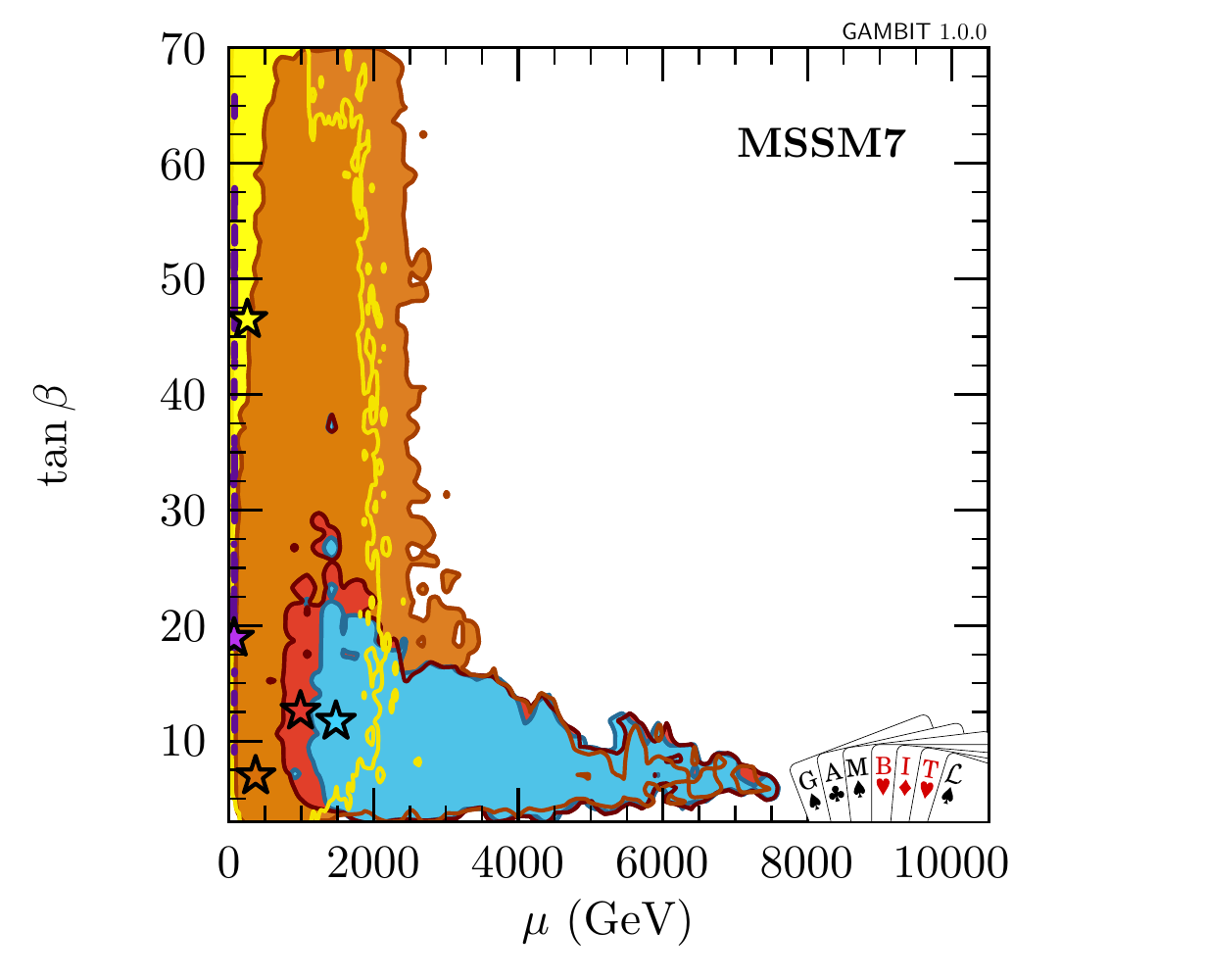}\\
  \includegraphics[width=0.49\textwidth]{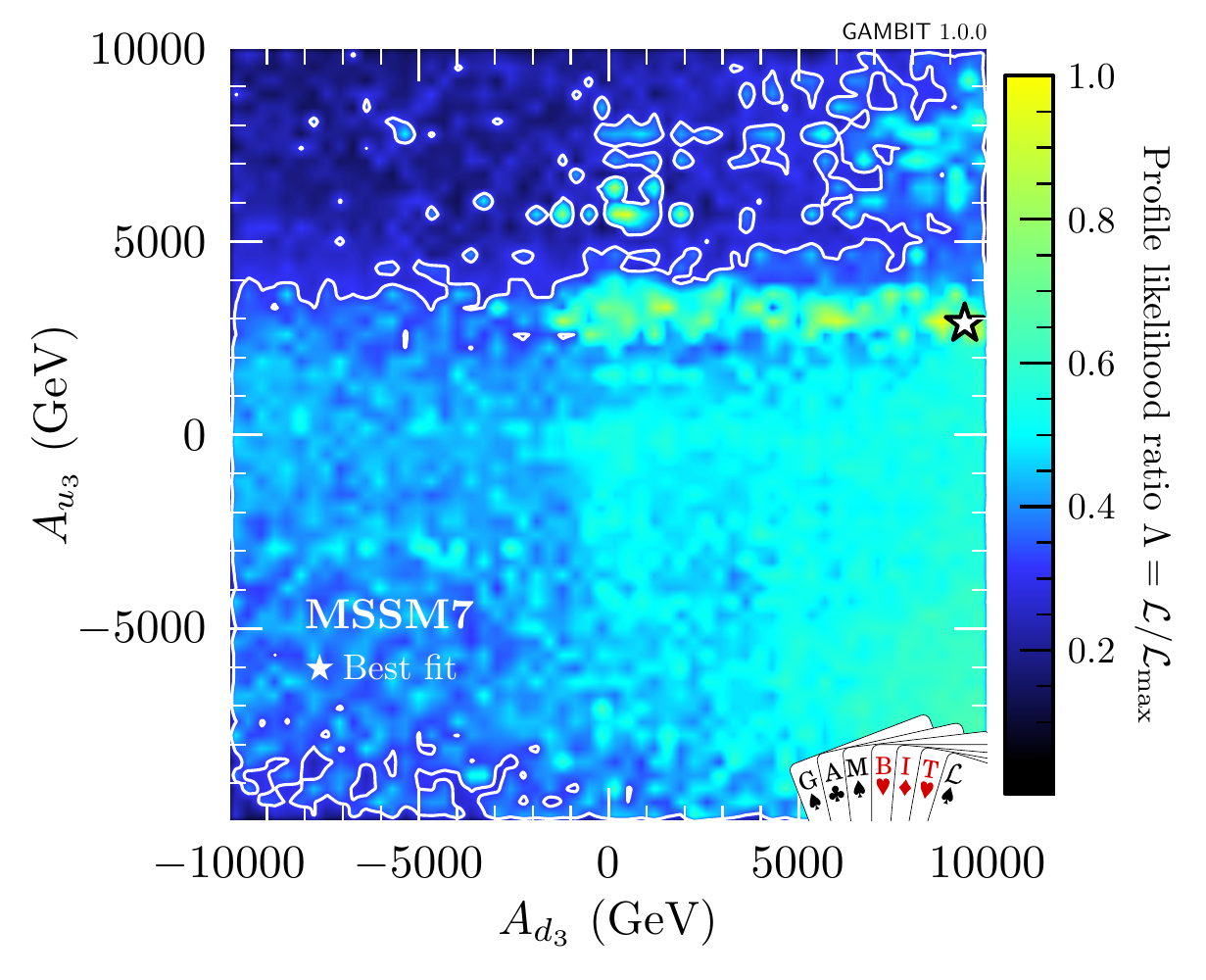}
  \includegraphics[width=0.49\textwidth]{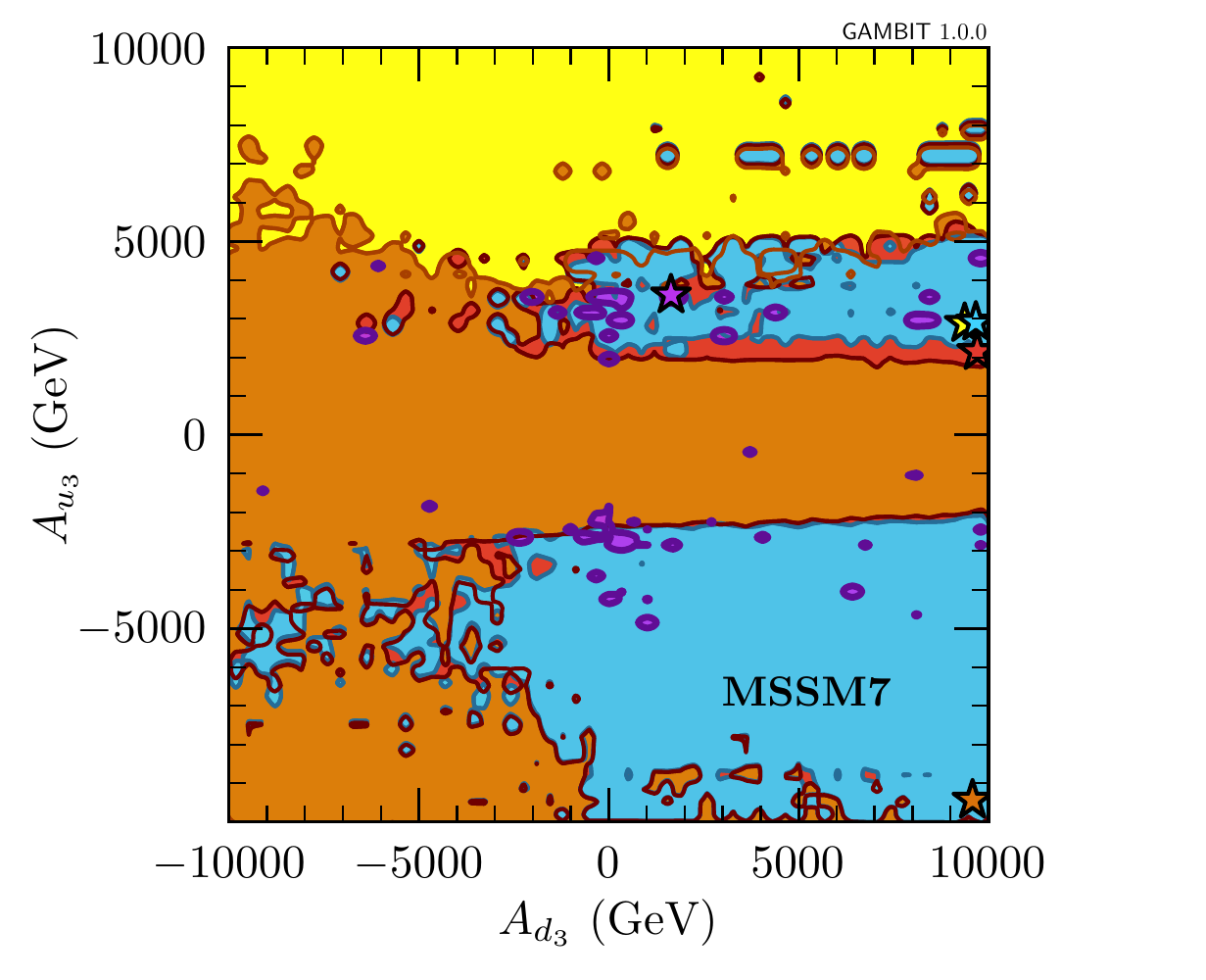}
  \includegraphics[height=4mm]{figures/rdcolours5.pdf}
  \caption{\textit{Left:} Joint profile likelihoods in $\mu$--$\tan\beta$ (top) and $A_{u_3}$--$A_{d_3}$ planes. \textit{Right:} Coloured regions indicating in which parts of the $2\sigma$ best-fit region different co-annihilation and funnel mechanisms contribute to the relic density. The best-fit point in each region is indicated by a star with the corresponding colour. In the bottom right plot the yellow chargino co-annihilation region covers the entire plane and the orange $A/H$ funnel region spans the entire plane below $A_{u_3} \sim 5$~\TeV.}
  \label{fig:2d_param_planes_2}
\end{figure*}

At very low masses, the chargino co-annihilation region reaches down far enough that resonant annihilation via the SM Higgs further boosts the annihilation cross-section, leading to a region of hybrid chargino co-annihilation-$h$ funnel models with neutralino masses as low as 61\,GeV.\footnote{For dedicated analyses of scenarios with a very light neutralino in MSSM parameterisations without a GUT relation on the gaugino masses, see for instance Refs.~\cite{Dreiner:2007fw, Profumo:2008yg, Dreiner:2009ic, Arbey:2012na, Boehm:2013gst}}
The best fit in this region (Table\ \ref{tab:mssm7-bf-1}) has $m_{\tilde\chi_1^0} = 69.2$\,GeV, $m_{\tilde\chi_1^\pm}=71.6$\,GeV and $m_{\tilde\chi_2^0} = 73.7$\,GeV, while the other sparticles are fairly heavy. This leads to considerable cross sections for direct pair production of $\tilde\chi_1^0\tilde\chi_2^0$ and $\tilde\chi_1^+\tilde\chi_1^-$ at LEP.  Indeed, such masses would naively seem to be in contradiction with published limits, e.g. $m_{\tilde\chi_1^\pm} > 94$ GeV~\cite{Olive:2016xmw,Abdallah:2003xe}.  However, this particular limit assumes $m_{\tilde\chi_1^\pm}-m_{\tilde\chi_1^0}>3$\,GeV, and does not strictly apply to our best fit. The \GB implementation of LEP limits in \colliderbit, detailed in Sec.\ 2.2 of Ref.\ \cite{ColliderBit}, takes into account the mass-dependent signal efficiency for the chargino and neutralino searches.  These are quite important for cases where the spectrum has some degenerate masses, as in our best fit.  In this case, the relevant search is the one for leptonic decays of the chargino at L3, with results shown in Fig.~2b of Ref.~\cite{L3:gauginos}.  Our treatment is a significant improvement on the hard lower limits that have often been used in the past.

Fig.\ \ref{mchi_vs_Oh2} shows that the heavy Higgs funnel can work for a wide range of neutralino masses in the MSSM7, from $\sim$200\,GeV up to many TeV. The lower limit here comes from the lower limit on the mass of the CP-odd Higgs boson, seen in the bottom-left corner of the $m_{A^0}$--$m_{\tilde\chi^0_1}$ plane (Fig.\ \ref{mchi_vs_Oh2}). This arises due to penalties from the flavour physics likelihoods and the LHC Higgs likelihood. Because $A^0$ is close in mass to $m_{H^+}$ ($m_{H^+}^2 = m_{A^0}^2 + m_W^2$ at tree level), having a light $A^0$ causes tension with the $BR(B \rightarrow X_s \gamma)$ likelihood, which in isolation requires $m_{H^+} \gtrsim 570$~\GeV at 95\% CL for type-II two Higgs doublet models such as the MSSM \cite{Misiak:2017bgg}. For large $\tan\beta$, the likelihoods for tree-level leptonic and semi-leptonic $B$ and $D$ decays also penalise low $A^0$ masses. The tension with these likelihoods at low masses is to some extent compensated for by an improvement in the fit to the electroweak penguin decay $B^0 \rightarrow K^{*0} \mu^+\mu^-$, but for $m_A^0 \lesssim 400$~\GeV, the combined restrictions imposed by flavour physics and measurements of the $125$~\GeV Higgs push the likelihood below the $95\%$ CL, as evident in Fig.~\ref{mchi_vs_Oh2}.

In this paper we have allowed neutralinos to be a sub-dominant component of DM.  Were we to instead require that they constitute all of DM, our fits would be concentrated in the area around the horizontal line in the upper panels of Fig.\ \ref{mchi_vs_Oh2}.  This would restrict the Higgsino-dominated DM models of the chargino co-annihilation region to $m_{\tilde{\chi}_1^0} \gtrsim 1$~\TeV, moving the best-fit point to the $A/H$ funnel and a mass of $m_{\tilde{\chi}_1^0} = 416$~\GeV.  In terms of the neutralino mass itself, this would rule out $m_{\tilde{\chi}_1^0} < 250$~\GeV at 95\% CL (1D).  As we discuss later in this section, the absence of light charginos would also degrade the (already poor) fit to $a_\mu$.

In Fig.\ \ref{fig:2d_param_planes_2}, we show the preferred regions and relic density mechanisms active in the $\mu$--$\tan\beta$ and $A_{d_3}$--$A_{u_3}$ planes.  The shape of the allowed region in the $\mu$--$\tan\beta$ plane can be understood as follows. For the scenario in Region 1 of the upper panels of Fig.\ \ref{fig:2d_param_planes}, $\mu \ll M_1$ and the lightest neutralino is dominantly Higgsino.  This leads to the relic density bound $\mu \lesssim 1.2$~\TeV.  In Region 2, where the lightest neutralino is a mixture of bino and Higgsino, this upper bound on $\mu$ increases to $\sim$$2.5$\,TeV.  This limit is where we see the edge of the chargino co-annihilation and $A/H$-funnel regions at intermediate and large $\tan\beta$ in the upper panels of Fig.\ \ref{fig:2d_param_planes_2}.

In Region 3, $\mu > |M_1|$ and the lightest neutralino is dominantly bino, so there is no upper bound on $\mu$ from the relic density.  In this case, the viable relic density mechanisms are stop/sbottom co-annihilation and the heavy Higgs funnel.  Stop/sbottom co-annihilation can only work if the bino mass ($M_1$) is similar to the mass of the lightest squark.  At large $\tan\beta$, the left-right mixing in the sbottom mass matrix is proportional to $\mu$, meaning that to keep the sbottom from becoming tachyonic, the diagonal entry ($m^2_{\tilde{f}}$) must be increased as $\mu$ is increased.  Pulling up $m^2_{\tilde{f}}$ therefore pulls up the mass of the lightest squark, which in turn requires pulling $|M_1|$ up in order to stay in the stop/sbottom co-annihilation region.  This is a delicate game, as $|M_1|$ needs to be kept below $\mu$ in order to remain in the bino LSP region (Region 3) at all.  Whether or not this is possible depends on small corrections from other parameters.  At smaller values of $\tan\beta$, the left-right mixing picks up an additional contribution proportional to $A_{d_3}$, and the adjustment can be pulled off with the help of some additional tuning in $A_{d_3}$.  The net result is that $|M_1|$ remains less than $\mu$, but not by more than a factor of a few.  Because the heavy Higgs bosons receive their dominant mass contribution from $|\mu|$, this sets their masses to be a factor of a few times that of the lightest neutralino, making stop/sbottom co-annihilation at higher $\mu$ in Region 3 appear mostly as a hybrid with the $A/H$ funnel.

At large $\mu$ and large $\tan\beta$, models in Region 3 are \textit{also} impacted significantly by the Higgs likelihood. As discussed in Refs.\ \cite{Eberl:1999he,Carena:1999py}, the bottom Yukawa coupling receives important SUSY corrections proportional to $\mu\tan\beta$, coming from gluino--sbottom and charged Higgsino--stop loops. For large $\mu$ and $\tan\beta$, this increases the decay rate $\Gamma(h^0 \rightarrow \bar{b} b)$, which reduces the signal strengths for all other Higgs channels.  The gluino--sbottom contribution is generally dominant, and for $\mu > 0$ it is always positive. On the other hand, the Higgsino--stop contribution is proportional to $A_{u_3}$, so that for large and negative $A_{u_3}$ it can compensate the gluino--sbottom correction. Thus, the good-fit region extending out towards large $\mu$ is dominantly associated with large, negative $A_{u_3}$.

Large $|A_{u_3}|$ may cause the scalar potential of the MSSM to develop a minimum that breaks gauge invariance.  We checked this in the same way as described in Sec.\ 4.1 of the companion paper \cite{CMSSM}, finding even less impact in the MSSM7 than in the CMSSM or NUHM: whilst a small number of individual points are potentially affected by colour- or charge-breaking vacua, the overall preferred regions of the model remain unaffected.  We naively carried out the same tests for $|A_{d_3}|$ as well, swapping all up-type parameters for their down-type equivalents.  We found that a few more models were affected than in the up-type tests, in particular those at large $\mu$ and small $\tan\beta$ discussed in the context of Fig.\ \ref{fig:2d_param_planes_2} above, where $A_{d_3}$ helps to prevent the sbottoms becoming tachyonic.  However, the impact was still quite isolated and had no impact on the overall inference.

\begin{figure}[tbp]
\centering
\includegraphics[width=\columnwidth]{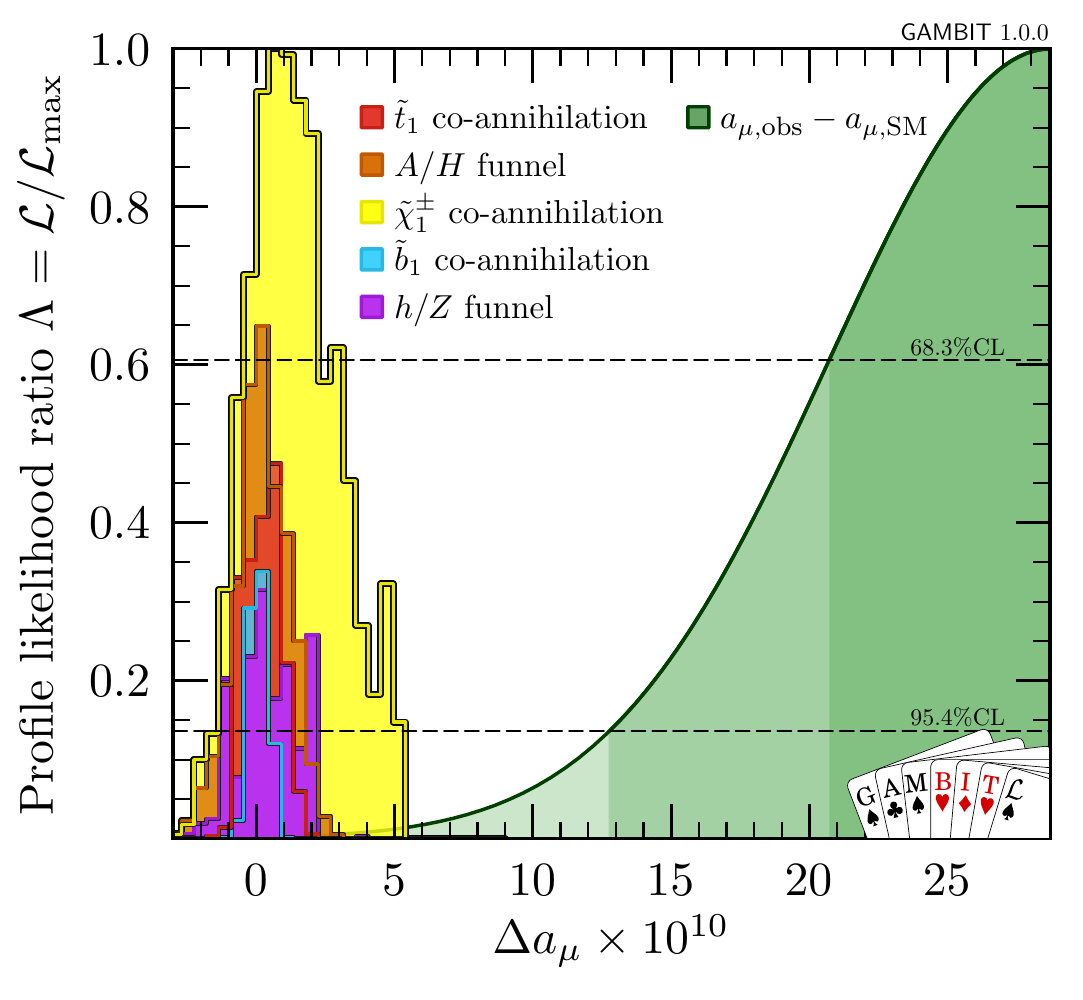}
\caption{1D profile likelihood ratio for the SUSY contribution $\Delta a_{\mu}$ to the anomalous magnetic moment of the muon. In green we show a Gaussian likelihood for the observed value $a_{\mu,\text{obs}} - a_{\mu,\text{SM}} = (28.7 \pm 8.0)\times10^{-10}$, where we have combined the experimental and Standard Model (SM) theoretical uncertainties in quadrature.}
\label{fig:amu}
\end{figure}

\begin{figure*}[tbh]
  \centering
  \includegraphics[width=0.32\textwidth]{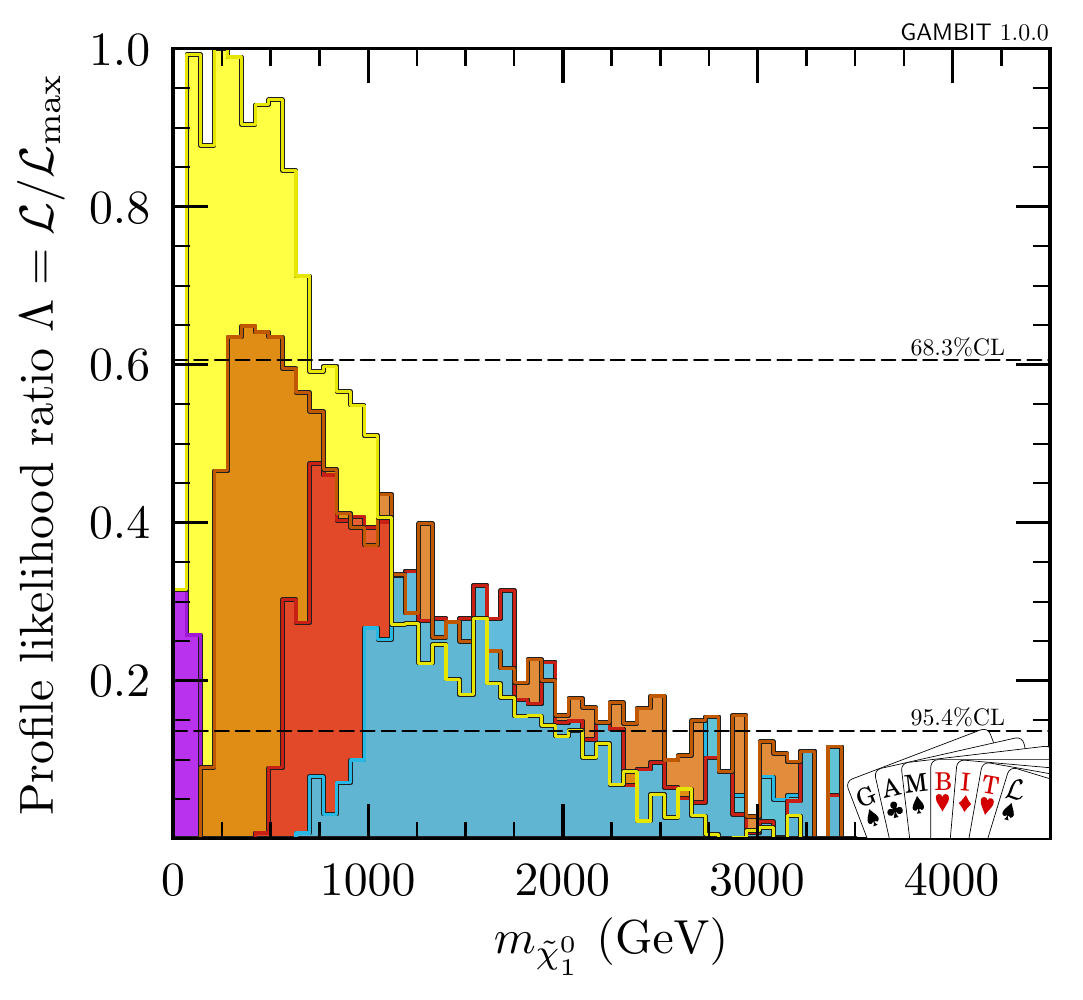}
  \includegraphics[width=0.32\textwidth]{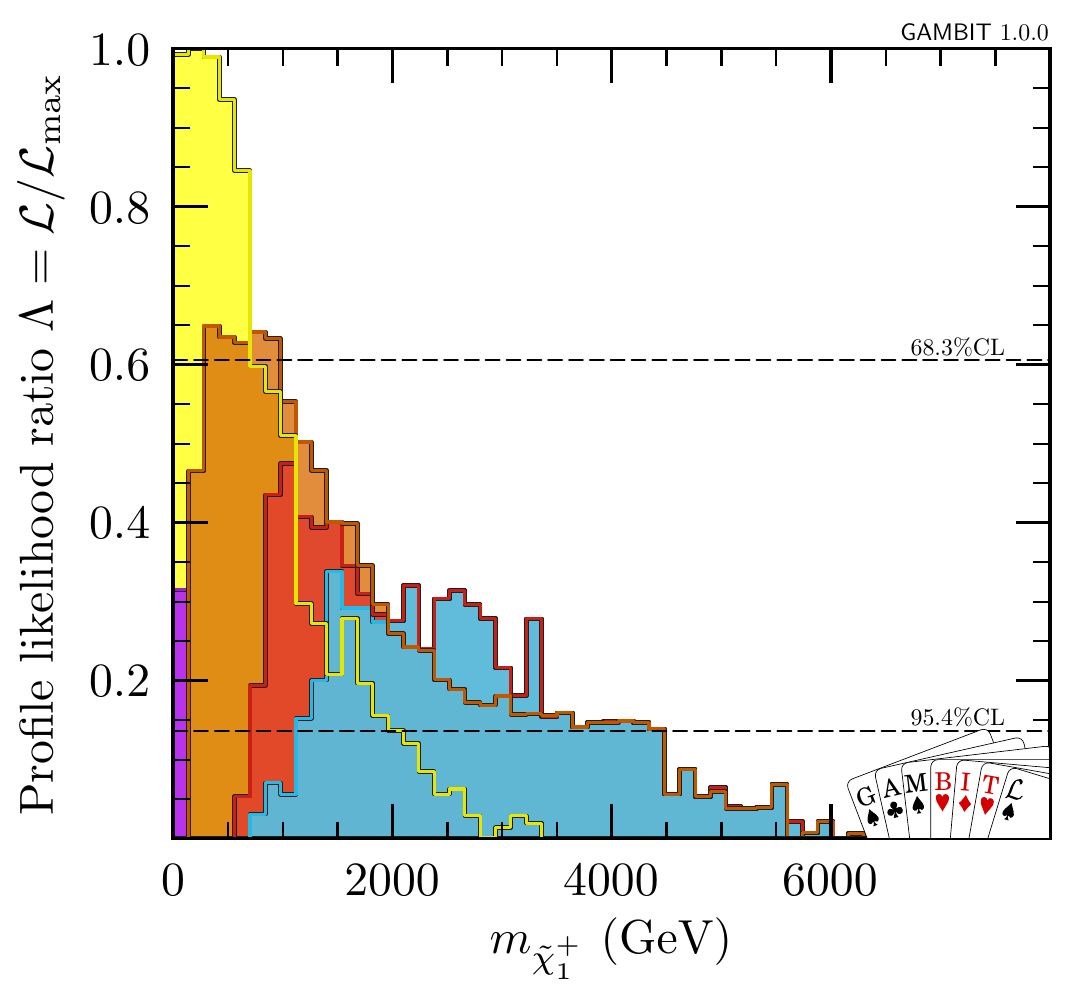}
  \includegraphics[width=0.32\textwidth]{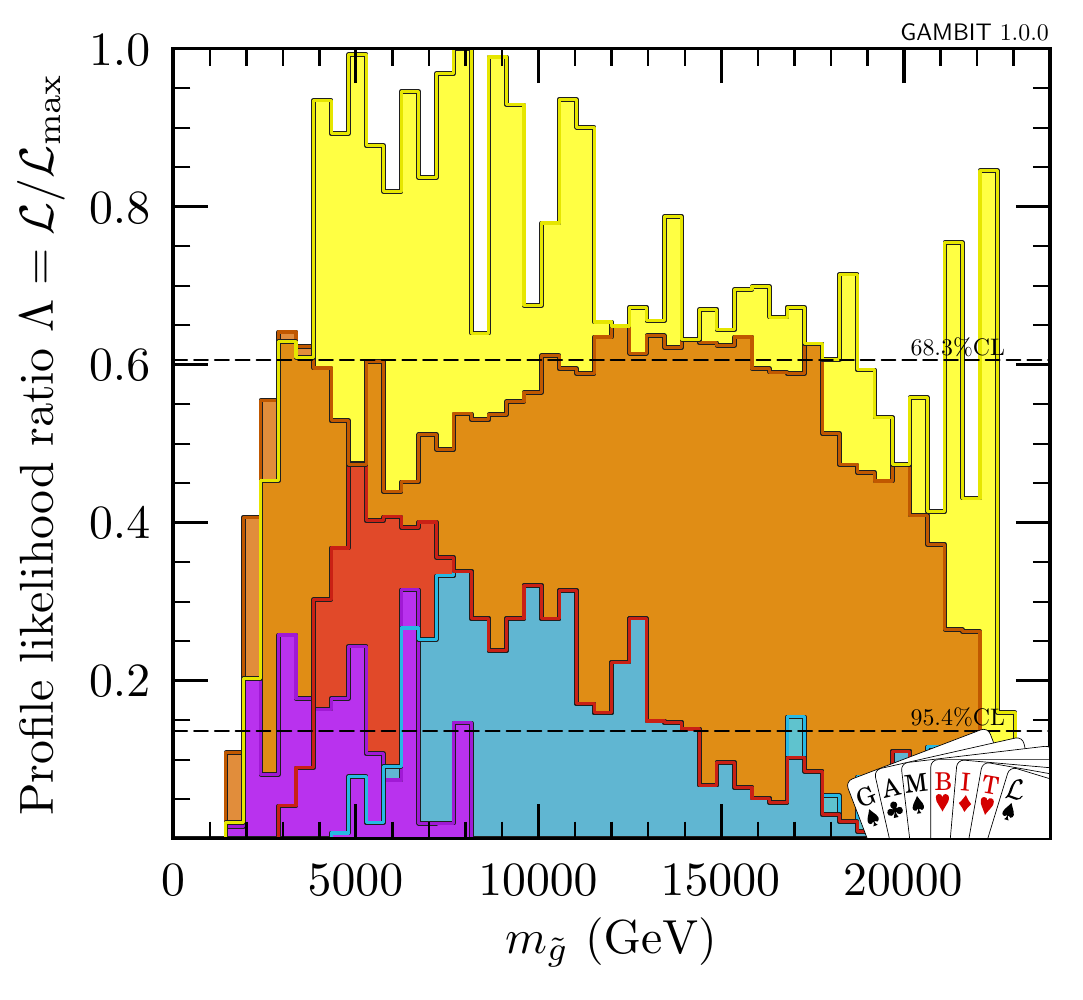}\\
  \includegraphics[width=0.32\textwidth]{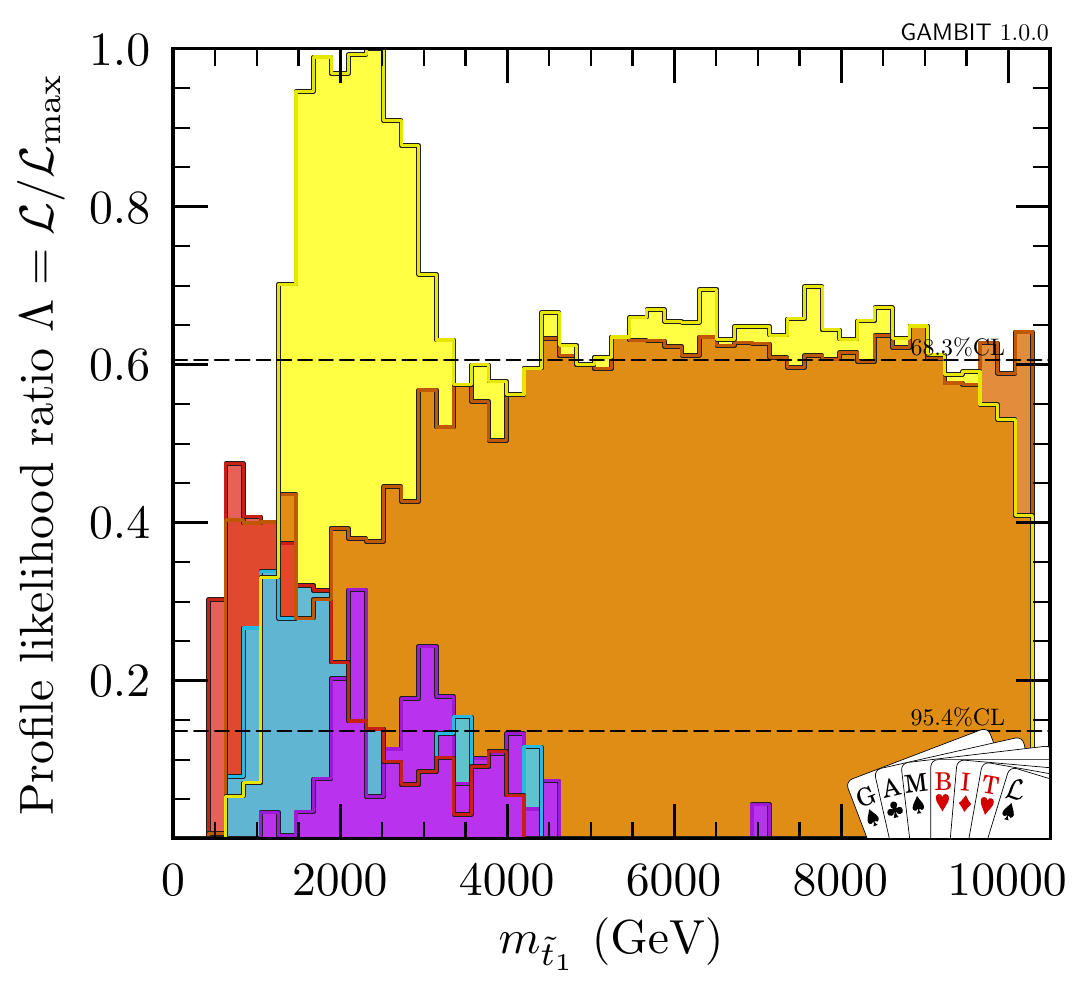}
  \includegraphics[width=0.32\textwidth]{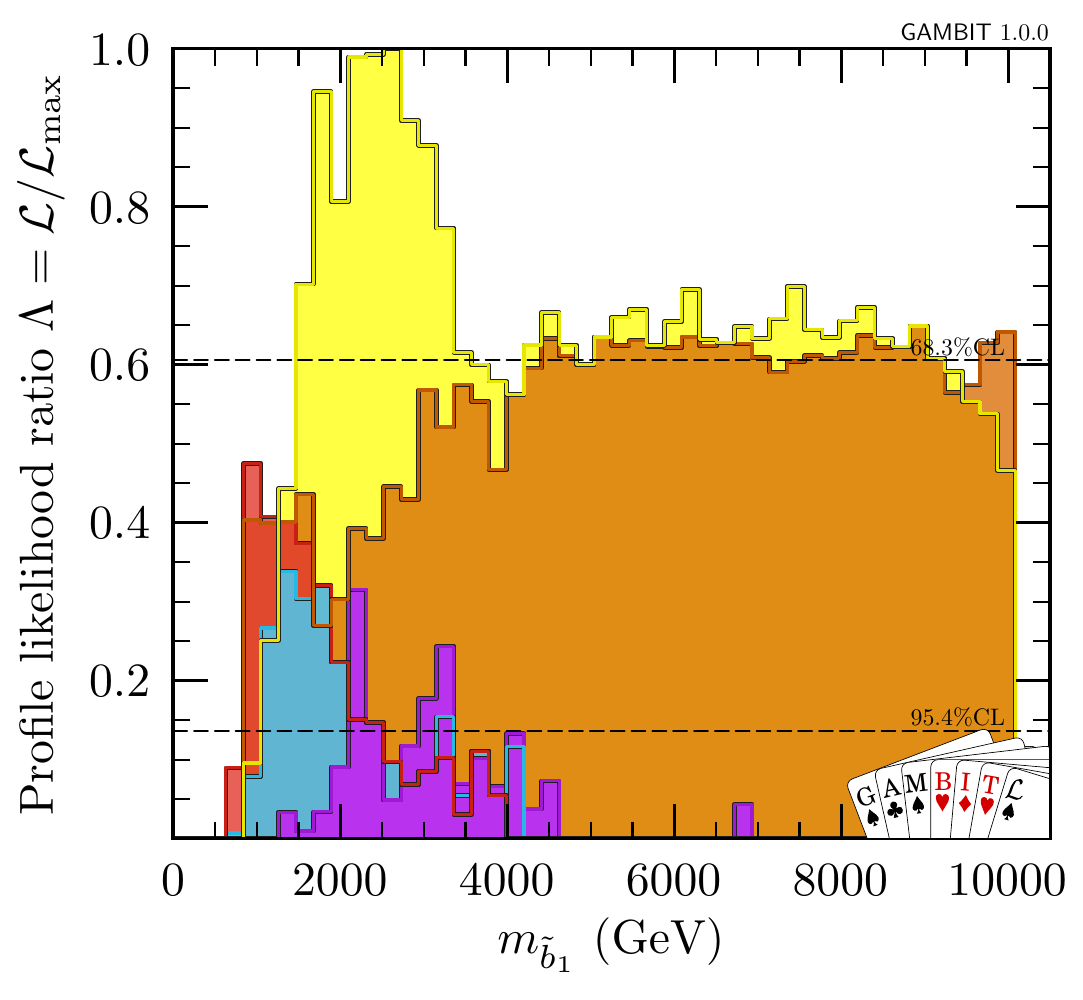}
  \includegraphics[width=0.32\textwidth]{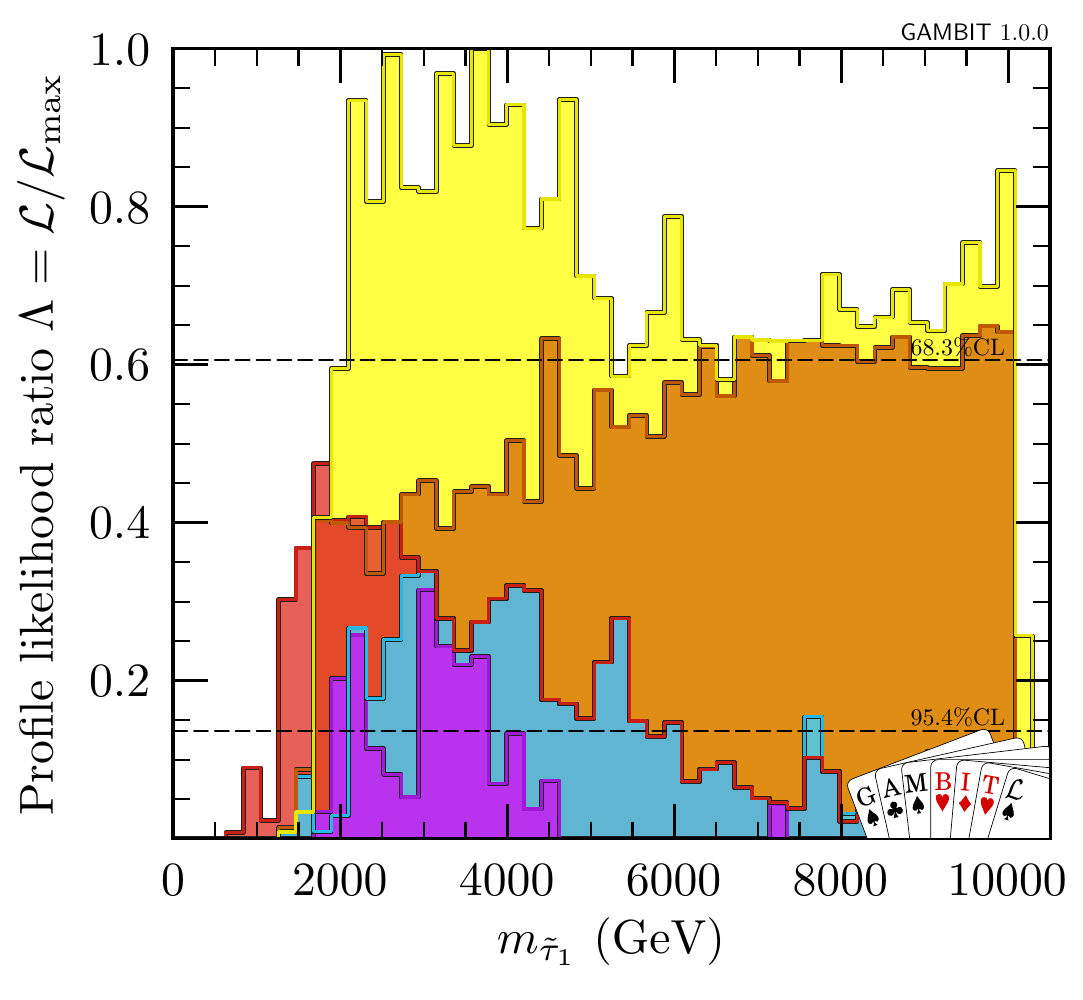}\\
  \includegraphics[height=4mm]{figures/rdcolours5.pdf}
  \caption{1D profile likelihood ratios for the masses $m_{\tilde{\chi}_1^0}$, $m_{\tilde{\chi}_1^+}$, $m_{\tilde{g}}$, $m_{\tilde{t}_1}$, $m_{\tilde{b}_1}$ and $m_{\tilde{\tau}_1}$. We show separate distributions for each mechanism that allows the models to obey the relic density constraint.}
  \label{fig:1d_like_observables}
\end{figure*}

In Fig.\ \ref{fig:amu}, we show the profile likelihood for the SUSY contribution $\Delta a_\mu$ to the magnetic moment of the muon, compared with the experimental likelihood function for the observed discrepancy $a_{\mu,\text{obs}} - a_{\mu,\text{SM}} = (28.7 \pm 8.0)\times10^{-10}$. Chargino co-annihilation models give the largest SUSY contributions, as they exhibit lighter charginos than other models.  However, due to the relatively large values preferred for $\msf$, which governs the masses of $\tilde{\mu}$ and $\tilde{\nu}_\mu$, it is essentially impossible to fit $a_\mu$ simultaneously with all other observables even in the chargino co-annihilation region.

Compared to the MSSM10 results discussed in Ref.\ \cite{MasterCodeMSSM10}, we see broadly similar and consistent phenomenology, up to differences expected from the slightly different models being scanned.  Both studies find the light Higgs funnel, chargino co-annihilation and stop/sbottom co-annihilation in essentially the same areas.  As already discussed, we find that the MSSM7 does not permit stau co-annihilation, and we see a preference for larger neutralino and sfermion masses than Ref.\ \cite{MasterCodeMSSM10}, a consequence of the unified gaugino and sfermion mass parameters in the MSSM7 and our inclusion of constraints from Run II of the LHC.  We also see stop/sbottom co-annihilation extend to higher masses than in Ref.\ \cite{MasterCodeMSSM10}, reflecting either a lower likelihood for such models relative to the best fit in the MSSM10 than in the MSSM7, or improved sampling in the current paper.  Unlike in the MSSM10, we find that it is not possible to consistently explain $a_\mu$ in the MSSM7.

\section{Future prospects}
\label{sec:future}

\begin{figure*}[tbh]
  \centering
  \includegraphics[height=0.3\textwidth]{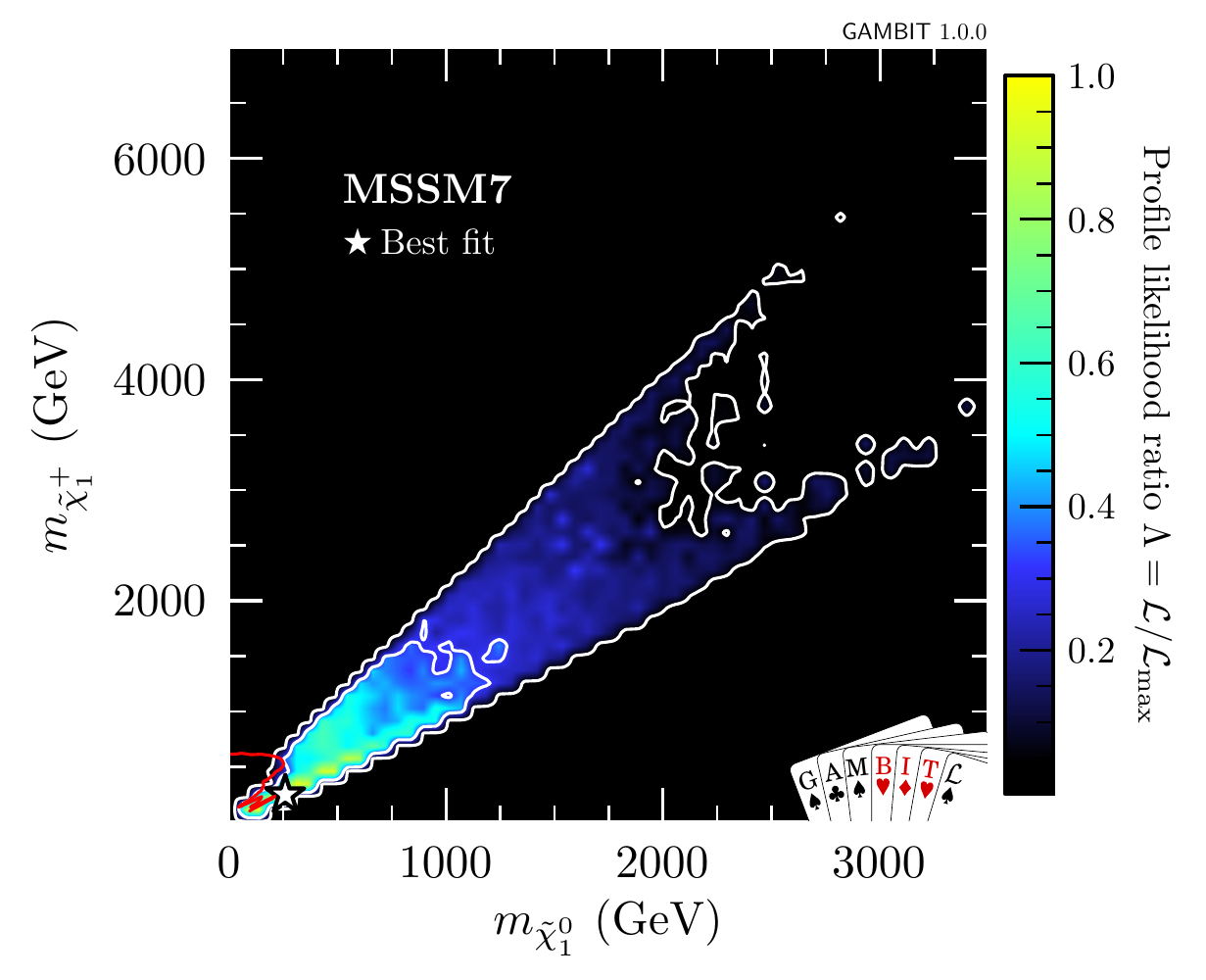}%
  \includegraphics[height=0.3\textwidth, clip=true, trim=0 0 50 0]{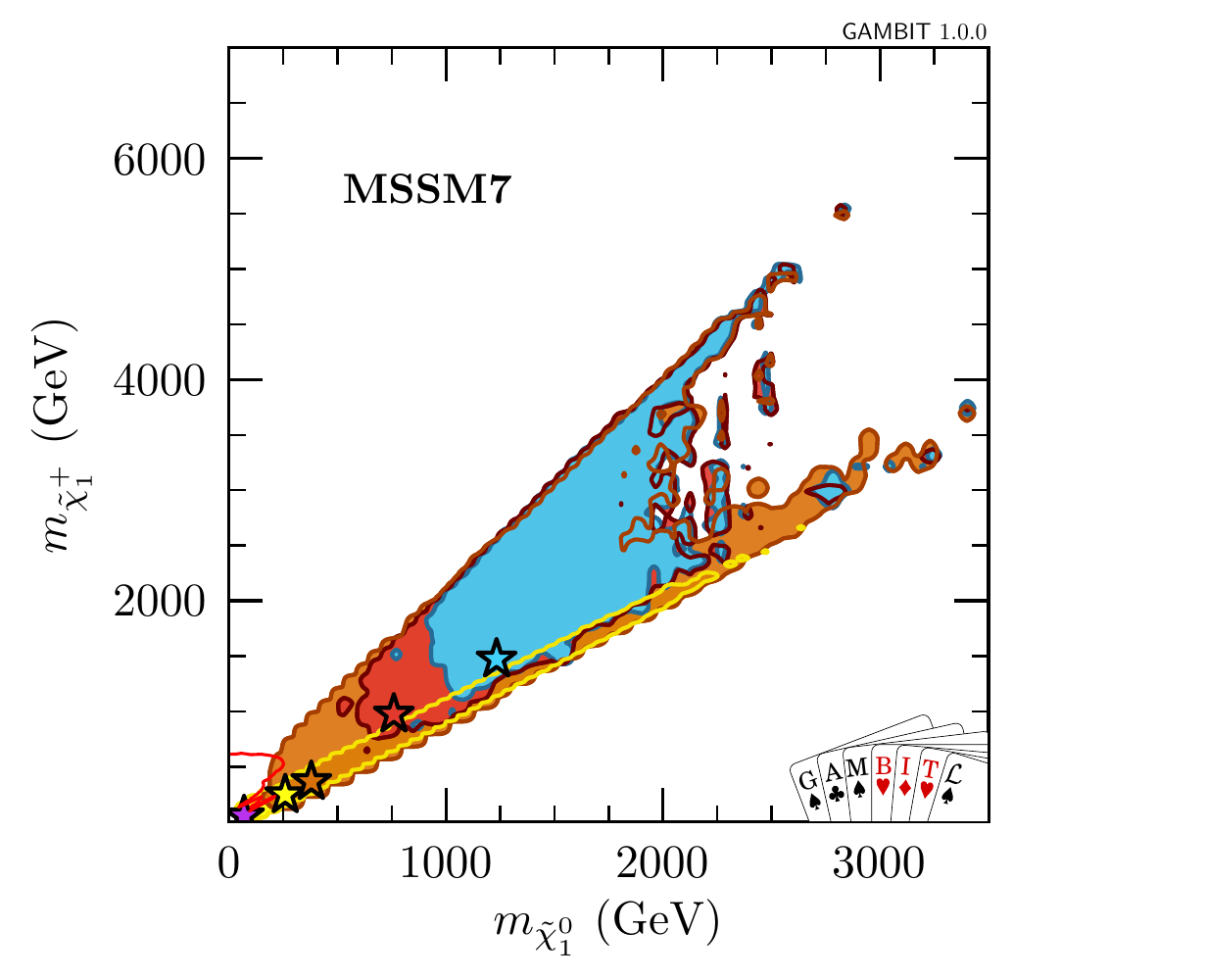}%
  \includegraphics[height=0.3\textwidth, clip=true, trim=0 0 50 0]{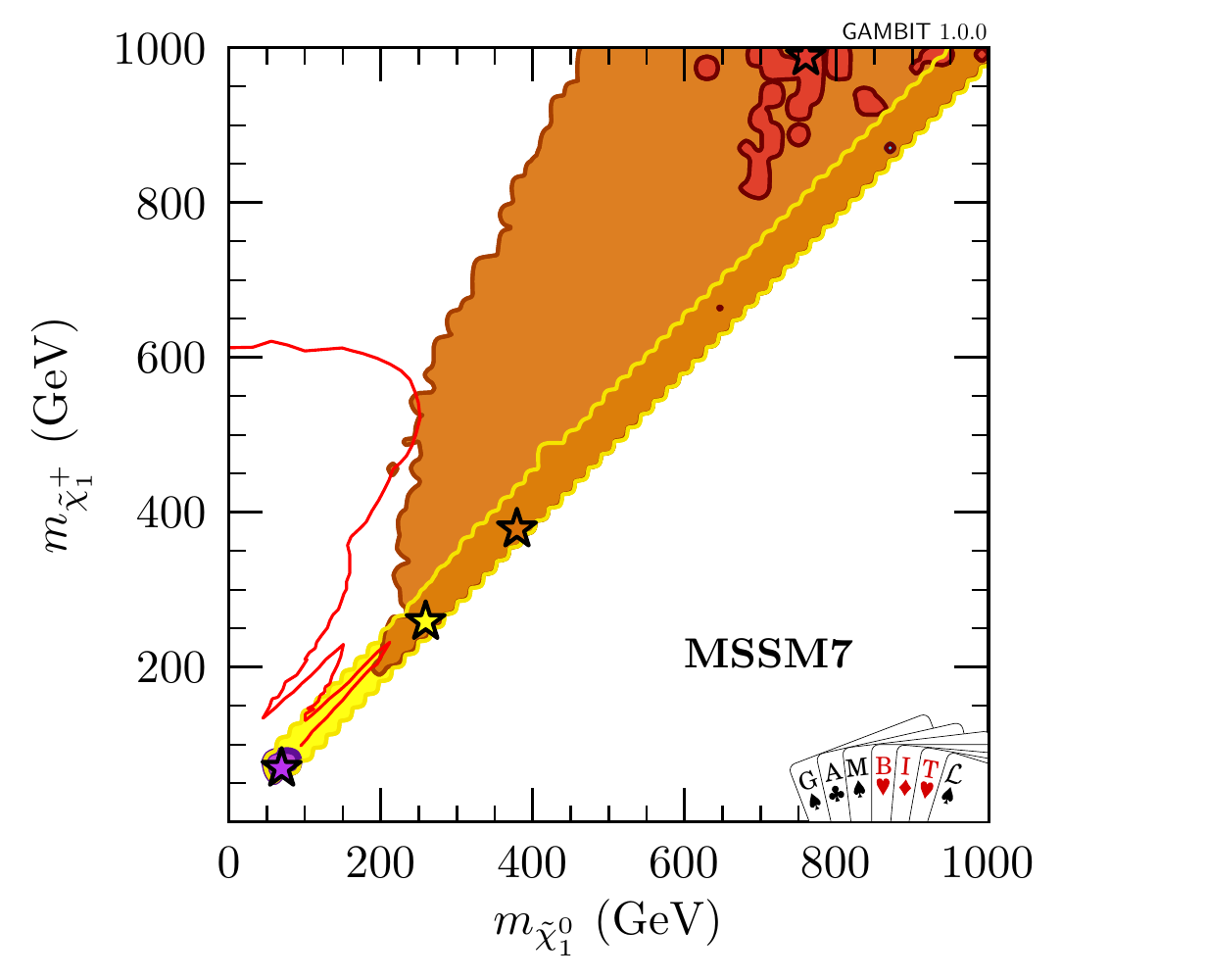}\\
  \includegraphics[height=4mm]{figures/rdcolours5.pdf}
  \caption{\textit{Left:} Profile likelihood in the $\tilde{\chi}^\pm_1-\tilde{\chi}^0_1$ mass plane.  \textit{Centre:}  Sub-regions within the 95\% CL profile likelihood region, coloured according to mechanisms by which the relic density constraint is satisfied. The regions shown correspond to neutralino co-annihilation with charginos, stops or sbottoms, and resonant annihilation through the light or heavy Higgs funnels.  Superimposed in red is the latest CMS Run II simplified model limit for  $\tilde{\chi}^\pm_1 \tilde{\chi}^0_1$ production and decay with decoupled sleptons~\cite{CMSEWSummary}. This limit should be interpreted with caution (see main text for details). \textit{Right:} The same information as the central plot, but zoomed into the low-mass region. Note that, although the CMS limit appears to have excluded part of the chargino co-annihilation region, this is a binning effect. One should instead refer to the plot of the $\tilde{\chi}^\pm_1-\tilde{\chi}^0_1$ mass difference in Fig.~\ref{fig:1d_like_observables}, which provides finer resolution on the mass difference in this region.}
  \label{fig:chargino_masses}
\end{figure*}

\begin{figure*}[tbh]
  \centering
  \includegraphics[width=0.49\textwidth]{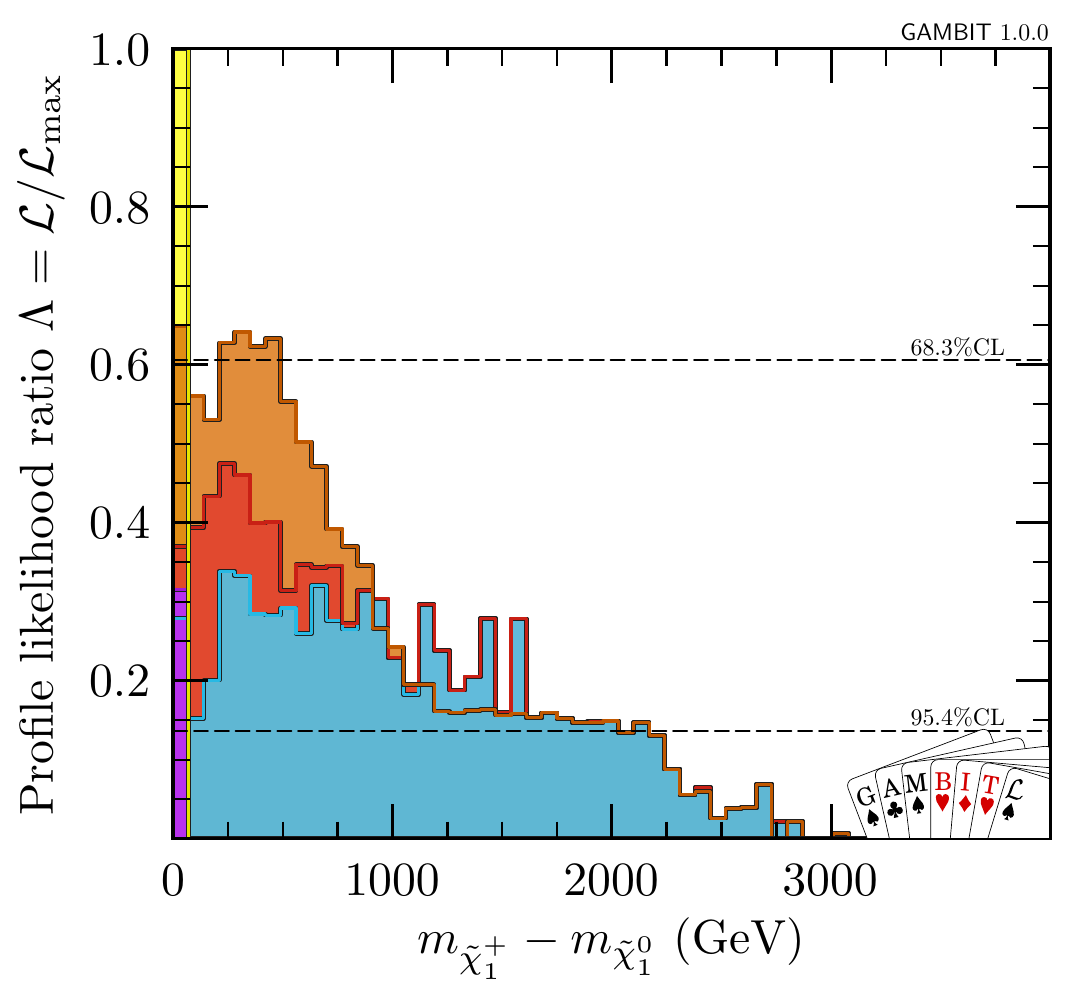}
  \includegraphics[width=0.49\textwidth]{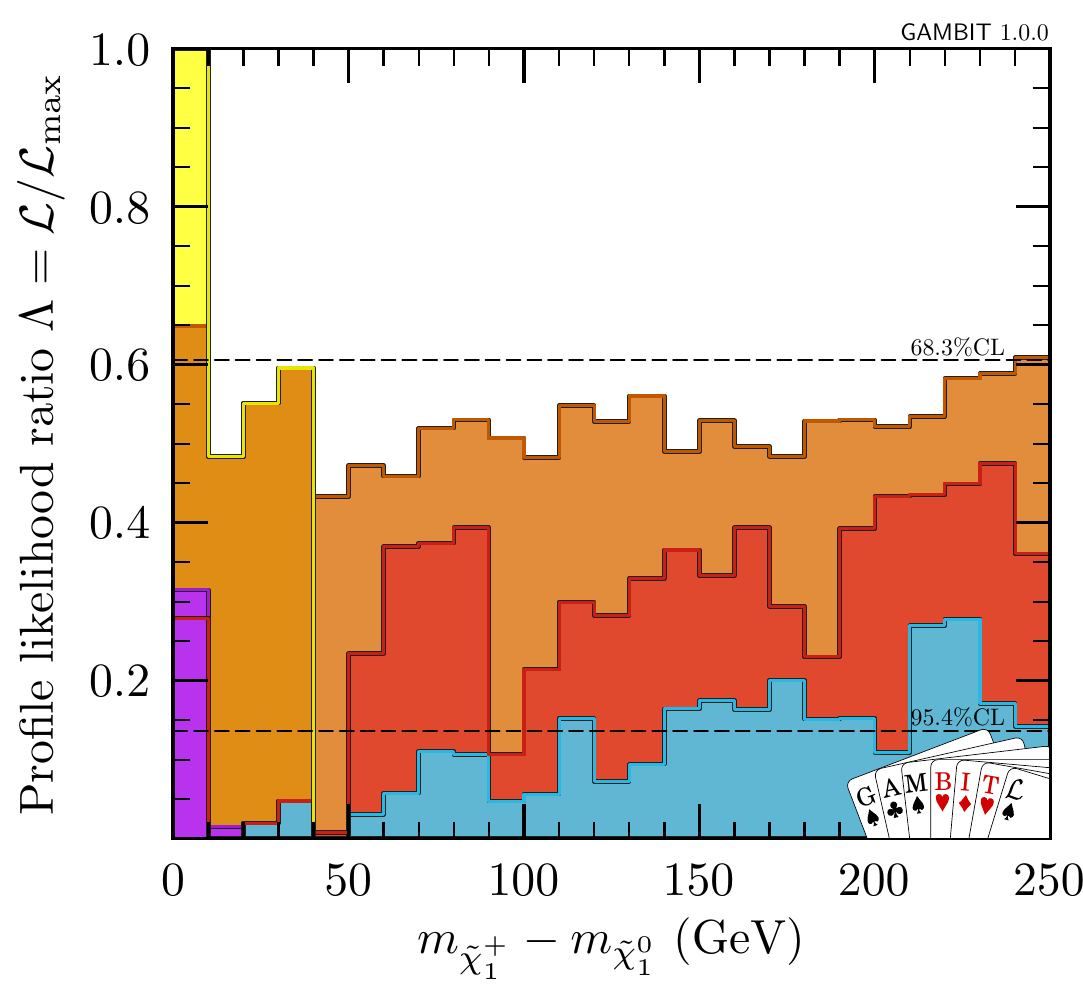}\\
  \includegraphics[width=0.49\textwidth]{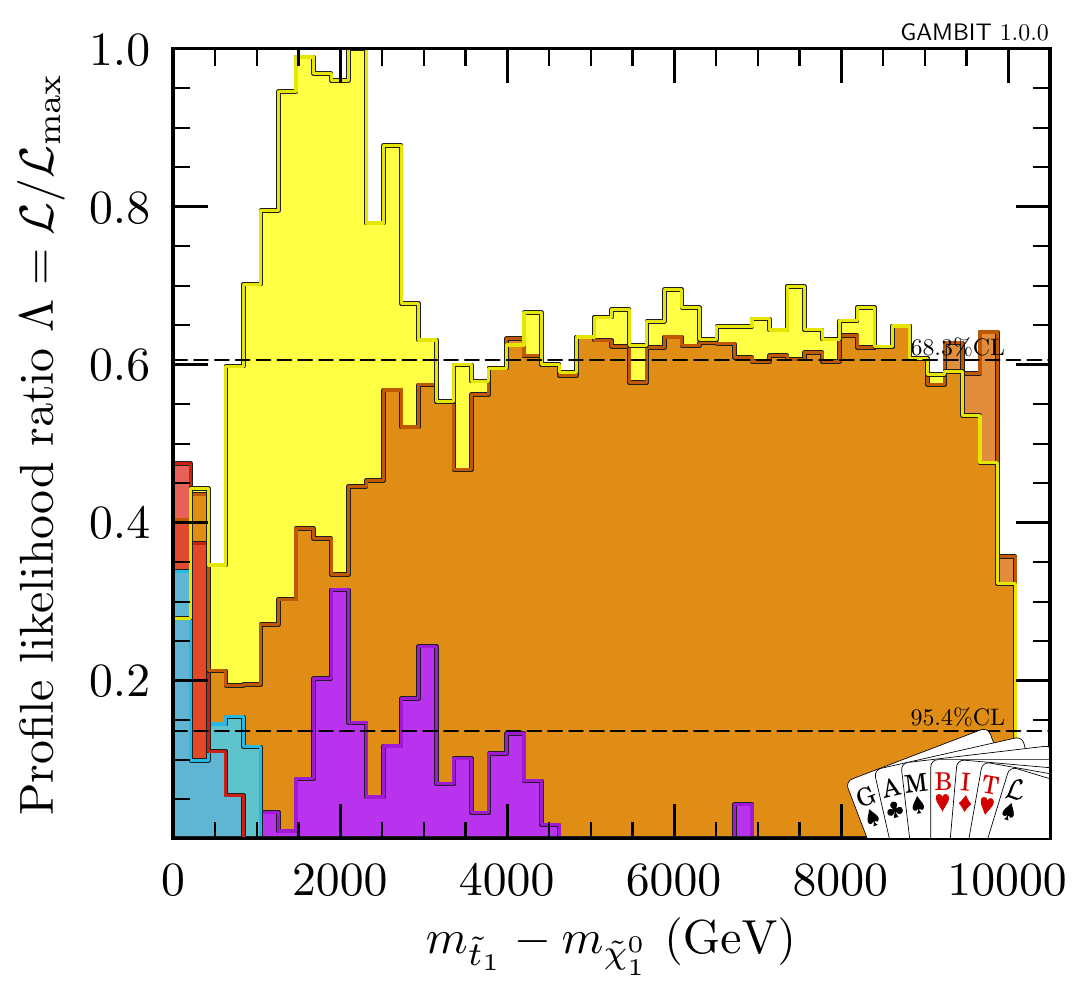}
  \includegraphics[width=0.49\textwidth]{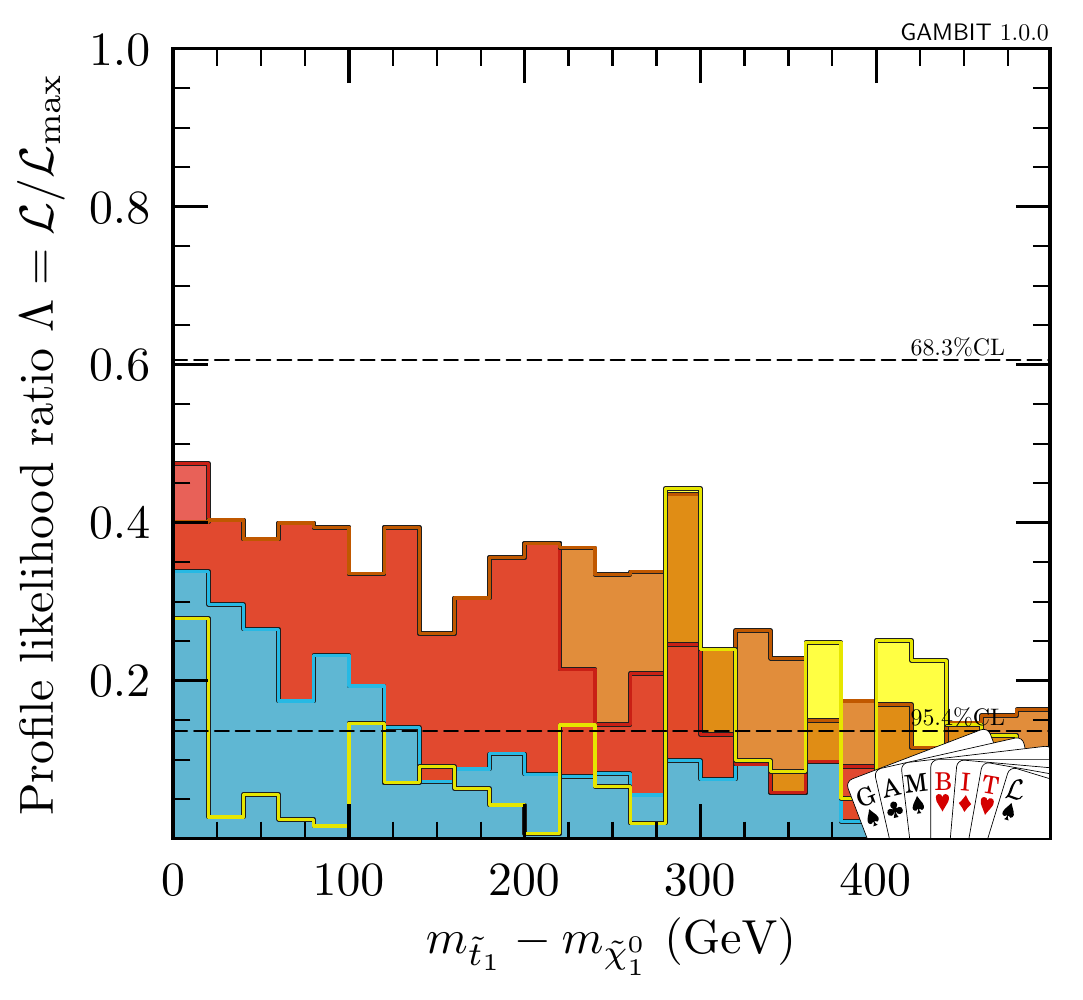}\\
  \includegraphics[height=4mm]{figures/rdcolours5.pdf}
  \caption{1D profile likelihood ratios for the $\tilde{\chi}^\pm_1-\tilde{\chi}^0_1$ mass difference (top) and the $\tilde{t}_1-\tilde{\chi}^0_1$ mass difference (bottom). \textit{Left:} separate distributions for each mechanism allowing models to obey the relic density constraint.  The regions correspond to neutralino co-annihilation with charginos, stops or sbottoms, and resonant annihilation through the light or heavy Higgs funnels. \textit{Right:} as per the left, but zoomed in to small mass differences.}
  \label{fig:1d_like_mass_differences}
\end{figure*}

\begin{figure*}[tbh]
  \centering
  \includegraphics[height=0.3\textwidth]{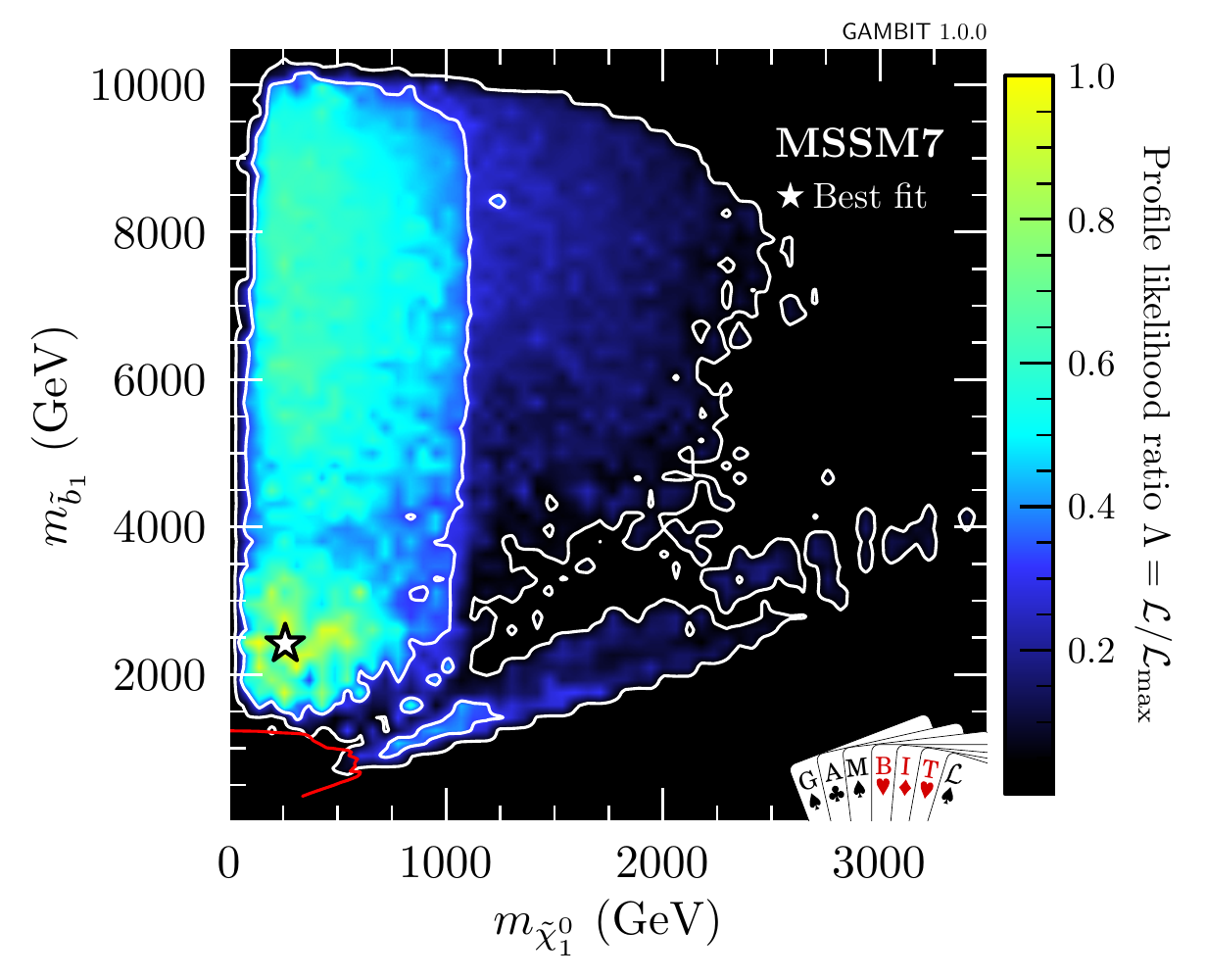}%
  \includegraphics[height=0.3\textwidth, clip=true, trim=0 0 50 0]{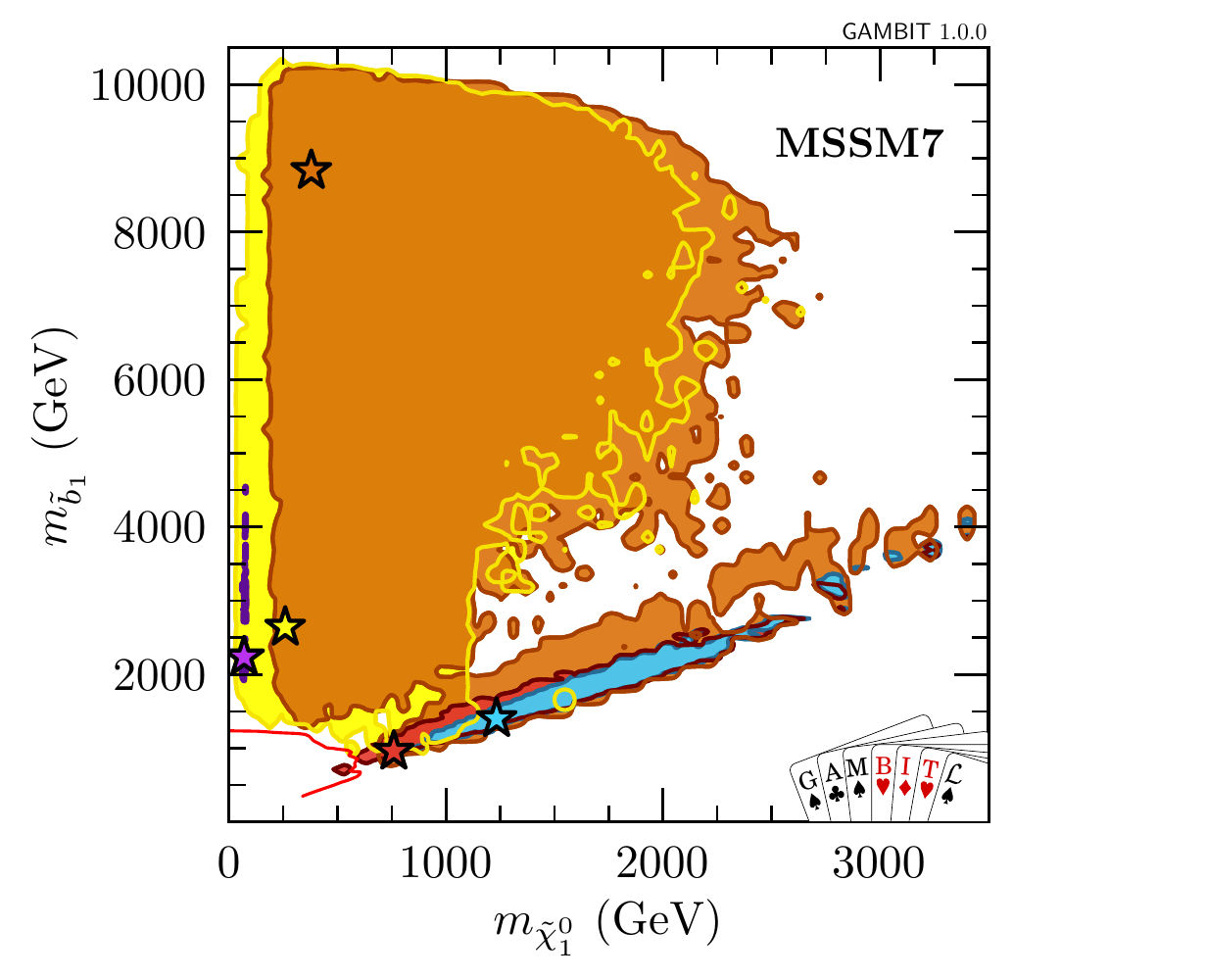}%
  \includegraphics[height=0.3\textwidth, clip=true, trim=0 0 50 0]{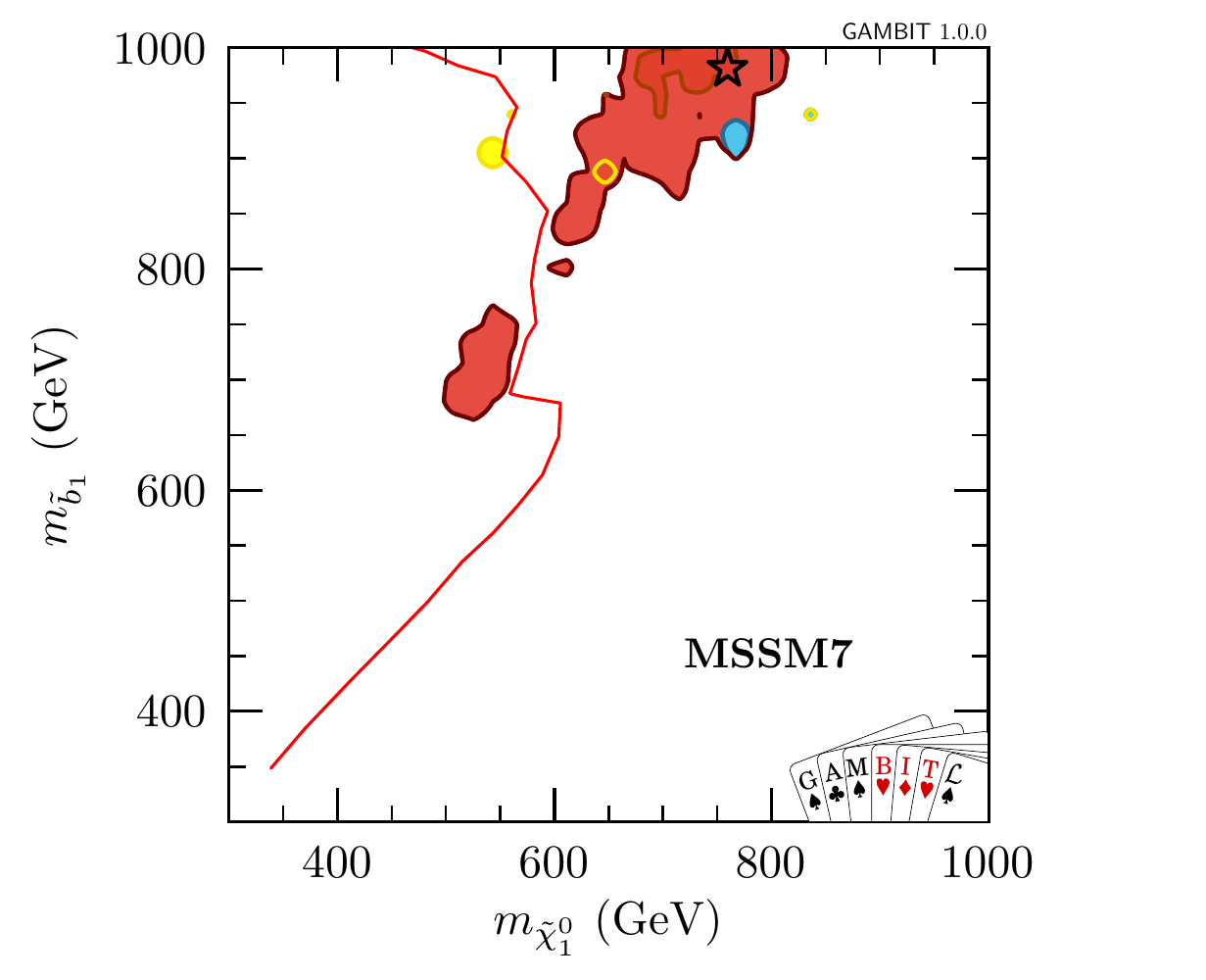}\\
  \includegraphics[height=4mm]{figures/rdcolours5.pdf}
  \caption{\textit{Left:} The profile likelihood ratio in the  $\tilde{b}_1-\tilde{\chi}^0_1$ mass plane. \textit{Centre:} Colour-coding shows mechanism(s) that allow models within the 95\% CL region to avoid exceeding the observed relic density of DM. The regions shown correspond to neutralino co-annihilation with charginos, stops or sbottoms, and resonant annihilation through the light or heavy Higgs funnels. \textit{Right:} The same information as the central plot, zoomed into the low-mass region.}
  \label{fig:sbottom_masses}
\end{figure*}

\begin{figure*}[tbh]
  \centering
  \includegraphics[height=0.3\textwidth]{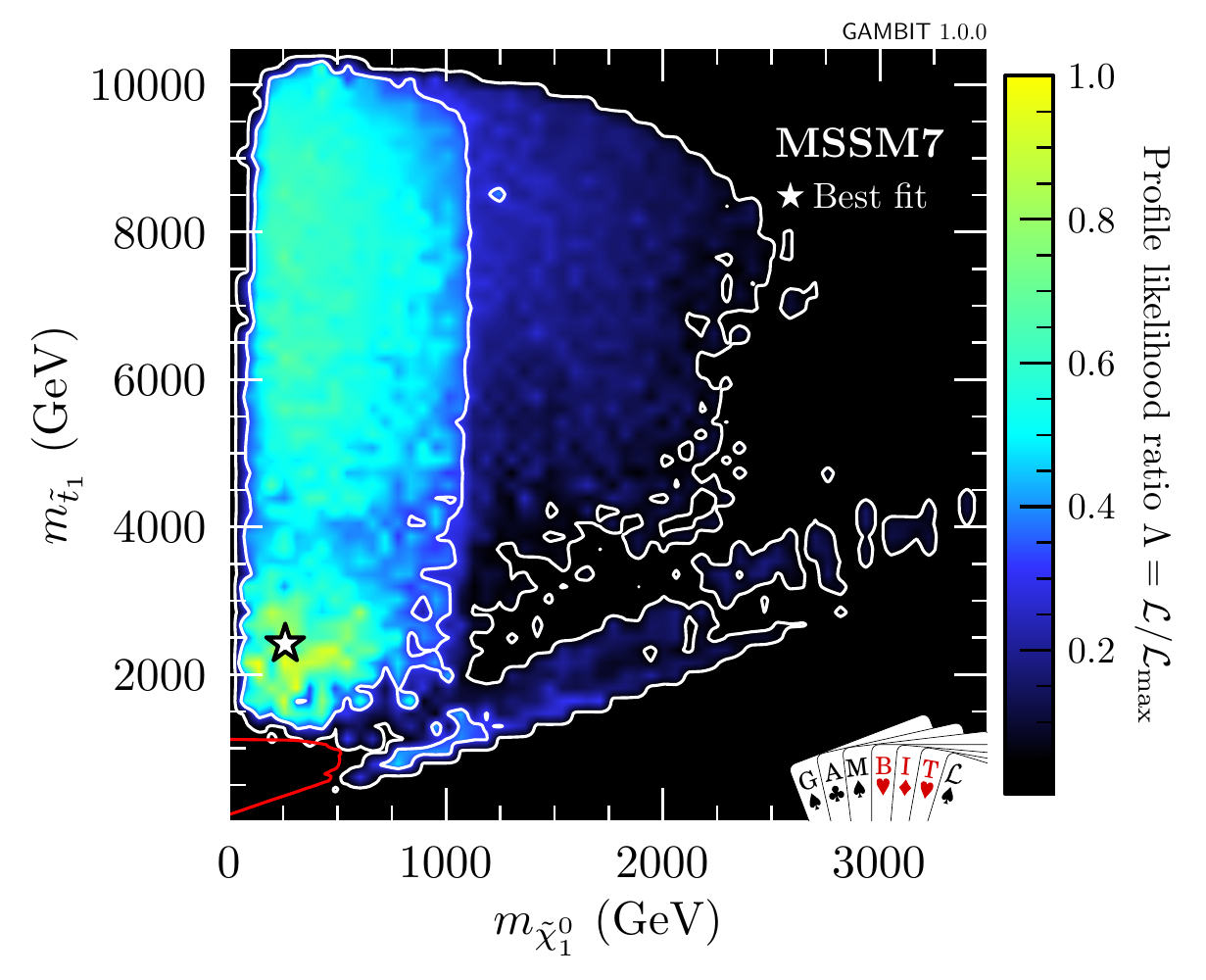}%
  \includegraphics[height=0.3\textwidth, clip=true, trim=0 0 50 0]{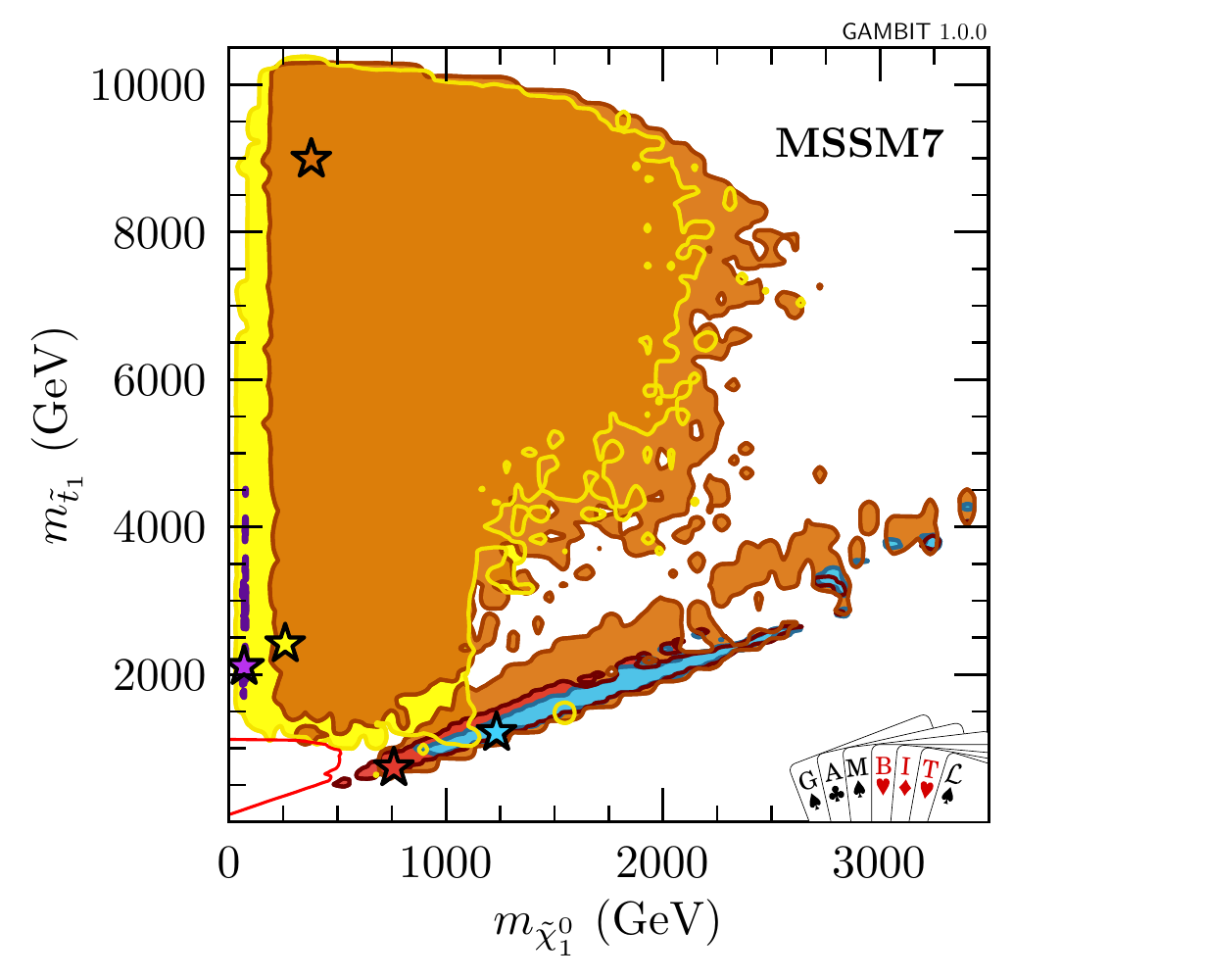}%
  \includegraphics[height=0.3\textwidth, clip=true, trim=0 0 50 0]{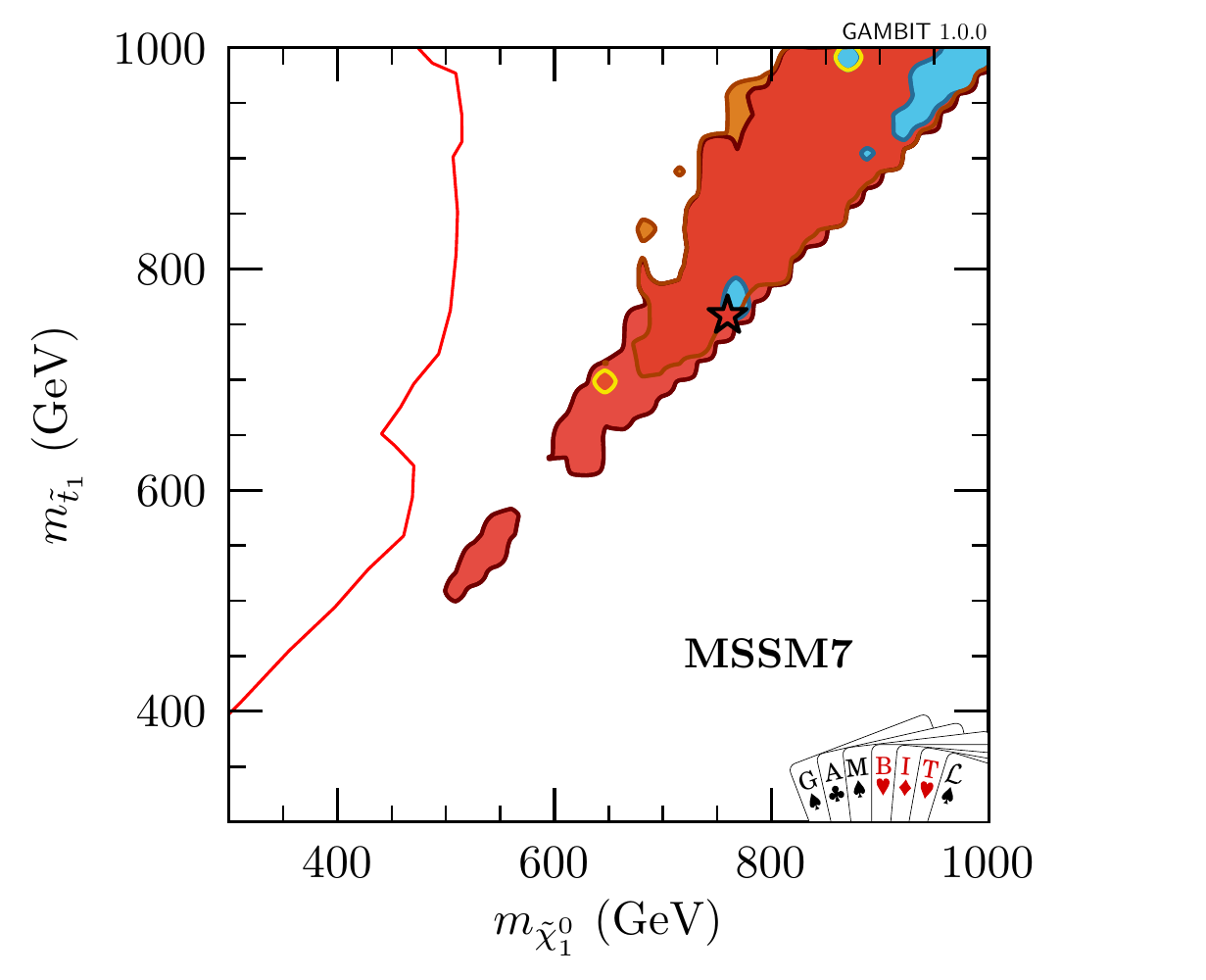}\\
  \includegraphics[height=4mm]{figures/rdcolours5.pdf}
  \caption{\textit{Left:} The profile likelihood ratio in the  $\tilde{t}_1-\tilde{\chi}^0_1$ mass plane. \textit{Centre:} Colour-coding shows mechanism(s) that allow models within the 95\% CL region to avoid exceeding the observed relic density of DM. The regions shown correspond to neutralino co-annihilation with charginos, stops or sbottoms, and resonant annihilation through the light or heavy Higgs funnels. Superimposed in red is the latest CMS Run II simplified model limit for stop pair production~\cite{CMSStopSummary}. \textit{Right:} The same information as the central plot, zoomed into the low-mass region.}
  \label{fig:stop_masses}
\end{figure*}

\subsection{LHC}
\label{sec:LHC}

In Fig.~\ref{fig:1d_like_observables} we show the 1D profile likelihoods for the masses of $\tilde{\chi}_1^0$, $\tilde{\chi}_1^\pm$, $\tilde{g}$, $\tilde{t}_1$, $\tilde{b}_1$ and $\tilde{\tau}_1$.
The $2\sigma$ preferred region for the gluino mass extends upwards from $\sim2$ \TeV, which is on the border of exclusion by current LHC searches for 0-lepton final states, to $\sim20$ \TeV, well beyond the reach of the LHC. Similarly for $m_{\tilde{\tau}_1}$, where the small, weak production cross-section ensures that the predicted mass range is currently unobservable at the LHC.

More interesting are the $\tilde{\chi}_1^0$ and $\tilde{\chi}_1^\pm$ profile likelihoods, which are both peaked at low values. Given that these are naively within range of both LEP and the LHC Run I analyses, it is worth examining the properties of these low mass points in detail. Fig.~\ref{fig:chargino_masses} shows our profile likelihood function in the $\tilde{\chi}^\pm_1$--$\tilde{\chi}^0_1$ mass plane, zoomed into the low-mass region, along with colour-coding indicating which mechanisms help to satisfy the relic density constraint. For the part of our $2\sigma$ region with $m_{\tilde{\chi}^\pm_1} \lesssim 275$~\GeV, an acceptable relic density is mostly generated via chargino co-annihilation, leading to very degenerate $\tilde{\chi}^\pm_1$ and $\tilde{\chi}^0_1$ masses. This explains the lack of exclusion by the LEP and LHC analyses included in our scan (which lose sensitivity for compressed spectra). Notably, our more careful treatment of the LEP limits than in previous studies has allowed models within the naive LEP reach to emerge unscathed.

One might wonder if other LHC analyses will soon (or have already) probed this low-mass region. The most recent EW gaugino limits are from CMS~\cite{CMS-PAS-SUS-16-039,CMS-PAS-SUS-16-043,CMS-PAS-SUS-16-034,CMS-PAS-SUS-16-048}, using 36 fb$^{-1}$ of 13 TeV data. A detailed study of the impact of these results would require the addition of the relevant analyses to the \colliderbit module, and the calculation of a complete likelihood similar to the equivalent Run I analyses already included in \colliderbit. In the present case, however, we can already obtain some insight from a more basic analysis of the simplified model limits presented by the CMS Collaboration. CMS interpreted their results for each final state in a range of simplified models of chargino and neutralino production, in which they set the branching fractions for specific decays to $100\%$, fixed the gaugino content, and set a 95\% CL exclusion limit in the $\tilde{\chi}^\pm_1$--$\tilde{\chi}^0_1$ mass plane. Fig.~\ref{fig:1d_like_observables} demonstrates that the sleptons are heavy across our entire preferred $2\sigma$ region, which is a natural consequence of having a unified sfermion mass in our parameterisation of the MSSM7. At least one stop mass must be high to induce large radiative corrections to the Higgs mass, which has the effect of dragging up the sfermion mass scale. In addition, and as mentioned previously, the $\tilde{\tau}_1$ will typically be heavier than the $\tilde{t}_1$ and $\tilde{b}_1$ due to the smaller Yukawa coupling. Thus, the relevant CMS simplified models are those featuring decoupled sleptons~\cite{CMSEWSummary}. We caution that these limits do not apply in general, and do not directly translate to limits on our model points without a detailed check that the neutralino and chargino mixing matrices and decay branching fractions match the CMS assumptions. One can, however, treat the CMS limits as the most optimistic possible exclusion in the $\tilde{\chi}^\pm_1$--$\tilde{\chi}^0_1$ plane, to obtain a rough guide to the CMS sensitivity.

Proceeding in this spirit, we see that the current CMS limits just barely touch our $2\sigma$ contour in regions where the spectrum is not compressed (Fig.\ \ref{fig:chargino_masses}). Indeed, the highest likelihood region looks to be out of reach in the near future. Note that if the GUT-inspired constraint on $M_2$ is relaxed, more solutions would fall within the CMS exclusion limit, so these searches will be important for global fits with more parameters. For compressed spectra, the details are less clear, as the ability of the CMS soft dilepton search to exclude the lightest models depends crucially on the precise $\tilde{\chi}^\pm_1$--$\tilde{\chi}^0_1$ mass splitting. This is shown in the top of Fig.~\ref{fig:1d_like_mass_differences}, where it is apparent that the chargino co-anihilation points appear as a peak in the likelihood at $\tilde{\chi}^\pm_1$--$\tilde{\chi}^0_1$ mass differences of less than 10\,GeV. This is too small to be probed by the recent CMS results. The chargino co-annihilation region remains free from LHC exclusion, assuming prompt decays of the chargino. We note, however, that for very small mass differences (approaching the pion mass), long-lived particle searches might provide additional constraints.  We defer a detailed analysis of these to future work.

We now look at whether it is possible to probe the squark sector of the MSSM7 at the LHC in the near future.  The lightest squarks are the $\tilde{t}_1$ and $\tilde{b}_1$. Fig.~\ref{fig:1d_like_observables} shows that the peak of the sbottom profile likelihood lies out of reach of the LHC in the near future, and that masses below $\sim$800\,GeV are disfavoured at the 2$\sigma$ level.
Fig.~\ref{fig:sbottom_masses} shows the $\tilde{b}_1-\tilde{\chi}^0_1$ mass plane, revealing that the lower sbottom masses are associated with a small $\tilde{b}_1-\tilde{\chi}^0_1$ mass difference.
This arises from the fact that stop and/or sbottom co-annihilation often account for the generation of an acceptable relic density in this low-mass region. However, there are also low-mass regions in which resonant $A/H$ annihilation or chargino co-annihilation contribute to DM annihilation, giving a wider range of mass differences. As above, comparison with recent CMS simplified model limits provides some insights into the ability of the LHC to probe these models in the near future. A variety of CMS searches for sbottom production have been interpreted in the context of a simplified model of sbottom pair production and decay to a bottom quark and the lightest neutralino \cite{CMS-PAS-SUS-16-032,Sirunyan:2017cwe,CMS-PAS-SUS-16-036}. We again treat these limits as a rough guide to the most favourable possible LHC exclusion potential, and compare our results to the CMS summary plot given in Reference~\cite{CMSSbottomSummary}. The current analyses have potentially probed a small region of Fig.~\ref{fig:sbottom_masses} (with $\tilde{\chi}^0_1$ masses below 600 GeV and $\tilde{b_1}$ masses below $\approx$ 1 TeV). However, almost our entire $2\sigma$ preferred region remains unconstrained. Directly ruling out sbottom co-annihilation as a viable contributor to an acceptable relic density would require probing compressed spectra in sbottom decays up to a mass of $\sim$4\,TeV, an impossible task at the LHC.  Nonetheless, the fact that current limits are nearing the tip of the stop co-annihilation strip means that discovery prospects even in the next run of the LHC are quite promising (although more so for models that exhibit only stop co-annihilation than those that display both stop and sbottom co-annihilation).

\begin{figure*}[tbh]
  \centering
  \includegraphics[width=0.49\textwidth]{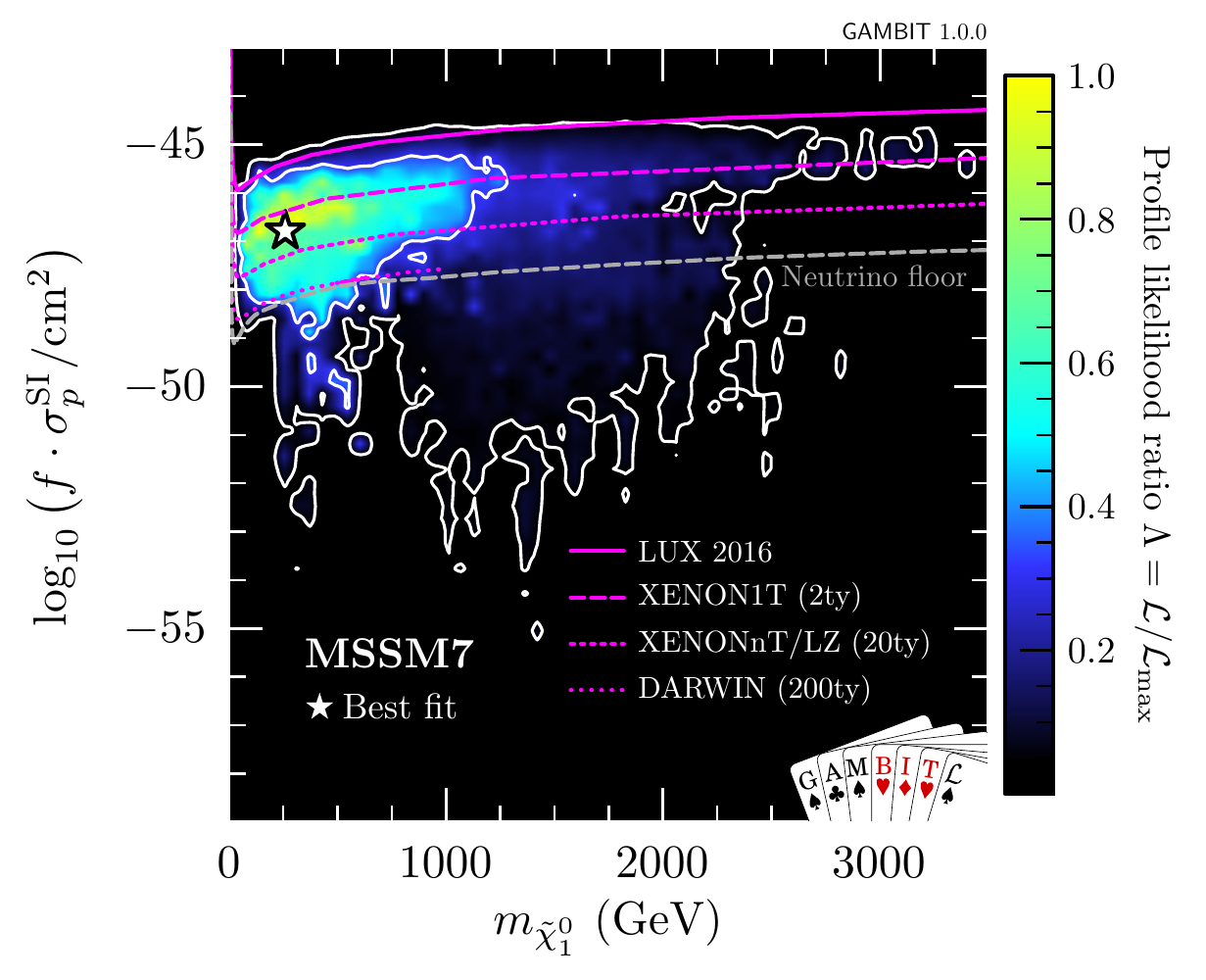}%
  \includegraphics[width=0.49\textwidth]{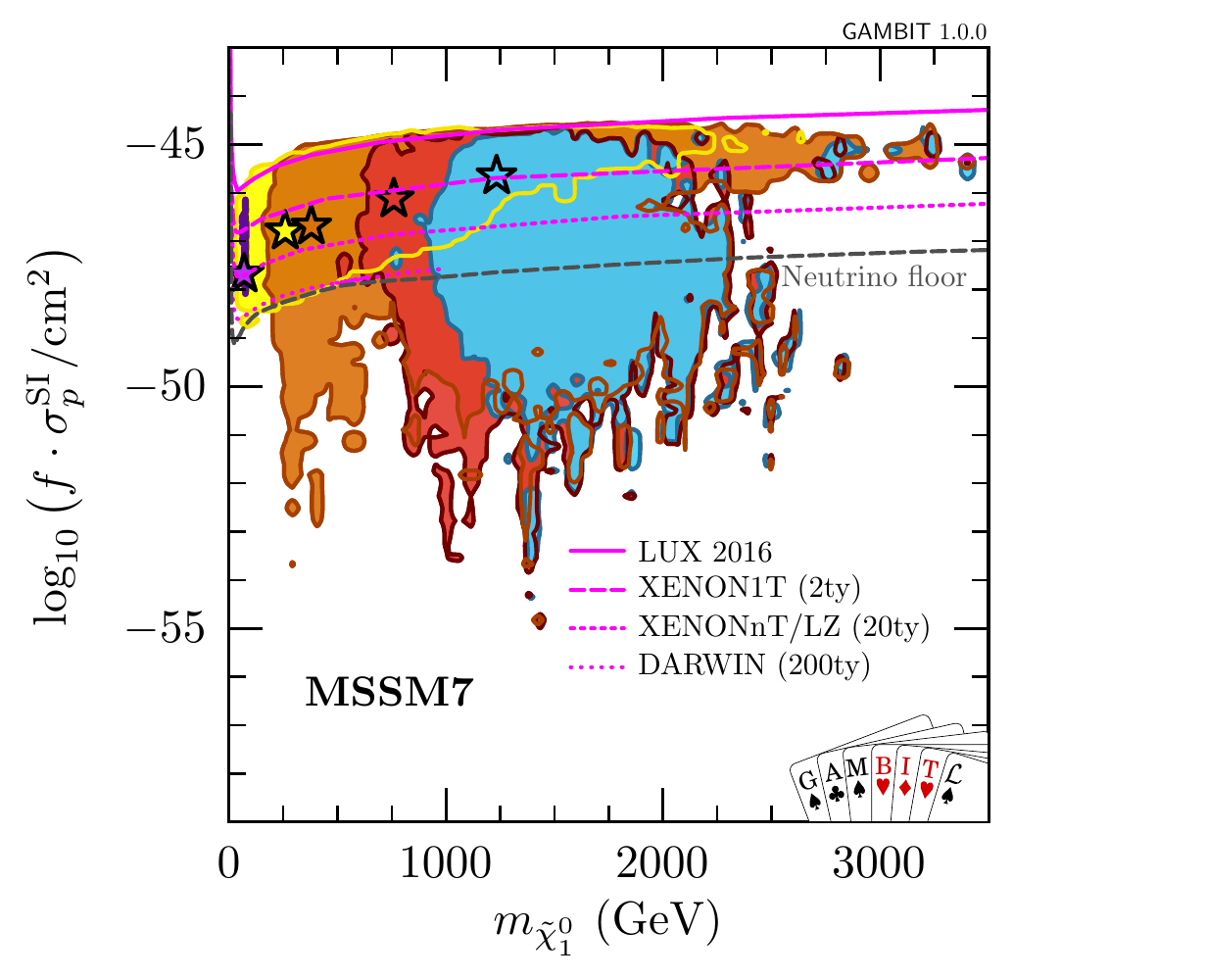}\\
  \includegraphics[height=4mm]{figures/rdcolours5.pdf}
  \caption{Spin-independent neutralino-proton scattering cross-sections in the MSSM7, rescaled by the fraction $f$ of the observed relic density predicted by each model.  \textit{Left:} Profile likelihood, showing 68\% and 95\% CL contours. \textit{Right:} Colour-coding shows mechanism(s) that allow models within the 95\% CL region of the profile likelihood to avoid exceeding the observed relic density of DM, corresponding to neutralino co-annihilation with charginos, stops or sbottoms, and resonant annihilation through the light or heavy Higgs funnels.  Overplotted are 90\% CL constraints from LUX, \cite{LUXrun2}, and projections for the reach of XENON1T after two years of exposure, XENONnT/LZ, assuming 1--3 years of data and an exposure of 20 tonne-years \cite{XENONnTLZ}, and DARWIN, assuming 3--4 years of data and 200 tonne-years of exposure \cite{DARWIN}.
  The dashed grey line indicates the ``neutrino floor'' where background events from coherent neutrino scattering start to limit the experimental sensitivity \cite{Billard:2013qya}. The exact placement of this limit is subject to several caveats; see \cite{Billard:2013qya} for further details.}
  \label{fig:DM_SI}
\end{figure*}

The stop mass has a marginally higher likelihood at lower masses (Fig.~\ref{fig:1d_like_observables}). Fig.~\ref{fig:stop_masses} shows the profile likelihood ratio in the $\tilde{t}_1$--$\tilde{\chi}^0_1$ mass plane, along with colour-coded regions illustrating the relevant relic density mechanisms. As for the sbottom mass, points with a $\tilde{t}_1$ mass below 1\,TeV show a strong mass correlation with the lightest neutralino, as they lie in the stop co-annihilation region. Comparison with the most recent CMS Run II simplified model results~\cite{Sirunyan:2017cwe,CMS-PAS-SUS-16-036,CMS-PAS-SUS-16-049,CMS-PAS-SUS-16-051,CMS-PAS-SUS-17-001} reveals that the lowest-mass points in the stop co-annihilation region remain unprobed, as do the chargino co-annihilation and $A/H$-funnel points. The $\tilde{t}_1-\tilde{\chi}^0_1$ mass difference is shown in the bottom panels of Figure~\ref{fig:1d_like_mass_differences}. Although this is of course small for the stop and sbottom co-annihilation region points, it is not, contrary to the chargino case, sharply peaked at sufficiently low values that decay products can be assumed to be hard to reconstruct at the LHC. This offers hope that LHC searches for compressed spectra (sensitive to smallish mass differences) can eventually tackle these models.

\subsection{Direct detection of dark matter}
\label{sec:DD}

\begin{figure*}[tbh]
  \centering
  \includegraphics[width=0.49\textwidth]{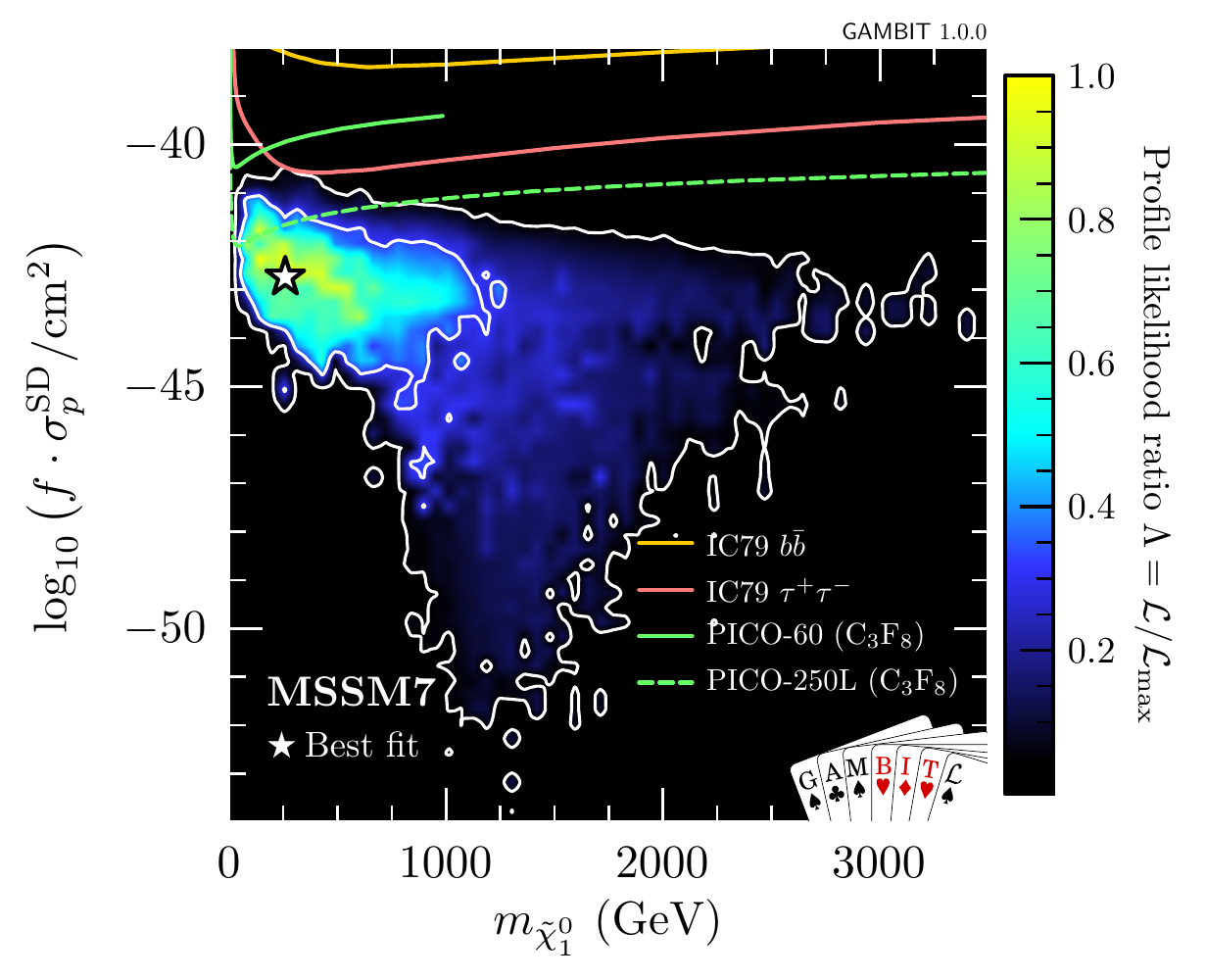}%
  \includegraphics[width=0.49\textwidth]{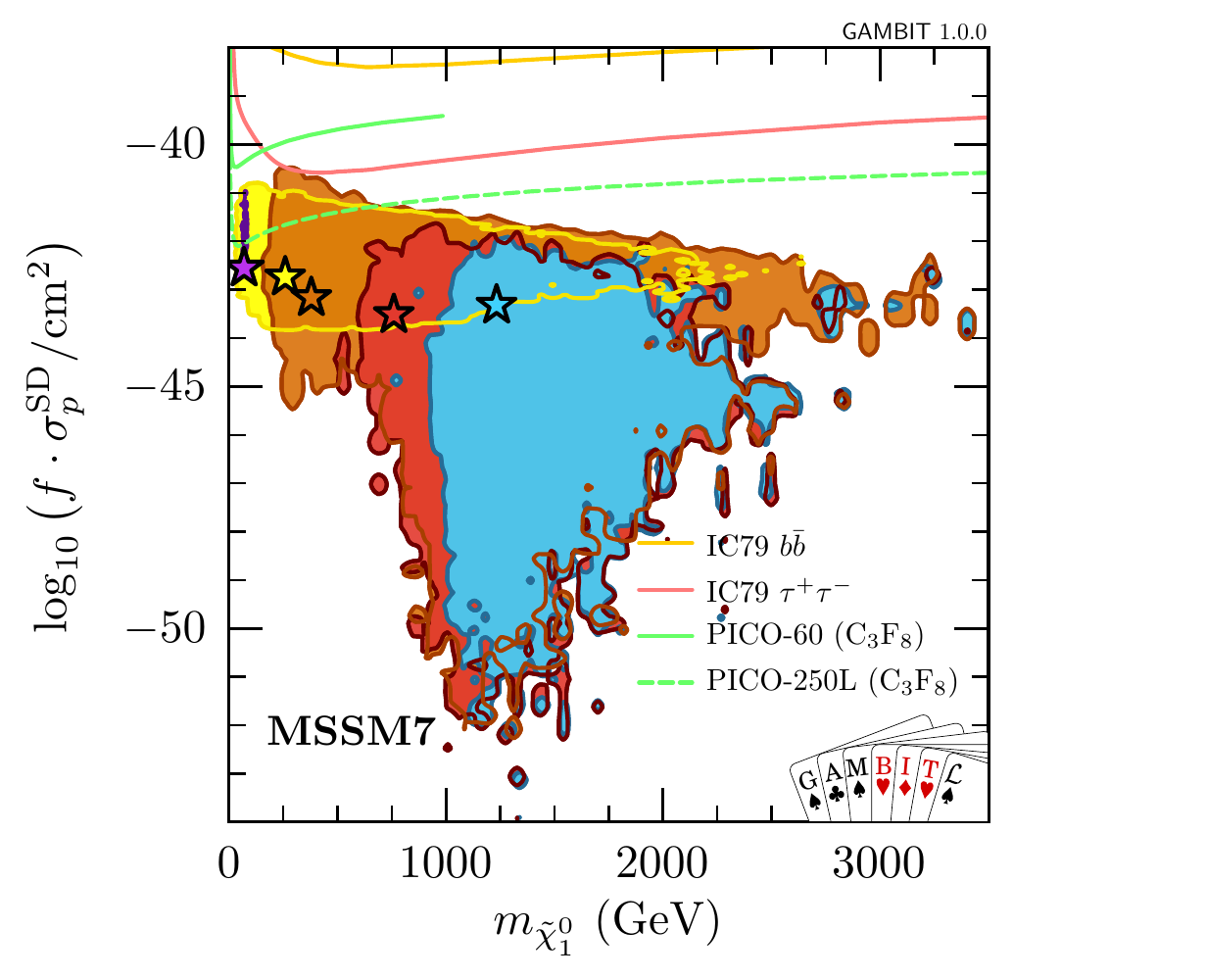}\\
  \includegraphics[height=4mm]{figures/rdcolours5.pdf}
  \caption{Spin-dependent neutralino-proton scattering cross-sections in the MSSM7, rescaled by the fraction $f$ of the observed relic density predicted by each model.  \textit{Left:} Profile likelihood, showing 68\% and 95\% CL contours. \textit{Right:} Mechanism(s) that allow models within the 95\% CL region of the profile likelihood to avoid exceeding the observed relic density of DM.  Overplotted are 90\% CL constraints from IceCube \cite{IC79,IC79_SUSY}, assuming that dark matter annihilates exclusively via the $\bar b b$ or $\tau^+\tau^-$ channel, PICO-60 \cite{PICO60_2}, and projections for the reach of PICO-250 \cite{PICO250}.}
  \label{fig:DM_SD}
\end{figure*}

In this section, we examine the preferred spin-independent (SI; Fig.\ \ref{fig:DM_SI}) and spin-dependent (SD; Fig.\ \ref{fig:DM_SD}) neutralino-proton scattering cross-sections in the MSSM7.  Here we rescale the scattering cross-sections by the ratio $f$ of the predicted to the observed relic density, so as to ease comparison with various experimental limits and projections.  Fig.\ \ref{fig:DM_SI} shows that SI limits from direct detection are already highly constraining, with many models with high likelihoods lying just below the current sensitivity of LUX \cite{LUXrun2}, and very soon to be probed by XENON1T \cite{XENONnTLZ} and its successors.  Eventually, DARWIN \cite{DARWIN} looks set to probe the entirety of the light Higgs funnel and the chargino co-annihilation region, as well as large parts of the heavy funnel and stop/sbottom co-annihilation regions.

In the SD sector, IceCube already constrains mixed gaugino-Higgsino models in the MSSM, as noted in Refs.\ \cite{IC22Methods,Silverwood12,IC79_SUSY}.  PICO \cite{PICO60_2} is not yet competitive for MSSM models, but its future upgrades appear set to make significant inroads into both Higgs funnels and the chargino co-annihilation region.  However, it remains to be seen if XENON1T will probe such models on a shorter timescale.  Future neutrino telescopes such as KM3NeT \cite{Adrian-Martinez:2016fdl} and proposed upgrades to IceCube \cite{Aartsen:2014oha,Aartsen:2014njl} may also offer significantly improved sensitivity to models in the MSSM7, but to date the expected sensitivity to DM masses above 100\,GeV is not known.  Whilst not particularly constraining in terms of SD proton scattering, LUX \cite{Akerib:2017kat} already provides constraints on the SD neutralino-neutron cross-section, which are just beginning to touch on the allowed parameter space of the MSSM7 (not shown, but included in our scans via \ddcalc \cite{DarkBit}).

\begin{figure*}[tbh]
  \centering
  \includegraphics[width=0.49\textwidth]{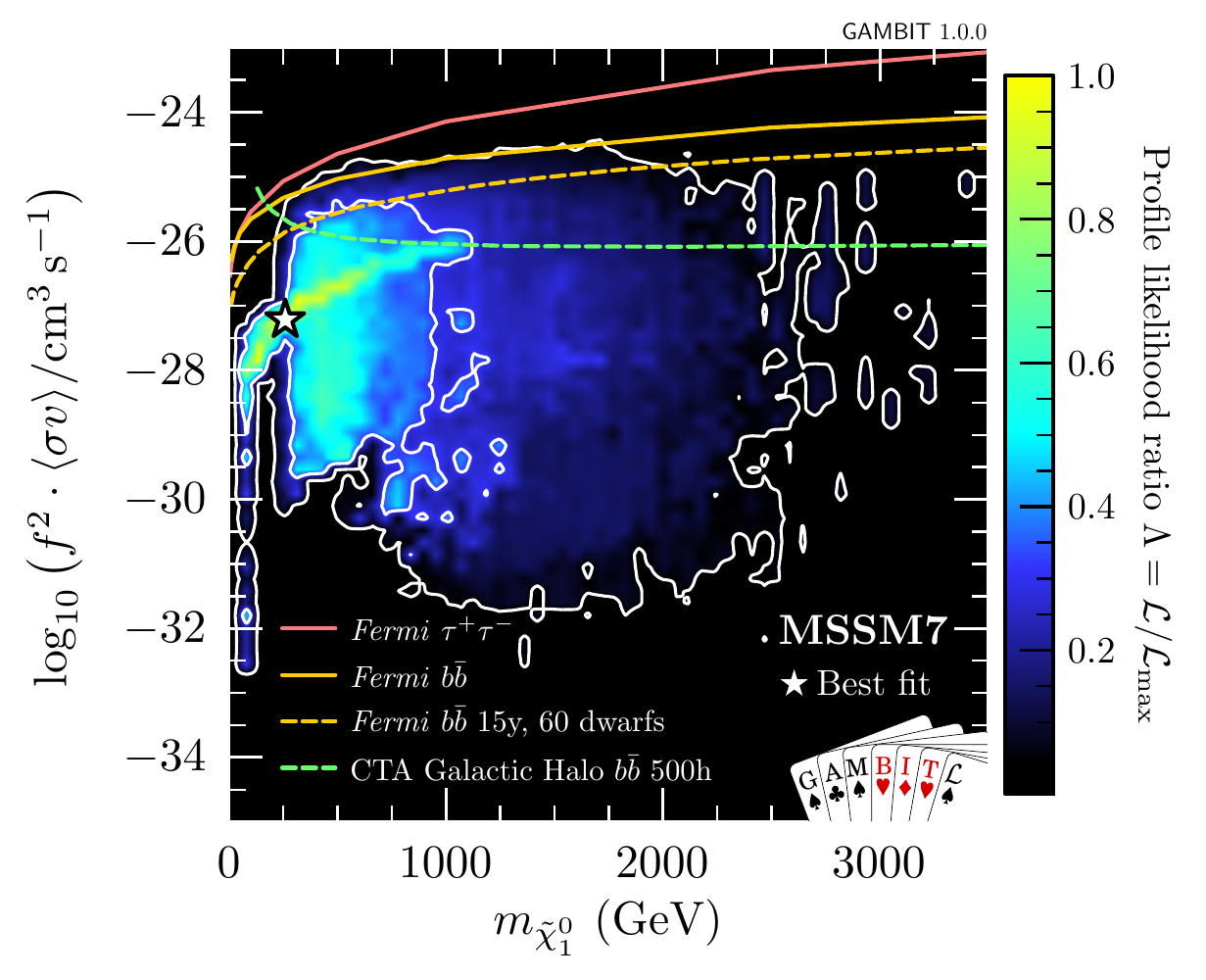}%
  \includegraphics[width=0.49\textwidth]{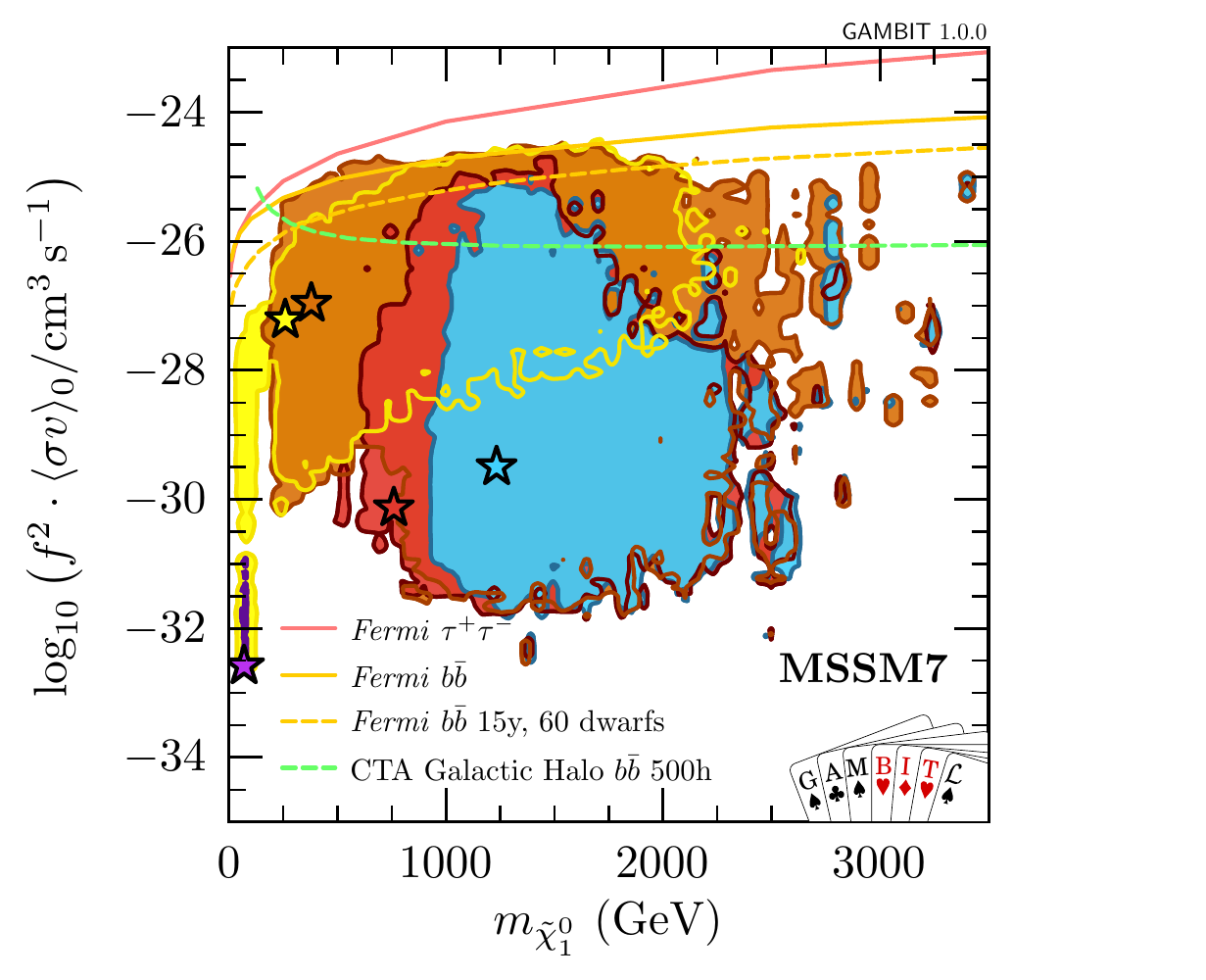}\\
  \includegraphics[height=4mm]{figures/rdcolours5.pdf}
  \caption{Zero-velocity neutralino self-annihilation cross-sections in the MSSM7, rescaled by the square of the fraction $f$ of the observed relic density predicted by each model.  \textit{Left:} Profile likelihood, showing 68\% and 95\% CL contours. \textit{Right:} Mechanism(s) that allow models within the 95\% CL region of the profile likelihood to avoid exceeding the observed relic density of DM.  Overplotted are 95\% CL constraints from the search for dark matter annihilation in 15 dwarf spheroidal galaxies by the \textit{Fermi}-LAT Collaboration \cite{LATdwarfP8}.  These limits are based on 6 years of {\tt Pass 8} data, and are given for two different assumed annihilation final states ($\bar b b$ and $\tau^+\tau^-$).  We also show the projected improvement in the $\bar b b$ channel after 15 years, if the number of known dwarfs were to quadruple in that time \cite{Charles:2016pgz}.  The final curve is the best-case projected sensitivity of the Cherenkov Telescope Array to annihilation in the Galactic halo, computed assuming $\bar b b$ final states, neglecting systematic errors, and assuming 500\,hrs of observation \cite{Carr:2015hta}.}
  \label{fig:DM_ID}
\end{figure*}

Although models exist down to SI and SD cross-sections of $10^{-55}$\,cm$^2$ in the stop/sbottom co-annihilation and $A/H$ funnel regions of our fits, the large cancellations required to produce such cross-sections may be spoilt by loop corrections \cite{Djouadi:2000ck,Djouadi:2001kba}.  This raises hope that future direct detection experiments will discover neutralino DM in the MSSM7 or a similar model.  However, specific investigations in the MSSM7 suggest that this is not necessarily expected for all parameter combinations, so some parts of the parameter space should still be expected to lie well below any future sensitivity, even after applying higher-order corrections \cite{Mandic:2000jz}.

\subsection{Indirect detection of dark matter}
\label{sec:ID}

Let us finally address discovery prospects of the MSSM7 with indirect DM searches. To this end, we show in
Fig.~\ref{fig:DM_ID} the velocity-weighted annihilation cross section, $\sigma v$, in the limit of vanishing relative
velocity of the annihilating DM particles, as a function of the lightest neutralino mass. We rescale this quantity
by the square\footnote{Here we assume that all DM clumps just like neutralinos; see Sec.\ 4.4.3 of Ref.\ \cite{CMSSM} for further discussion.} of the fraction $f$ of the calculated neutralino relic density to the observed DM abundance, thereby
accounting for the possibility of the lightest neutralino making up only a fraction of DM. In the left panel, we show the
profile likelihood, while in the right panel we indicate the mechanism(s) responsible for increasing the
(co-)annihilation rate in the early Universe, and hence decreasing the present neutralino relic density to or below the
observed DM abundance.
For comparison, we also indicate the same current and projected future limits from selected indirect detection
experiments as in Fig.\ 21 of the companion paper \cite{CMSSM}, namely present \textit{Fermi}-LAT \cite{LATdwarfP8}
limits for $\bar b b$ and $\tau^+\tau^-$ final states from observations of 15 dwarf galaxies, projected \textit{Fermi}-LAT
limits for $\bar b b$, and the projected sensitivity of the Chrerenkov Telescope Array (CTA) for $\bar b b$, assuming 500
hours of Galactic halo observations \cite{Carr:2015hta}.

Across almost the entire neutralino mass range, we find models within the 95\% CL region of the profile likelihood that
exhibit present-day annihilation rates above the canonical thermal value of $3\times10^{-26}$\,cm$^3$\,s$^{-1}$.
Those models are a subset of the $A/H$ funnel region, where the pseudoscalar Higgs is almost exactly
twice as heavy as the lightest neutralino, $m_A\simeq 2 \tilde\chi^0_1$.  This leads to resonant enhancement of the annihilation rate as $v\to 0$, as is the case today --- but not in the early Universe, where thermal effects mean that $v\ne0$ in general.  For some models in this part of the parameter space, current \textit{Fermi} limits are already quite constraining.
Projected \textit{Fermi} limits, assuming 15 years of data on 60 dwarf galaxies (vs.~6
years and 15 dwarfs for the current limits), will start to cut into the (current) 68\% CL region.  For neutralino
masses above around 300\,GeV, CTA will be even more sensitive than this. Large parts of the MSSM7 parameter space, however, will be impossible
to probe with any planned indirect detection experiment; this includes, unfortunately, both the global best fit
point of the MSSM7 and the best-fit points of all of the individual parameter regions corresponding to different
mechanisms of lowering the relic density.

We emphasise that even though the CTA limits shown here are rather optimistic, in that they neglect the effect of
systematic uncertainties \cite{Silverwood:2014yza}, the above discussion somewhat underestimates the
prospects of indirect DM searches. One reason is that we have neglected in our discussion other detection channels than
gamma-rays, in particular charged cosmic rays.
As discussed in some more detail in Sec.\ 4.4.3 of the companion paper~\cite{CMSSM}, radiative corrections in particular,
e.g.~\cite{Bergstrom:1997fh, Bringmann:2007nk,Bringmann:2017sko}, as well as Sommerfeld
enhancement \cite{Hisano:2002fk,Hisano:2004ds,Hryczuk:2010zi,Catalan:2015cna}, are further
effects that we have not taken into account here. For parts of the parameter space this leads to increased annihilation rates and/or distinct spectral features, which
are much easier to constrain or identify with experiments than the featureless gamma-ray spectra from the final
states that we have considered here.  A full discussion of these effects, and their impact on indirect DM searches within the MSSM7, is beyond
the scope of this study, although we plan to return to this in future work.

\section{Conclusions}
\label{sec:conc}

We have carried out an extensive global fit of the 7-parameter, weak-scale phenomenological MSSM, using the newly-released \GB global fitting framework.  Our fit takes into account updated experimental data, improved theoretical calculations and more advanced statistical sampling methods than previous studies of similar models.  We have also considered leading uncertainties from the Standard Model, the dark matter halo of the Milky Way, and the quark content of the nucleon, fully scanning over the relevant parameters and profiling them out in the final fit. Finally, we have explored the full range of experimentally-allowed parameters, by allowing neutralinos to constitute any fraction of the observed cosmological dark matter.

The MSSM7 shows quite a rich selection of phenomenology across its parameter space, ranging from Higgsino-dominated dark matter annihilating through co-annihilations with other Higgsinos in the early Universe, to resonant annihilation via the light and heavy Higgs funnels, to co-annihilation of neutralinos with both light stops and sbottoms.  We find a preference for light, Higgsino-dominated neutralinos, with $m_{\tilde\chi_1^0}\lesssim750$\,GeV at 68\% CL and $m_{\tilde\chi_1^0}\lesssim2.5$\,TeV at 95\% CL.  We have shown that stop/sbottom co-annihilation models lie just out of reach of current LHC searches, with stops and sbottoms as light at 500\,GeV.  This makes the prospects for probing at least some such models at the LHC in the near future quite promising.  Both direct and indirect searches for dark matter place significant constraints on the allowed parameter ranges in the MSSM7, and the next generation of these experiments will probe large parts of the highest-likelihood areas of its parameter space.

The current study is essentially a starting point for detailed, modular scans of supersymmetric models defined at the weak scale with \GB.  \GB's hierarchical model database already contains many generalisations of the MSSM7, which would themselves make very interesting targets for global analyses similar to this one.

To ensure reproducibility and encourage further exploration of our results, we provide a set of supplementary data online through \textsf{Zenodo} \cite{the_gambit_collaboration_2017_801640}.  This includes all \GB input files, generated likelihood samples and best-fit benchmarks for this paper.

\begin{acknowledgements}
We thank Andrew Fowlie, Tom\'as Gonzalo, Julia Harz, Sebastian Hoof, Felix Kahlhoefer, James McKay, Roberto Ruiz, Roberto Trotta and Sebastian Wild for useful discussions, and Lucien Boland, Sean Crosby and Goncalo Borges of the Australian Centre of Excellence for Particle Physics at the Terascale for computing assistance and resources.  \gambitacknospmare
\end{acknowledgements}

\bibliography{R1}

\end{document}